\documentclass[a4paper,fleqn,usenatbib]{mnras}

\usepackage{newtxtext,newtxmath}

\usepackage[T1]{fontenc}
\usepackage{ae,aecompl}

\usepackage{graphicx}	
\usepackage{amsmath}	
\usepackage{amssymb}	
\usepackage{multicol}
\usepackage{graphics}
\usepackage{wrapfig}
\usepackage{url}
\usepackage{cleveref}
\usepackage{caption}
\usepackage{pgfplotstable}
\usepackage{booktabs}
\usepackage{array}
\usepackage{colortbl}
\usepackage{blindtext}
\usepackage{tabularx}
\usepackage[english]{babel}
\usepackage{longtable}

\title{Probing the Jets of Blazars Using the Temporal Symmetry of Their Multi-Wavelength Outbursts}

\author[N. Roy et al.]{
Namrata Roy,$^{1,2}$\thanks{E-mail: naroy@ucsc.edu}
Ritaban Chatterjee,$^{1}$
Manasvita Joshi$^{3}$
and Aritra Ghosh$^{1}$\thanks{Current address: Astronomy Department, Yale University, New Haven, CT 06520, USA}
\\
$^{1}$Department of Physics, Presidency University, 86/1 College Street, Kolkata-700073, India\\
$^{2}$Department of Astronomy and Astrophysics, University of California Santa Cruz, 1156 High Street, CA 95064, USA\\
$^{3}$Institute for Astrophysical Research, Boston University, 725 Commonwealth Avenue, Boston, MA 02215, USA
}

\date{Accepted 2018 October 09. Received 2018 October 03; in original form 2018 May 10}

\begin{document}
\label{firstpage}
\pagerange{\pageref{firstpage}--\pageref{lastpage}}
\maketitle

\begin{abstract}
We compare the rise and decay timescales of $\sim$200 long-term ($\sim$weeks-months) GeV and R-band outbursts and $\sim$25 short-term ($\sim$hr-day) GeV flares in a sample of 10 blazars using light curves from the Fermi-LAT and the Yale/SMARTS monitoring project. We find that most of the long-term outbursts are symmetric, indicating that the observed variability is dominated by the crossing timescale of a disturbance, e.g., a shock. A larger fraction of short-term flares are asymmetric with an approximately equal fraction of longer and shorter decay than rise timescale. We employ the MUlti-ZOne Radiation Feedback (MUZORF) model to interpret the above results. We find that the outbursts with slow rise times indicate a gradual acceleration of the particles to GeV energy. A change in the bulk Lorentz factor of the plasma or the width of the shocked region can lead to an increase of the cooling time causing a faster rise than decay time. Parameters such as the luminosity or the distance of the broad line region (BLR) affects the cooling time strongly if a single emission mechanism, e.g., external Compton scattering of BLR photons is considered but may not if other mechanisms, e.g., synchrotron self-compton and external Compton scattering of the torus photon are included. This work carries out a systematic study of the symmetry of flares, which can be used to estimate relevant geometric and physical parameters of blazar jets in the context of the MUZORF model.
\end{abstract}




\section{Introduction}

Active galactic nuclei (AGN) are the extremely luminous ($\sim \rm 10^{42}-10^{48}~erg~s^{-1}$) central core of the so called ``active galaxies'', which are powered by the accretion of matter by a supermassive black hole of mass $>\sim 10^6~M_{\sun}$ \citep{lyn69,sha73} at their center. Some AGN are associated with a bipolar relativistic outflow termed ``jet''. According to the orientation-dependent unification theory of AGN \citep{urr95,will92}, when these jets are aligned within a few degrees of our line of sight, they are categorised as a particular sub-class of AGN, called blazars.

One of the prominent characteristics of blazars is their rapid high amplitude variability. The variability timescale ranges from hours to years, and it is often observed over the entire spectrum of electromagnetic radiation from radio to $\gamma$-rays in some cases. The detailed nature of the acceleration of particles to ultra-relativistic energies in the jet and subsequent cooling through radiation as well as the structure and location of the emission region is not clearly known. The broadband emission in a jet is mostly of nonthermal origin. Emission from the lower part of the energy spectrum (from radio to optical wavelengths, sometimes extending to X-rays) is likely produced by the relativistic electrons in the jet via synchrotron radiation \citep{mar98,bre81,urr82}, while origin of the high energy component (X-rays and $\gamma$-rays, sometimes extending to TeV energies) is debated. The \emph{hadronic model} stresses on the interaction of relativistic protons in presence of radiation fields \citep{muc01,muc03}, while the \emph{leptonic model} interprets the X-ray and $\gamma$-ray emission to be a result of the inverse-Compton (IC) scattering of low energy photons by relativistic electrons present in the jet. The source of the said low energy ``seed" photons may either be the synchrotron emission in the jet itself \citep[known as synchrotron self-Compton process or SSC;][]{mar92,chi02,arb05}, or originating external to the jet \citep[external Compton process or EC;][]{sik94,der09}. The precise location of the EC seed photons can be from the broad emission line region (BLR), accretion disk or dusty
torus (DT). 

Analyzing the multiwavelength variability of blazars is one of the most powerful tools to probe the physics of jets \citep[e.g.,][]{cha08,mar08}. Many blazars emit a large fraction of their total radiative power in the GeV band. Hence, analyzing GeV variability is important for obtaining insights on the physical parameters and acceleration processes in blazars. Before 2008, detailed GeV light curves were available for only a handful of blazars. Since then \textit{Fermi Gamma-Ray Space Telescope} has detected close to 1700 blazars \citep{3FGL} and their light curves are publicly available. In addition, many ground and space-based telescopes have monitored blazars that are detected by \textit{Fermi} at multiple wave bands. 

Emission variability in blazars is caused by various physical processes, e.g., moving shocks, radiation cooling, turbulence, etc., each of which has typical timescales associated with it. Amplitude of variability in blazars at all wave bands, including GeV, is larger at longer than at shorter timescales \citep[e.g.,][]{cha08, cha12, abd10}. In accordance with this, higher-amplitude variability of blazars --- in which the flux changes by a factor of a few or more --- takes place at $\sim$ weeks-months timescale. Light curves containing high-amplitude outbursts in many blazars have been accumulated by \textit{Fermi}.  

However, a closer look into the long-term light curves reveals that in addition to the big prominent flares at longer timescales, there are small flares that are relatively short-spanned, e.g., within a day. \citet{gio94, sai13, abd10, ack16, abd10c, hay15, kus14} have shown that some of the blazars show strong variability even within $\sim$hours. Hence, physical processes responsible for shorter-timescale flares may be probed by analyzing the $\sim$ hr-day timescale variability of blazars. For example, \citet{kus14} has shown that 6~hr-binned lightcurves of two bright $\gamma$-ray outbursts  of PKS1222+216 have asymmetric profile, with decay time unequal to the rise time.  It may be expected that the short-term flares are asymmetric as the decay is related to radiative cooling of the emiting particles and is longer than the rise time. The rise time may be due to acceleration of particles and is assumed to be effectively instantaneous.

However, recently, \citet{sai13} have claimed that the relevant cooling timescales are shorter than the decay timescales of even the short-term ($\sim$ hr-day) GeV flares. On the other hand, \citet{nal13} has stated that an overall asymmetric profile could be a result of the superposition of a number of individual outbursts stemming from different emission regions that lie at different azimuthal angles in the jet cross-section. 

In this paper, we focus on the symmetry properties of the outbursts in the $\gamma$-ray and optical light curves of a chosen sample of blazars from \textit{Fermi}-LAT and a supporting monitoring program by the Yale/SMARTS group, respectively. \citet{cha12} showed that in a sample of six blazars monitored over $\rm2~yr$, most long-term ($\sim$ months) GeV and optical outbursts were symmetric. They concluded that the rise and decay of long-term outbursts were dominated by the crossing time of a shock front or of a disturbance through the emission region. Here, at first we carry out a similar analysis for a much larger sample of 10 blazars in GeV and 9 blazars in optical monitored over $\rm\sim8~yr$ and exhibiting close to 200 outbursts. Furthermore, there have been several instances of high-amplitude GeV flares at shorter characteristic timescales during the \textit{Fermi} era. We analyze those short-term flares as well, to investigate their symmetry properties.  

Details of the sample chosen are given in \S2. In \S3, we decompose the light curves of each of those blazars into individual outbursts and determine their rise and decay timescales at both wavebands, and quantify the symmetry in terms of a parameter based on the similarity between the rise and decay times. 
To facilitate the interpretation of our results on the symmetry properties of the outbursts, we use a MUlti-ZOne Radiation Feedback \textit{MUZORF} model \citep{jos14} to reproduce the spectral variability patterns of a generic blazar at various wavebands included in our analysis. Through our theoretical approach, we discuss the impact of various physical jet parameters on the overall profile of the light curves. These are described in \S4. Finally we summarize the results and conclusion in \S5.

\section{Data}
We have selected our targets of interest based on the following factors. We have chosen only those objects whose $\gamma$-ray light curves have flux values above the monitoring threshold ($\rm 10^{-6}~cm^{-2}~s^{-1}$) throughout the major part of the time interval of interest (55000 MJD$-$57550 MJD), thus giving us more than 500 data points for each object. Our chosen sample is also biased more towards those objects which have shown a significant number ($>$ 10) of prominent distinct outbursts throughout the observed time range. We narrowed down to our final sample of blazars based on their presence in both \textit{Fermi} and Yale/SMARTS monitored blazar lists.

For investigating the short-timescale GeV flares we have chosen the events, in which the flux values increased by a factor of two or more in less than a day.

\subsection{$\gamma$-Ray Data} 
We have used the data from the \textit{Fermi} Gamma Ray Space Telescope, which was launched in 2008 and detects $\gamma$-rays in the energy range 0.1--300~GeV. Large Area Telescope (LAT) onboard \textit{Fermi} detects $\gamma$-ray photons through the electron-positron pair production in a silicon tracker. It has a spatial resolution of $\sim$few tens of arcseconds (for $E>1$~GeV), a very wide field of view ($\sim \rm2.4~Sr$), and an effective area $> 8000~\rm cm^{2}$ (for $E>100$~MeV) \citep{atw09}. It performs a full scan of the entire sky every 3 hours. This observing strategy of \textit{Fermi}-LAT has enabled us to study the variability property in the $\gamma$-ray waveband at shorter ($\sim\rm hours$) as well as longer ($\sim\rm years$) timescales.

To obtain the light curves binned over sub-day timescales during the short-term GeV flares, we carry out unbinned likelihood analysis of the data using the standard analysis tool, namely, Fermi Science tool version v10r0p5. We specify the required time and energy range in the first step of the analysis, {\tt gtselect}, which creates a filtered file consisting of user specified cuts. Only events classified as {\tt evclass}~=~128 and {\tt evtype}~=~3 within the region of interest (ROI) of 20$^{\circ}$ around the co-ordinates of the blazar of our interest are used for further analysis. We use {\tt gtmktime} to select good time intervals (GTI) using the filter ``DATA\_QUAL$==$1" and ``LAT\_CONFIG$==$1". The $\gamma$-ray photons from the ROI surrounding the source blazar coordinates are selected using the 4-year LAT catalog \citep[3FGL;][]{3FGL} given in \textit{gll\_psc\_v16.fit} to create a model of all possible $\gamma$ ray sources that may contribute to the flux observed. But in our model, we generally discard the sources which are far away (greater than 3$^{\circ}$) from the blazar of our interest. We model the sources with power law, keeping the parameters of the source as free. The instrument response function CALDB is used for the Pass 8 data. We incorporate the Galactic and extragalactic diffuse emission and istropic background emission in the model via the templates \textit{gll\_iem\_v06} and \textit{iso\_P8R2\_SOURCE\_V6\_v06}. Finally, we carry out the unbinned likelihood analysis using the \textit{gtlike} task. We use a value of the Test Statistic (TS) $>25$ as the detection criteria for the source. We generate the light curves by carrying out these analyses for each time bin of interest. To probe the long-term ($\sim$ days $-$ months) variability, we use the daily binned light curves obtained from the Fermi Science Support Center\footnote{https://fermi.gsfc.nasa.gov/ssc/data/access/lat/msl$\_$lc/}.

\subsection{Optical Data} 
To study the variability properties of the blazars in our sample in the optical wavelengths, we use data from the Yale/SMARTS blazar monitoring program\footnote{http://www.astro.yale.edu/smarts/glast/home.php}. All the blazars that are detected by \textit{Fermi}-LAT and are accessible from the Cerro-Tololo observatories in Chile are monitored by this program with a variable cadence that depends on the $\gamma$-ray brightness of the source at a given time. The data are publicly available. The details of the data acquisition and analysis are given in \citet{bon12}.

\section{Analysis and Results} 
\subsection{Long-Term Outbursts} \label{analysis:decomp} 
\begin{figure}
\hspace*{-0.1in}
\includegraphics[angle=270,width=\textwidth]{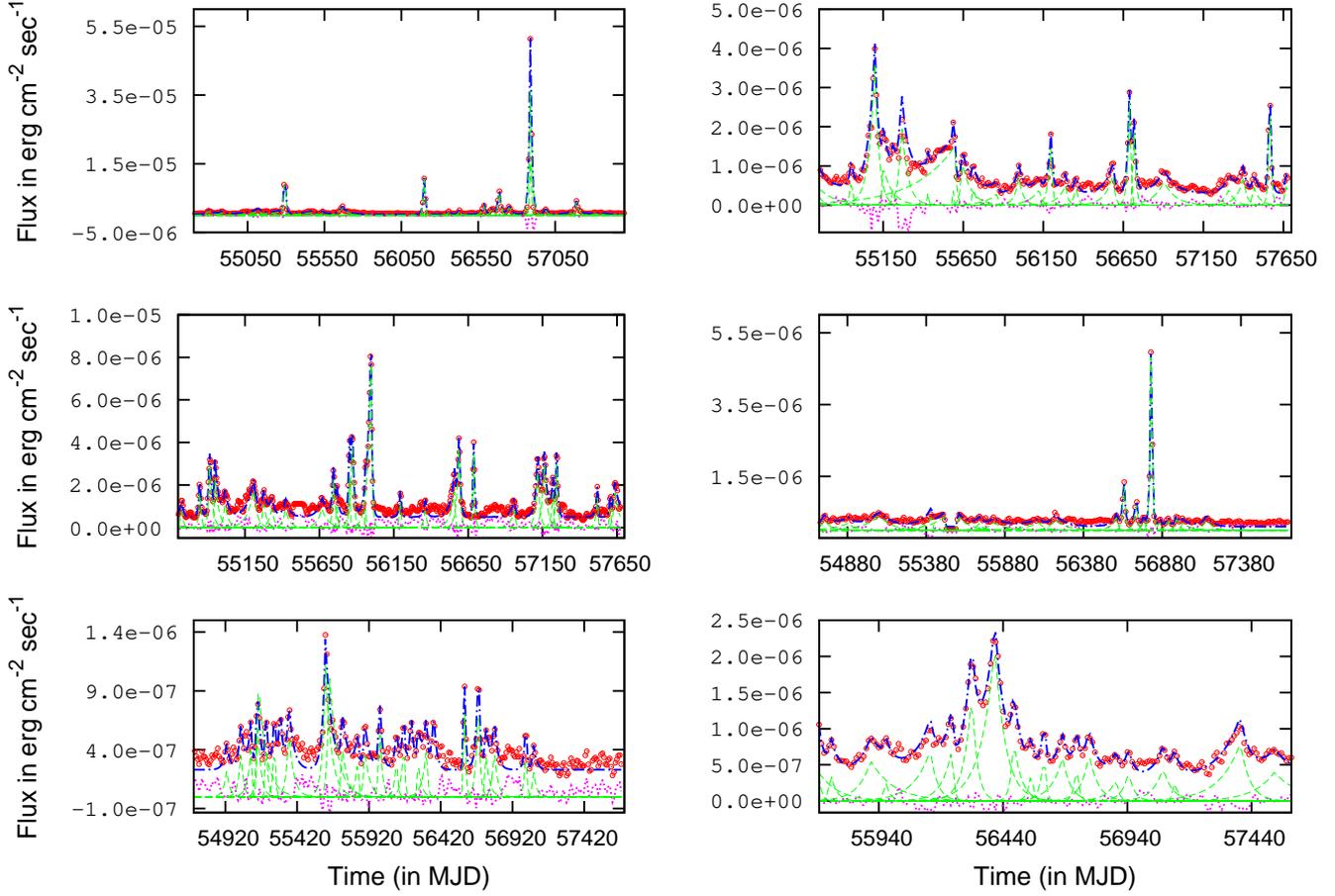}
\caption{\textit{Top left}$- $ 3C 279, \textit{Top right}$- $ 3C 273, \textit{Middle left}$- $ 1510-089,  \textit{Middle right}$- $ Mrk 501, \textit{Bottom left}$- $ PKS 2155-304, \textit{Bottom right}$- $ PKS 1424-41. The red open circles denote the \textit{Fermi-LAT} light curves of the above blazars at the energy range 0.1-300 GeV, which are smoothed with a Gaussian function of width 10 days, green long-dashed lines represent the individual decomposed flares (see text), the blue dot-dashed line is the best-fit to the model function given in \S\ref{analysis:decomp} , which is the sum of the individual flares, while the magenta dotted line is the residue after the fit.}
\label{fig1}
\end{figure}

\begin{figure}
\centering
\hspace*{-0.1in}
\includegraphics[angle=270,width=\textwidth]{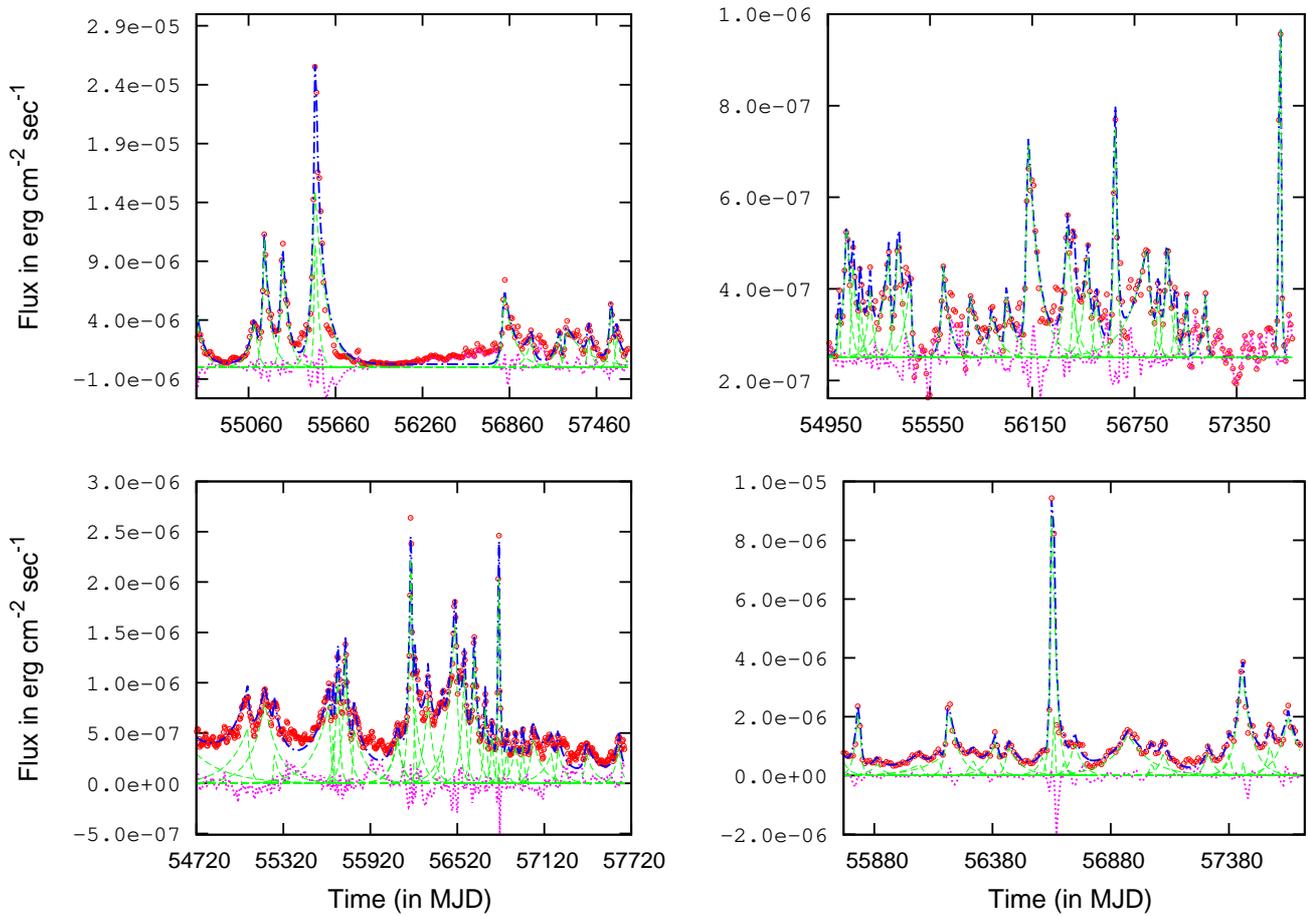}
\caption{\textit{Top left}$- $ 3C454.3, \textit{Top right}$- $ Mrk 421, \textit{Bottom left}$- $ 1633+382, \textit{Bottom right}$- $ CTA 102. The  symbols and representations are the same as in Fig \ref{fig1}.}
\label{fig2}
\end{figure}

The long-term $\gamma$-ray light curves (\Cref{fig1,fig2}), which are binned over a day, show large (factor of $\sim$5-10) variation over the timescale of weeks to months with the average flux-doubling timescale $\sim$100 days. Short-timescale ($\sim$days) flares are often present in between or during the large outbursts but their amplitude is much smaller with peak flux value $<1.5$ times the average quiescent level. In order to probe the high-amplitude long-term outbursts efficiently and to ignore any small fluctuations, we smooth the light curves using a Gaussian function of full-width half maxima of 10 days. 

In order to study the properties of the individual long-term outbursts, we decompose the light curves following \citet{val99, cha12}. Subsequent outbursts often start before the previous one has fully decayed. So to extract each individual flare, we fit the highest one with an exponential rise and decay function, subtract it from the light curve, and fit the next highest outburst in the residual. We repeat the procedure many times till adding another flare does not change the residue by more than 10\%. The model functional form f(t) we use is given by: \\ $ \rm f(t)=f_{0}+f_{max}~e^{\frac{(t-t_0)}{T_r}}$   for $\rm t<t_0$\\ \hspace*{3ex} $\rm =f_{0}+f_{max}~e^{\frac{-(t-t_0)}{T_d}}$  for $\rm t>t_0$,\\
where the various parameters obtained from this model are: the constant background flux f$_{0}$, the time at the peak of the flare t$_0$, the amplitude of the flare $\rm f_{max}$, rise and decay timescale of the flares T$\rm_r$ and T$\rm_d$, respectively. These parameter values obtained from each of the flares can be used to estimate the amount of symmetry present in them. We define a symmetry parameter following \citet{cha12} as $\rm \xi = \frac{(T_d-T_r)}{(T_d+T_r)}$, where $\xi = 0$ for exactly symmetric flares. No constraints are imposed in the modeling process on the rise and decay timescales. There are a few cases where some artificial data points have been created by the smoothing program due to lack of real data in those specific parts in the light curve. Any flare fit to those points has been tracked from the co-incidence of the flare peak and missing data points in the actual light curve and completely removed from further analyses. 

To estimate the errors in the $\xi$ value obtained from the flare-decomposition analysis described above, we have repeated the above fitting procedure by keeping the fitting parameters free within a range of values. The best-fit parameters have been obtained from the chi-squared minimization technique. We have calculated the uncertainty of each of those parameters (4 parameters in total) for each flare by fixing the other 3 parameters to their best-fit value and noting how much the 4th free parameter has to change to increase absolute chi-square value by 1.0. That gave us the 1 $\sigma$ error on that parameter on the particular flare. The uncertainty in $\xi$ is obtained from the uncertainties of the 4 free parameters via simple error-propagation. The value of the best fit parameters of the decomposed flares with 1$\sigma$ errors in $\gamma$-ray and optical wavebands are given in Table \ref{long_param} and \ref{long_param_opt} respectively, full versions of which are available online. 

\begin{figure}
\hspace*{-0.1in}
\includegraphics[angle=270,width=\textwidth]{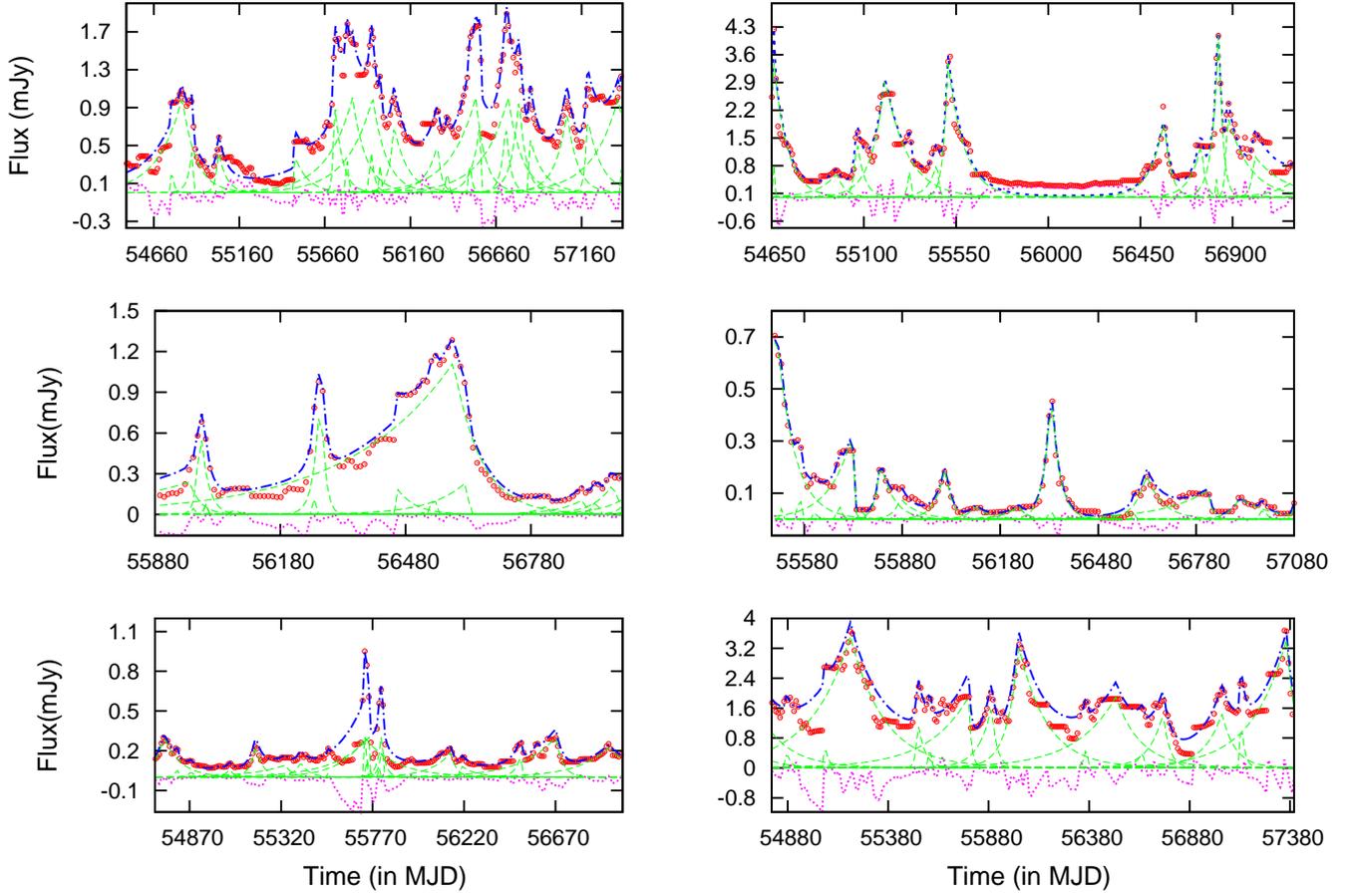}
\caption{\textit{Top left}$- $ 3C 279, \textit{Top right}$- $ 3C454.3, \textit{Middle left}$- $ PKS 0454-234, \textit{Middle right}$- $ CRATES J0531-4827, \textit{Bottom left}$- $ 0208-512, \textit{Bottom right}$- $ OJ 287. The red open circles denote the SMARTS $R$-band light curves of the above blazars, which are smoothed with a Gaussian function of width 10 days, green long-dashed lines represent the individual decomposed flares (see text), the blue dot-dashed line is the best-fit to the model function given in \S\ref{analysis:decomp} , which is the sum of the individual flares, while the magenta dotted line is the residue after the fit.}
\label{fig3}
\end{figure}

\begin{figure}
\hspace*{-0.1in}
\includegraphics[angle=270,width=\textwidth]{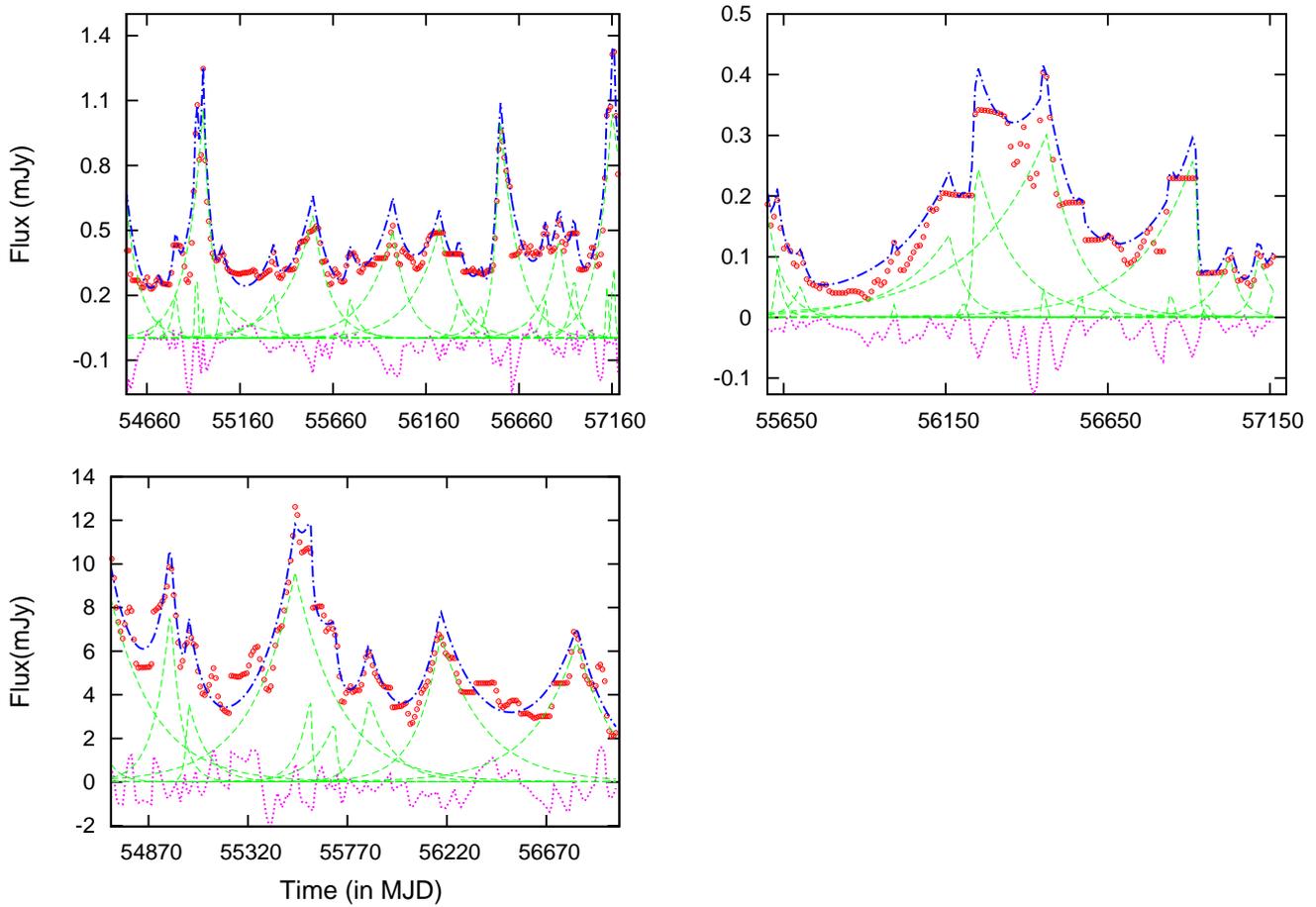}
\caption{\textit{Top left}$- $ 1510-089, \textit{Top right}$- $ 1144-379, \textit{Bottom left}$- $ PKS 2155-304. The symbols and representations are the same as in Fig \ref{fig3}.}
\label{fig4}
\end{figure}

\begin{figure}
\centering
\hspace*{-0.1in}
\includegraphics[angle=270,width=0.5\textwidth]{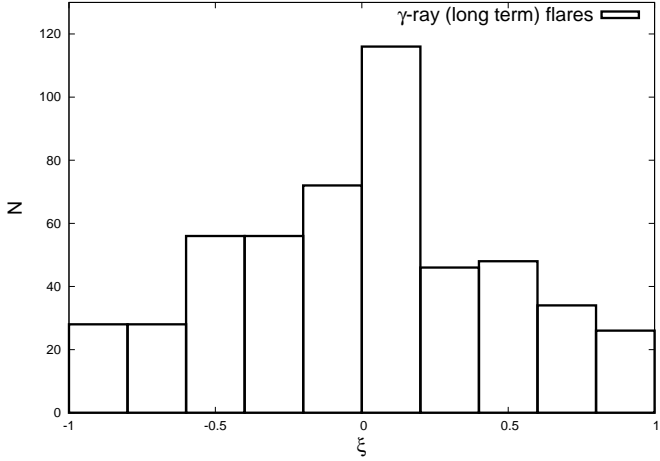}
\caption{The histogram of the symmetry parameter $\xi$ obtained from the decomposition of the $\gamma$-ray light curves shown in Fig \ref{fig1} \& \ref{fig2} into individual flares. Approximately, $\rm \xi>0.3$ indicates faster rise than decay, $\rm \xi<-0.3$ indicates slower rise than decay, while $\rm -0.3<\xi<+0.3$ indicates a symmetric profile.}
\label{fig5}
\end{figure}

\begin{figure}
\centering
\hspace*{-0.1in}
\includegraphics[angle=270,width=0.5\textwidth]{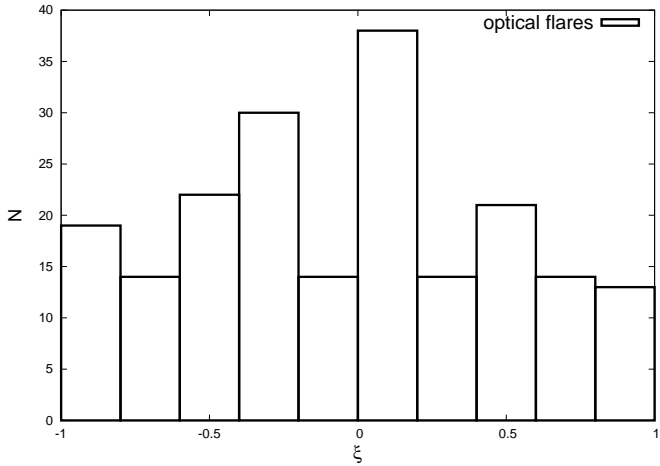}
\caption{The histogram of the symmetry parameter $\xi$ obtained from the decomposition of the $R$-band light curves shown in Fig \ref{fig3} \& \ref{fig4} into individual flares. }
\label{fig6}
\end{figure}

The decomposition of the long-term \textit{Fermi-LAT} light curves of our sample are shown in  \Cref{fig1,fig2}. The Symmetry parameter $\xi$ of most of the large outbursts are within $\pm$ 0.3, which indicates that their growth and decay timescales are similar. The variability in the optical data is comparatively slower, resulting in a smaller number of individual outbursts than the $\gamma$-ray data (\Cref {fig3,fig4}). The $\xi$ parameter distribution in the optical wavelength shows nearly symmetric nature of the outbursts with comparable rise and decay timescales, similar to that at the $\gamma$-ray energies. The distribution of $\xi$ parameters in $\gamma$-ray and optical wavebands are shown in \Cref{fig5,fig6}.
\clearpage
\onecolumn

\clearpage
\twocolumn

	%
\clearpage
\onecolumn
\begin{center} 
	\begin{longtable}{|l|l|l|l|l|l|l|}
	\caption{
	Values of the Parameter of $\gamma$-ray long-term Outbursts (Full version available in ancillary files).}
	\label{long_param}\\
		\hline
		\hline
\textbf{Blazar} & \textbf{Flare\#} & \textbf{f$_{max}$} & \textbf{t0} & \textbf{T$\rm _r$} & \textbf{T$\rm _d$} & \textbf{$\rm \xi$} \\
		\hline
 & 1 & 36.00 $\pm$ 1.00 & 55098 $\pm$ 2 & 37.5 $\pm$ 2.5 & 26.0 $\pm$ 1.0 & -0.18 $\pm$ 0.03  \\ 
 &2 & 36.00 $\pm$ 1.00 & 56692 $\pm$ 0 & 12.5 $\pm$ 0.5 & 6.0  $\pm$ 0.5 & -0.35 $\pm$ 0.04  \\
 &3 & 35.00 $\pm$ 1.00 & 57564 $\pm$ 0 & 7.5 $\pm$ 0.5 & 9.0 $\pm$ 0.5 & 0.09 $\pm$ 0.05  \\
3C 273 &4 & 21.00 $\pm$ 0.50 & 55265 $\pm$ 1 & 20.0 $\pm$ 1.5 & 38.0 $\pm$ 2.0 & 0.31 $\pm$ 0.05 \\ 
 &5 & 15.50 $\pm$ 0.50 & 55597 $\pm$ 4 & 250.0 $\pm$ 35.0 & 11.0 $\pm$ 1.5 & -0.91$\pm$ 0.01 \\ 
 &6 & 33.00 $\pm$ 1.00 & 56711 $\pm$ 1 & 2.5 $\pm$ 0.5 & 12.0 $\pm$ 0.5 &0.65 $\pm$ 0.05  \\
 &7 & 16.00 $\pm$ 0.50 & 56200 $\pm$ 1& 20.0 $\pm$ 1.0 & 9.0 $\pm$ 1.5  & -0.38 $\pm$ 0.08 \\
 &8 & 11.00 $\pm$ 0.50 & 55643 $\pm$ 1 & 12.5 $\pm$ 2.0 & 56.0 $\pm$ 7.5 & 0.63 $\pm$ 0.06 \\ 
 &... & ... & ... & ... & ... & ... \\

           \hline
	\end{longtable}
\end{center}

\begin{center} 
	\begin{longtable}{|l|l|l|l|l|l|l|}
	\caption{
	Values of the Parameter of optical long-term Outbursts (Full version available in ancillary files).}
	\label{long_param_opt}\\
		\hline
		\hline
\textbf{Blazar} & \textbf{Flare\#} & \textbf{f$_{max}$} & \textbf{t0} & \textbf{T$\rm _r$} & \textbf{T$\rm _d$} & \textbf{$\rm \xi$} \\
		\hline

           & 1 &     1.00$\pm$0.15 &      56731$\pm$9 &      150.0 $\pm$25.0 &      80.0$\pm$1.0 &    -0.30 $\pm$  0.06\\
           &2  &       1.00$\pm$0.10 &      55821$\pm$10 &      150.0$\pm$5.0 &      80.0$\pm$1.0 &    -0.30 $\pm$    0.06\\
           &3  &       1.00$\pm$0.10 &      56541$\pm$7 &      150.0$\pm$18.0 &      56.0$\pm$6.5 &    -0.45 $\pm$   0.05\\
 3C 279  &4  &        1.00$\pm$0.15 &      55939$\pm$3 &      150.0$\pm$20.0 &      80.0$\pm$1.0 &    -0.30 $\pm$   0.07\\
           &5  &      1.00$\pm$0.20 &      57374$\pm$6 &      150.0$\pm$15.0 &      80.0$\pm$1.0 &    -0.30 $\pm$  0.07\\
           &6  &      1.00$\pm$0.10 &      54825$\pm$4 &      141.0$\pm$16.0 &      80.0$\pm$1.5 &    -0.27 $\pm$    0.06\\
           &7  &   1.00$\pm$0.10 &      56794$\pm$1 &      45.0$\pm$4.5 &      40.0$\pm$5.0 &     -0.05 $\pm$   0.07\\
           &8  &    1.00$\pm$0.20 &      55715$\pm$1 &      9.0$\pm$1.0 &      45.0$\pm$6.5 &      0.66$\pm$    0.04\\
            &... & ... & ... & ... & ... & ... \\

           \hline
	\end{longtable}
\end{center}
\twocolumn

\subsection{Short-Term Outbursts} \label{analysis:short}
We investigate the characteristic properties of the variability with a higher time resolution, i.e. shorter time bins (6 hours). It has been possible only for those sources which have shown considerable brightening ($\rm 10^{-6}~cm^{-2}~s^{-1}$) in short timescales with good photon statistics ($>25$).

However, the above modeling is not robust in the case of short-term variability. Even in the brighter flares, the shortest width of a bin for which TS $>25$ may be obtained is $\sim \rm 6~hr$. Hence, a portion of a light curve of duration $\sim$ 5 days, containing a short-term flare, have a total of $\sim$ 20 data points. Consequently, the parameters obtained from the above model have significant statistical uncertainty and strong dependence on the functional form we use. In order to get around the above difficulty, we define a flux-doubling timescale ($\tau$) as a measure to find $\xi$. This is to ensure the results depend only on the observed data and not on any assumed model.

We use $\rm \tau_d=\Delta t \frac{ln2}{ln\frac{F_2}{F_1}}$, where $\rm F_1$ and $\rm F_2$ are the flux values at two different times at an interval $\rm \Delta t$ in which there has been a significant change in the flux \citep{sai15}. We calculate the doubling (or halving) time in the rise or decay branch and determine the symmetry parameter as before.

\begin{figure}
\centering
\hspace*{-0.1in}
\includegraphics[angle=270,width=\textwidth]{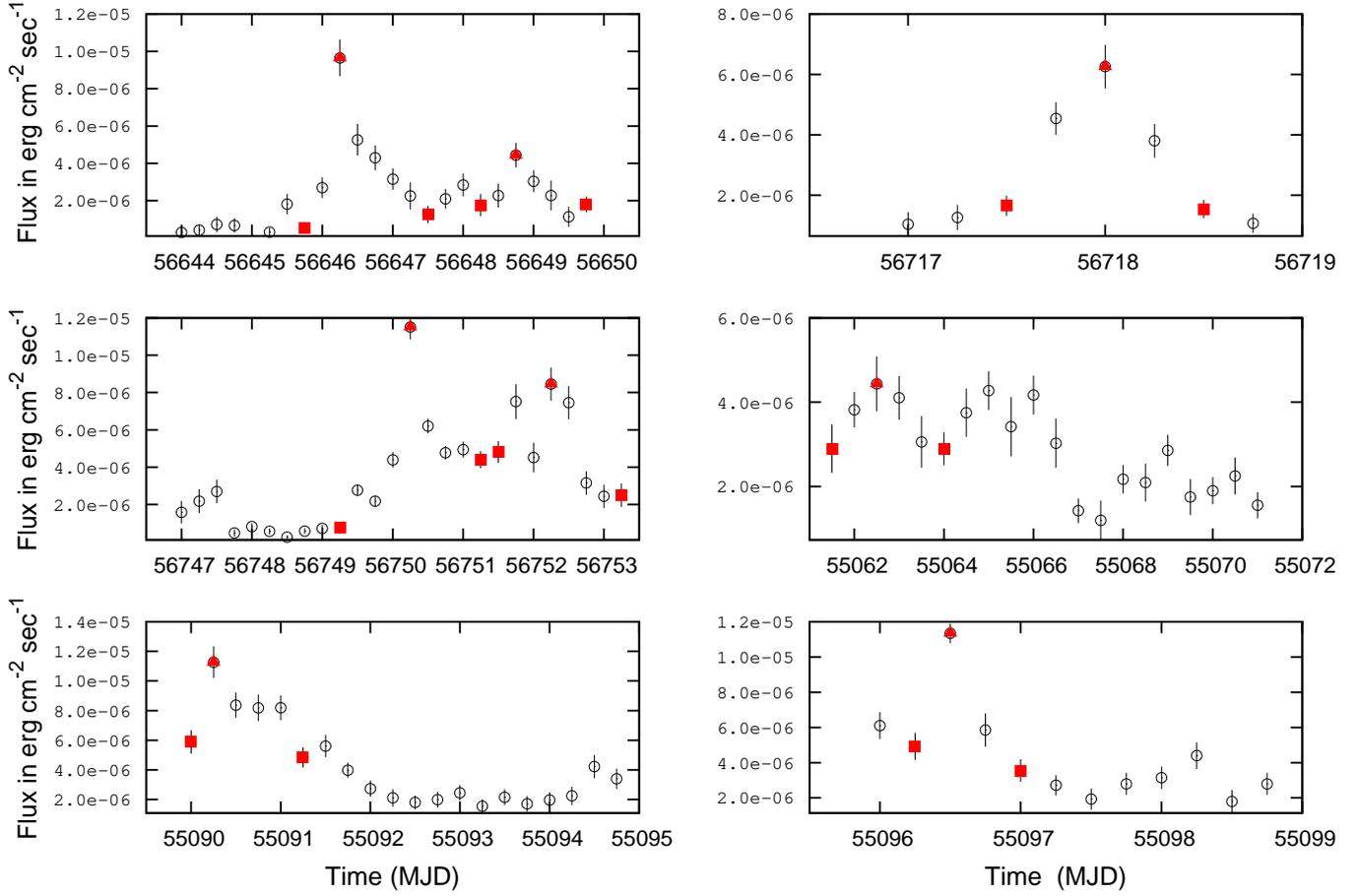}
\caption{\textit{Top left}$- $ 3C 279 flare 1, \textit{Top right}$- $ 3C 279 flare 2, \textit{Middle left}$- $ 3C 279 flare 3,  \textit{Middle right}$- $ 3C 273 flare 1, \textit{Bottom left}$- $ 3C 273 flare 2, \textit{Bottom right}$- $ 3C 273 flare 3. The open circles with the error bars denote the \textit{Fermi}-LAT light curves of the above blazars binned over 6 hours during short-term outbursts. The filled triangles represent the peak of the flares and the filled squares denote the data points used for calculating the flux-doubling time scale for a particular flare. If there are more than three points marked in red, we have considered more than one short-term flare within that time interval in that case. }
\label{fig7}
\end{figure}

\begin{figure}
\centering
\hspace*{-0.1in}
\includegraphics[angle=270,width=\textwidth]{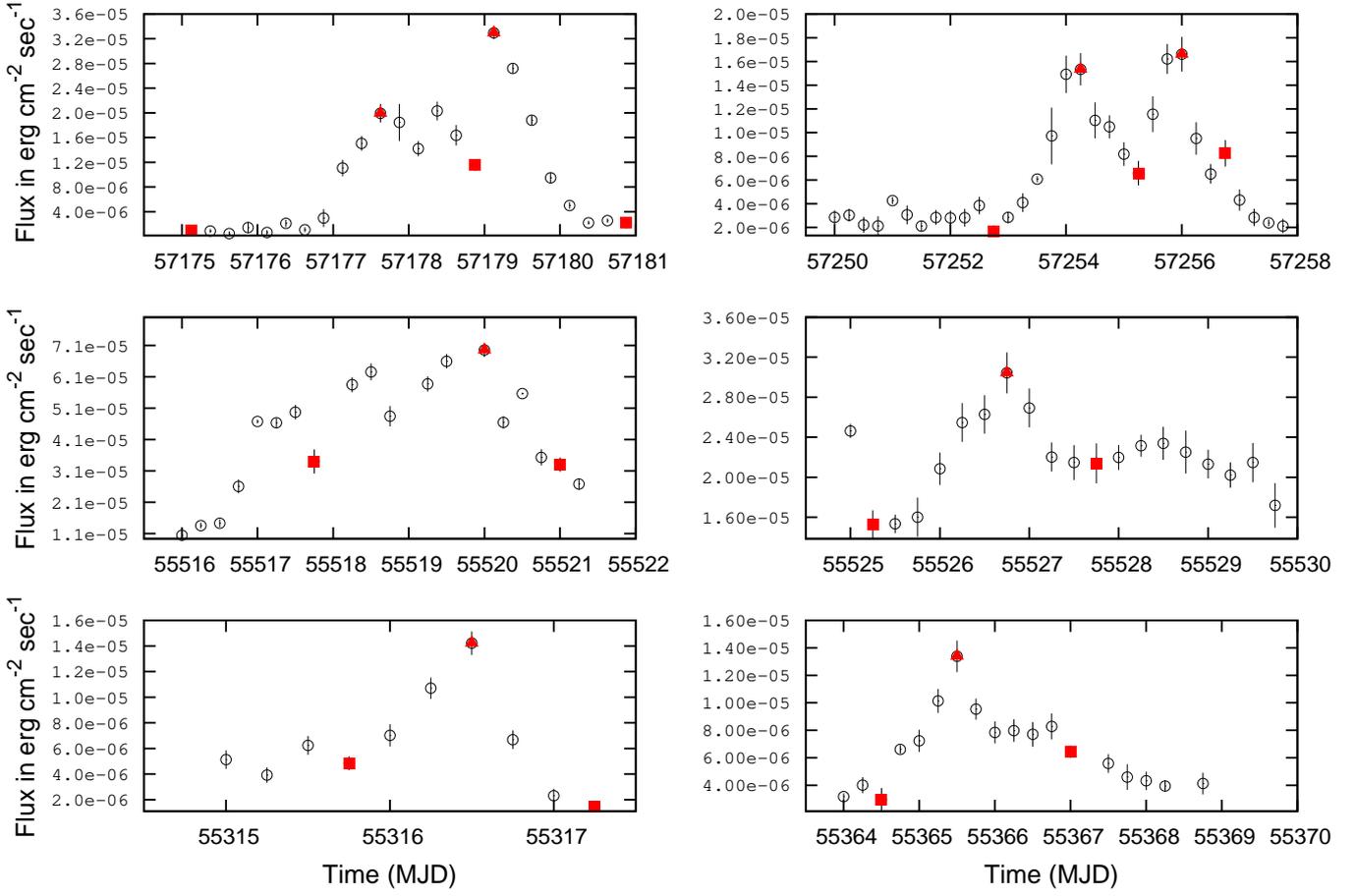}
\caption{\textit{Top left}$-$ 3C 279 flare 4, \textit{Top right}$- $ 3C 454 flare 1, \textit{Middle left}$- $ 3C 454 flare 2,  \textit{Middle right}$- $ 3C 454 flare 3, \textit{Bottom left}$- $ PKS B1222+216 flare 1, \textit{Bottom right}$- $PKS B1222+216 flare 2. The symbols are the same as in Fig \ref{fig7}.}
\label{fig8}
\end{figure}

\begin{figure}
\centering
\hspace*{-0.1in}
\includegraphics[angle=270,width=\textwidth]{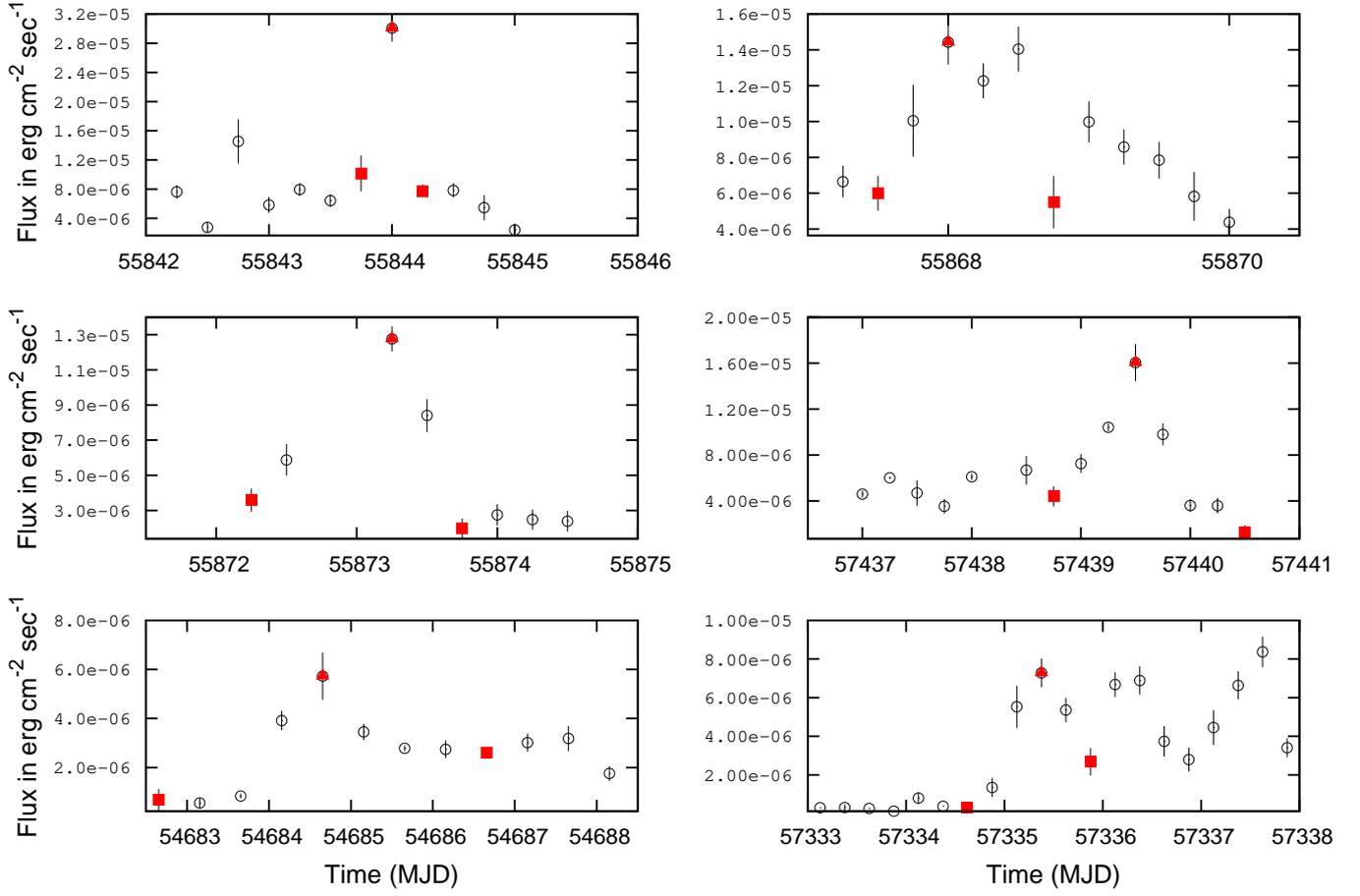}
\caption{\textit{Top left}$- $ 1510-089 flare 1, \textit{Top right}$- $ 1510-089 flare 2, \textit{Middle left}$- $ 1510-089 flare 3,  \textit{Middle right}$- $ CTA 102, \textit{Bottom left}$- $ PKS 1502+106, \textit{Bottom right}$- $ S50836+71. The symbols are the same as in Fig \ref{fig7}.}
\label{fig9}
\end{figure}

\begin{figure}
\centering
\hspace*{-0.1in}
\includegraphics[angle=270,width=0.5\textwidth]{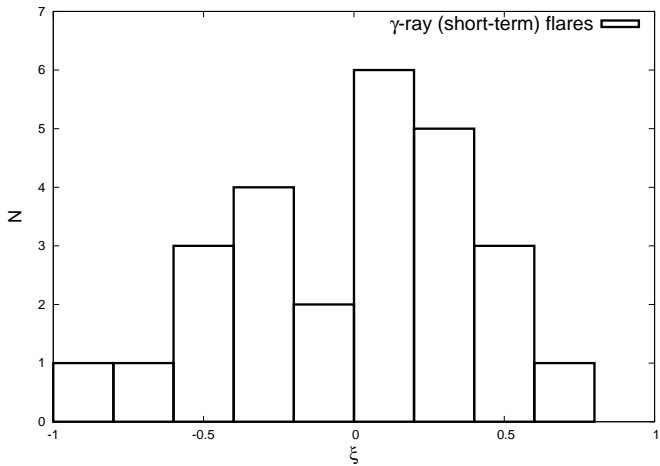}
\caption{The histogram of the distribution of the symmetry parameter $\xi$ of the short-term flares shown in Fig \ref{fig7}-\ref{fig9}.}
\label{fig10}
\end{figure}

 \citet{gio94, mar08, bon12, cha12, sai13, kus14} have previously reported hr-timescale flares in blazars. We select a sample of 26 short-term flaring states, including the recent prominent flares of 3C279, S5 0836+71, 3C454, CTA 102, where the flux values increase considerably within a few hours. Figure \ref{fig7}, \ref{fig8}, and \ref{fig9} present the light curves of those bright blazars, binned in the interval of 6 hr, during the time of the major short-term outbursts. Unlike the small amplitude spurious peaks observed in daily binned data in previous images, in this case the flux values increase $\sim$3 times within $\Delta \rm t$ $\sim \rm few~hr$. The flux doubling timescale ($\rm \tau_d$) has been calculated using the data points shown by filled triangles and squares in Figure \ref{fig7}, \ref{fig8}, and \ref{fig9}. Average $\rm \tau_d \sim 2$ hr instantly points to the extreme and violent acceleration processes happening in the jet environment that energize the particles to GeV energies in hours. The distribution of the values of symmetry parameter $\xi$ obtained from the short term flares is shown in Figure \ref{fig10}. The approximate uncertainty estimate of $\xi$ can be obtained from the expression $\rm \delta \xi = [2(3+\xi^2)]^{1/2}(\delta t/T)$, as given in \cite{nal13}, where $\delta t$ is the uncertainty of the rise and decay times and T is the total duration of the flare ($\sim 1$ day). If $\delta t$ is of the order of 2 hr (average doubling time), $\delta \xi \sim 0.2$.

\clearpage
\onecolumn

\clearpage
\twocolumn

The short-term flares with faster rise and slower decay (positive $\xi$) can be explained if the acceleration of emitting particles is effectively instantaneous and radiative cooling timescale is comparatively slow, e.g., $\sim$hr-day. As per the leptonic model of blazar jets, $\gamma$-ray emission from the blazars of our sample may be primarily contributed by the inverse-Compton scattering of low energy photons, coming either from the broad line region (BLR), or from the dusty torus. The energy loss rate equation $ \rm \frac{dE_{IC}}{dt}\simeq \frac{4}{3} \sigma_{T}cU_{rad}\Gamma^{2}\gamma^{2}$ leads to a cooling timescale of $\rm t_{cool}=\frac{3m_{e}c}{4\sigma_{T}\gamma U_{rad}\Gamma^{2}}$. Here, $\rm \sigma_{T}, U_{rad}, \Gamma$ and $\gamma$ refers to the Thomson scattering cross-section, the radiation energy density, the bulk Lorentz factor of the plasma and the Lorentz factor of electrons respectively. We assume the emitting region to be compact with observing angle $\theta_{obs}\backsimeq 1/\Gamma$, and hence used the approximation $\delta\simeq\Gamma$ \citep{nal13} where $\delta$ is the Doppler factor. We obtain a theoretical estimate of the cooling timescale $\rm t_{cool}= \frac{3m_{e}c^{2}\pi R_{IC}^{2}}{\sigma_{T} L \gamma  \Gamma^{2}}$ in terms of the physical parameters of the jet emission region, where $\rm R_{IC}$ is the location of the source of seed photons which are up-scattered, and $L$ is the luminosity of BLR/dusty torus, depending on the location of seed photons. We assume the $\gamma$-rays to be due to the IC scattering of optical/infrared photons (frequency $\rm \simeq 10^{14}~Hz$) originating from the accretion disk/BLR/dusty torus. For this, electrons with $\gamma\simeq10^{3}$ are required. Using $\Gamma=10$, L~=~L$\rm _{BLR}= \rm10^{44}-10^{46}~erg~s^{-1}$, and $\rm R_{IC}$ to be $\rm 0.1-10~pc$, we get $\rm t_{cool} \sim$ few hours. Hence, the cooling time may be probed in outbursts where the shortest time bins with $\rm TS>25$ are shorter than a few hours. If $\rm R_{IC}$ is more or $\rm L_{BLR}$ is less than what we assumed $\rm t_{cool}$ will be longer and probing it will be possible for fainter flares.

On the other hand, \citet{nal13} and \citet{sai13} have argued that $\rm t_{cool} \sim hr$, and GeV outbursts in blazars may have longer decay than rise timescale primarily due to geometric effects. If the emission region inside the jet consists of several zones at different angles with respect to the viewer, we may expect the profile of an observed outburst to be composed of a few smaller overlapping flares of gradually decreasing amplitude due to varying beaming effect, which would effectively increase the observed decay time \citep{nal13}. In order to interpret the results found above and to test the physical conditions that will give rise to asymmetric flare profiles we carry out theoretical modeling of jet emission, which is described in the next section.

\section{Theoretical Interpretation}
\subsection{The MUlti-ZOne Radiation Feedback (\textit{MUZORF}) Model}
The effect of rapid acceleraton of particles and subsequent non-thermal radiation has been investigated by various theoretical approaches based on shock propagation \citep[e.g.,][]{mar98, chi02, sik94, jos11}. Internal shock model is one such variant in which two shells of plasma travelling at different speeds with differing masses and internal energies collide with one another to give rise to a single episode of collision \citep{jos11}. The collision results in the formation of forward and reverse shocks internal to the jet. The shocks then propagate through the jet and continue to instantaneously accelerate particles to very high energies as they plough through the emission region. The accelerated particles then radiate via various radiation mechanisms, such as synchrotron, synchrotron self-Compton (SSC), and external Compton (EC). Here, we employ the theoretical approach of \citet{jos14} in order to relate the symmetry properties of the observed light curves with the physical parameters associated with jet emission. In what follows, the unprimed quantities refer to the AGN frame, primed quantities to the comoving frame, and starred quantities indicate the observer's frame.

The MUlti-ZOne Radiation Feedback (\textit{MUZORF}) model developed by \citet{jos11} is a time dependent radiation transfer model of blazar jets. As explained in \citet{jos11}, an outer shell moving at a slower speed with a bulk Lorentz factor (BLF) $\Gamma_{\rm o}$ collides with an inner shell moving at a higher speed with BLF $\Gamma_{\rm i}$ at a certain distance, $z_{\rm c}$, from the central engine (comprising black hole and an accretion disk). The widths and masses of the shells are $M_{\rm o}, \Delta_{\rm o}$ and $M_{\rm i}, \Delta_{\rm i}$, respectively. The collision results in the formation of forward and reverse shocks inside the merged shell (emission region), which is considered to be of a cylindrical geometry with radius $R$. The shocks move relative to each other, in an opposite direction in the plasma frame, with their respective BLFs given by $\Gamma^{\prime}_{\rm fs}$ and $\Gamma^{\prime}_{\rm rs}$. The plasma particles upon encountering these internal shocks undergo acceleration at the shock fronts, which amplifies their energies to high enough values to produce the observed nonthermal emission of blazars, including $\gamma$-rays. As described in \citet{jos11}, shocks convert part of the bulk kinetic energy of the plasma, in the emission region, into magnetic and electron energy densities. This fraction is quantified in terms of $\varepsilon^{\prime}_{\rm B}$ and $\varepsilon^{\prime}_{\rm e}$, respectively. The shocks exit their respective emission region according to their shock crossing times \citep{jos11}, which could be related to the rising time of the simulated flare profiles. We consider an electron population following a simple power-law distribution with an injection index $q^{\prime}$. Only a fraction, $\zeta^{\prime}_{\rm e}$, of the electron population is accelerated during the passage of the shocks. The entire electron and photon populations are evolved in time according to the prescription given in \citet{jos11}, which allows us to calculate the cooling timescale of electrons that could be related to the decay time of the simulated flare profiles.

As discussed in \citet{jos14}, \textit{MUZORF} considers all three radiation mechanisms mentioned above. The EC emission of blazars includes anisotropic radiation fields of the accretion disk, the broad line region (BLR), and the dusty torus (DT). The luminosity of the BLR is represented by $L_{\rm BLR}$ while the covering factor of the DT is given by $f_{\rm cov, DT}$. The fraction of the disk luminosity illuminating the DT is represented by $\xi$ and the viewing angle of the generic blazar is given by $\theta^{*}_{\rm obs}$. The evolution of particle and photon populations in the emission region is followed in a time-dependent manner to distances beyond the BLR and into the DT. The multi-zone feature of \textit{MUZORF}, with radiation transfer within each zone and in-between zones, lets us address the issue of inhomogeneity in the photon and electron populations throughout the emitting volume. The effects due to internal light travel time delays have also been incorporated in MUZORF which lets us correctly register the emanating radiation in the observer's frame.

In order to relate the symmetry of observed light curves with that of the simulated ones, we study the impact of varying various physical input parameters mentioned above on the symmetry of the simulated light curves. The symmetry of a light curve is discussed in terms of the equality between the rise and decay times of a pulse. These times could get affected by the geometry of the emission region or due to particle acceleration and particle cooling scenarios \citep{sas17}. We run a total of 30 simulations to study the effects of varying the values of each of the physical parameters on the symmetry of lightcurves. For all of our simulations, flux values are calculated for the frequency range $\nu^{\prime} = 7.5 \times 10^{7} - 7.5 \times 10^{24}$ Hz and electron energy distribution (EED) range $\gamma^{\prime} = 1.01 - 10^{7}$ with both ranges divided into 150 grid points. The entire emission region is divided into 100 slices with 50 slices in the forward and 50 in the reverse shock region. Table \ref{basesetlist} shows the values of the input parameters used to obtain our baseline model (run 1) for a generic blazar source. The parameters of the base set are representative of that of the FSRQ 3C~454.3 based on the value of the redshift adopted for our simulations.


\begin{table}
	\centering
	\scriptsize
	\caption{
	Parameter list of run 1 used to obtain the baseline model.}
	\label{basesetlist}
	\begin{tabular}{lcc} 
		\hline
		\hline
Parameter & Symbol & Value\\
		\hline
Kinetic Luminosity & $L_w$ & $1 \times 10^{48}$~erg/s\\
Event Duration & $t_w$ & $1.78 \times 10^{7}$~s\\
Outer Shell Mass & $M_o$ & $5.38 \times 10^{32}$~g\\
Inner Shell BLF & $\Gamma_i$ & 26\\
Outer Shell BLF & $\Gamma_o$ & 10.8\\
Inner Shell Width & $\Delta_i$ & $5.7 \times 10^{15}$~cm\\
Outer Shell Width & $\Delta_o$ & $8.3 \times 10^{15}$~cm\\
Inner Shell Position & $z_i$ & $7.8 \times 10^{15}$~cm\\
Outer Shell Position & $z_o$ & $1.65 \times 10^{16}$~cm\\
Electron Energy Equipartition Parameter & $\varepsilon^{\prime}_e$ & $0.3$\\
Magnetic Energy Equipartition Parameter & $\varepsilon^{\prime}_B$ & $1 \times 10^{-4}$\\
Fraction of Accelerated Electrons & $\zeta^{\prime}_e$ & $2.5 \times 10^{-2}$\\
Acceleration Timescale Parameter & $\alpha^{\prime}$ & $1 \times 10^{-6}$\\
Particle Injection Index & $q^{\prime}$ & 4.0\\
Zone/Jet Radius & $R^{\prime}_z$ & $3.43 \times 10^{16}$~cm\\
Observer Frame Observing Angle & $\theta^{*}_{\rm obs}$ & $1.3^{\circ}$\\
Disk Luminosity & $L_{\rm disk}$ & $2 \times 10^{46}$~erg/s\\
BH Mass & $M_{\rm BH}$ & $1 \times 10^{9} M_{\odot}$\\
Accretion Efficiency & $\eta_{\rm acc}$ & 0.1\\
BLR Luminosity & $L_{\rm BLR}$ & $8 \times 10^{44}$~erg/s\\
BLR inner radius & $R_{\rm in, BLR}$ & $6.17 \times 10^{17}$~cm\\
BLR outer radius & $R_{\rm out, BLR}$ & $1.85 \times 10^{18}$~cm\\
BLR optical depth & $\tau_{\rm BLR}$ & 0.01\\
BLR covering factor & $f_{\rm cov, BLR}$ & 0.03\\
DT inner radius & $R_{\rm in, DT}$ & $3.086 \times 10^{18}$~cm\\
DT outer radius & $R_{\rm out, DT}$ & $8.994 \times 10^{18}$~cm\\
Ldisk fraction & $\xi$ & 0.2\\
DT covering factor & $f_{\rm cov, DT}$ & 0.2\\
Redshift & $Z^{*}$ & 0.859\\
		\hline
	\end{tabular}

\end{table}


\begin{table}
	\centering
	\scriptsize
	\caption{
	Parameter list for other simulations.}
	\label{paramlist}
	\begin{tabular}{ccc} 
		\hline
		\hline
Run\# & Parameter Value\\
		\hline
2 & $\Gamma_{\rm i} = 36$ & $M_{\rm o} = 4.23 \times 10^{32}$\\
3 & $\Gamma_{\rm i} = 16$ & $M_{\rm o} = 7.38 \times 10^{32}$\\
4 & $\Gamma_{\rm o} = 20$ & $M_{\rm o} = 4.30 \times 10^{32}$\\
5 & $\Gamma_{\rm o} = 5$  & $M_{\rm o} = 6.38 \times 10^{32}$\\
6 & $\Delta_{\rm i} = 5.7 \times 10^{16}$~cm\\
7 & $\Delta_{\rm i} = 5.7 \times 10^{14}$~cm\\
8 & $\Delta_{\rm o} = 8.3 \times 10^{16}$~cm & $R_{\rm o} = 9.5 \times 10^{16}$~cm\\
9 & $\Delta_{\rm o} = 8.3 \times 10^{14}$~cm\\
10 & $z_{\rm c} = 6.82 \times 10^{17}$~cm\\
11 & $z_{\rm c} = 1.95 \times 10^{18}$~cm\\
12 & $\varepsilon^{\prime}_{\rm e} = 0.99$\\
13 & $\varepsilon^{\prime}_{\rm e} = 0.03$\\
14 & $\varepsilon^{\prime}_{\rm B} = 1.0 \times 10^{-3}$\\
15 & $\varepsilon^{\prime}_{\rm B} = 1.0 \times 10^{-6}$\\
16 & $\zeta^{\prime}_{\rm e} = 0.9$\\
17 & $\zeta^{\prime}_{\rm e} = 2.5 \times 10^{-3}$\\
18 & $q^{\prime} = 6.0$\\
19 & $q^{\prime} = 2.0$\\
20 & $R^{\prime}_{\rm z} = 1 \times 10^{17}$~cm\\
21 & $R^{\prime}_{\rm z} = 9.0 \times 10^{15}$~cm\\
22 & $L_{\rm BLR} = 8 \times 10^{45}$~erg/s\\
23 & $L_{\rm BLR} = 8 \times 10^{43}$~erg/s\\
24 & $f_{\rm cov, DT} = 0.9$ & $R_{\rm out, DT} = 5.038 \times 10^{18}$~cm\\
25 & $f_{\rm cov, DT} = 0.02$ & $R_{\rm out, DT} = 2.689 \times 10^{19}$~cm\\
26 & $\xi = 0.95$ & $R_{\rm out, DT} = 1.867 \times 10^{19}$~cm\\
27 & $\xi = 0.02$ & $R_{\rm out, DT} = 4.082 \times 10^{18}$~cm\\
28 & $\theta^{*}_{\rm obs} = 3.3^{\circ}$\\
29 & $\theta^{*}_{\rm obs} = 0.3^{\circ}$\\
		\hline
	\end{tabular}
\end{table}

\subsection{Baseline Model}
\begin{figure}[htb]
\hspace*{-0.4in}
{\includegraphics[width=0.5\textwidth]{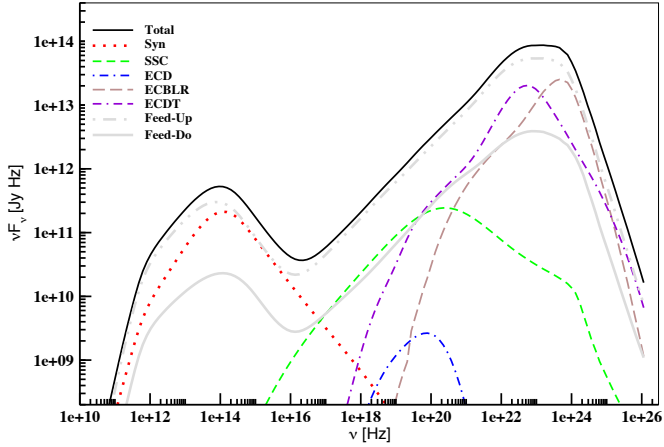}}
\caption{\small{Simulated time-averaged SED of our baseline model. The input parameters are chosen to mimic those of the FSRQ 3C~454.3 and are listed in Table \ref{basesetlist}. The contribution of various radiative components to the total SED is also shown here. The total time-averaged SED is depicted by the thick solid black line. The dotted line represents the contribution of the synchrotron emission to the low-energy component. The dashed line shows the contribution of the synchrotron self-Compton (SSC) to the X-rays and soft $\gamma$-ray regime. The dot-dashed line stands for the contribution of the external Compton (EC) component due to disk photons (ECD) to the hard X-ray regime. The dot-dot-dashed line is for the EC component due to the BLR photons (ECBLR) to the $\gamma$-ray regime. The dash-dash-dotted line shows the contribution of the EC component due to the dusty torus photons (ECDT) to the high-energy component of the total SED. The gray dash-double-dotted and solid lines, respectively, represent the contribution of the forward (Feed-Up) and backward (Feed-Do) radiation-feedback components \citep[see]{jos11} to the total time-averaged SED of the generic blazar source.}}
\label{1sed}
\end{figure}

Figure \ref{1sed} shows the resultant time-averaged spectral energy distribution (SED) of the baseline model, averaged over $\sim$1-day period, for a generic blazar source. The choice of input parameters for the base set as mentioned in Table \ref{basesetlist} yields a value for the BLF of the emission region as $\Gamma_{\rm sh} = 16$ while the relative values of the BLFs of the forward and reverse shocks in their respective emission regions are $\Gamma^{\prime}_{\rm fs} = 1.08$ and $\Gamma^{\prime}_{\rm rs} = 1.12$. A magnetic field value of $B^{\prime} = 1.43$ G is obtained for both emission regions along with $\gamma^{\prime}_{\rm min,fs} = 1.12 \times 10^{3}$, $\gamma^{\prime}_{\rm min,rs} = 1.82 \times 10^{3}$, and $\gamma^{\prime}_{\rm max;fs,rs} = 3.89 \times 10^{4}$. The derived width of the forward and reverse emission regions are $\Delta^{\prime}_{\rm fs} = 1.23 \times 10^{16}$ cm and $\Delta^{\prime}_{\rm rs} = 2.0 \times 10^{16}$ cm. The shock crossing time for each of the regions is derived to be $t^{\prime}_{\rm cr, fs} = 1.10 \times 10^{6}$ s and $t^{\prime}_{\rm cr, rs} = 1.45 \times 10^{6}$ s. In the observer's frame, this implies that the forward shock exits its region in $t^{*}_{\rm fs} = 7.3 \times 10^{4}$ s, i.e., $\sim$ 20 hours while the reverse shock exits the system in $t^{*}_{\rm rs} = 9.56 \times 10^{4}$ s, i.e. $\sim$ 26 hours. The emission region is placed at a distance of $z_{\rm c} = 1.20 \times 10^{17}$ cm from the central engine at the beginning of the simulation and covers a total distance $\sim z_{\rm c} = 1.54 \times 10^{18}$ cm from the central engine by the end of the simulation. As discussed in \citet{jos07}, for the choice of our particle injection index for the baseline model the corresponding particle acceleration scenario is most likely Fermi first-order.

As can be seen from Fig. \ref{1sed}, in terms of the radiative processes, the IR and optical emission of the base set is due to synchrotron radiation only. The 2.4 keV and 10 keV X-ray photons are SSC-dominated while the 50 MeV and 0.1 GeV emission are ECDT-dominated with some contribution from the ECBLR process. The 1 GeV emission is due to a combination of the peaks of both ECDT and ECBLR spectra while the 10 GeV photons are ECBLR-dominated. The 100 GeV emission, on the other hand, are due to a combination of the declining parts of the ECDT and ECBLR spectra.

\begin{figure}[htp]
\hspace*{-0.5in}
{\includegraphics[width=0.5\textwidth]{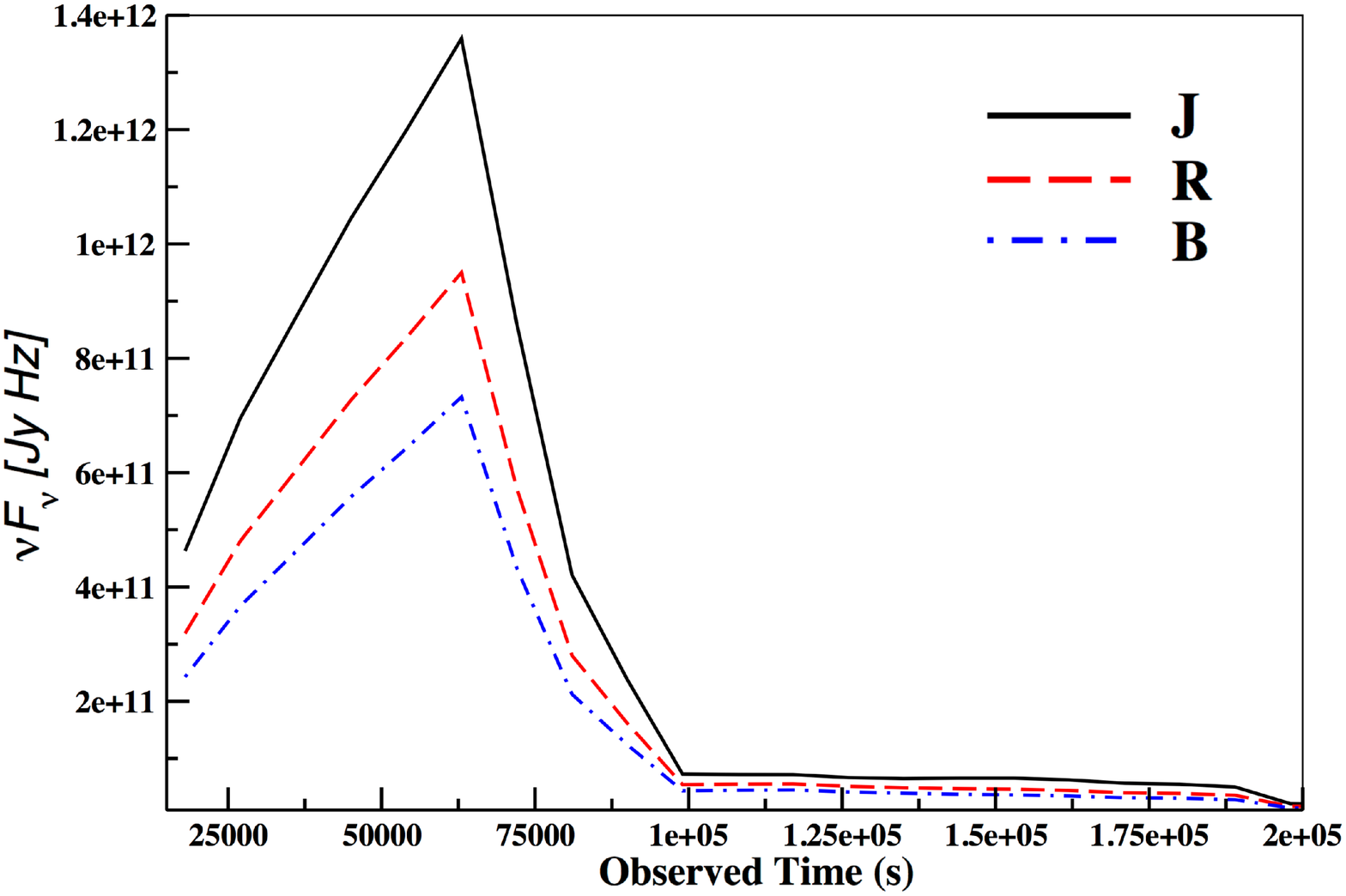}}\hfill
{\includegraphics[width=0.5\textwidth]{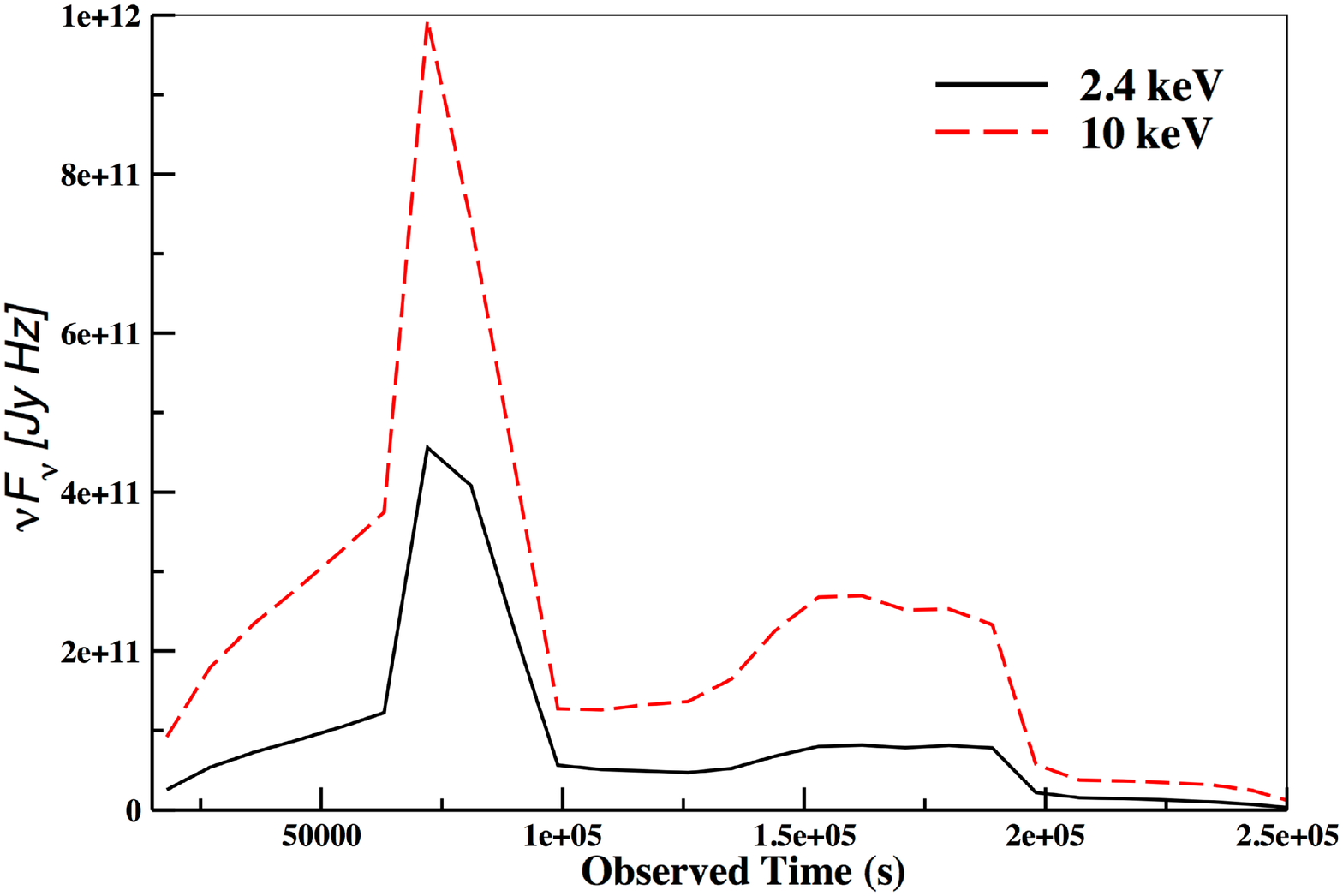}}\hfill
\hspace*{-0.2in}
{\includegraphics[width=0.5\textwidth]{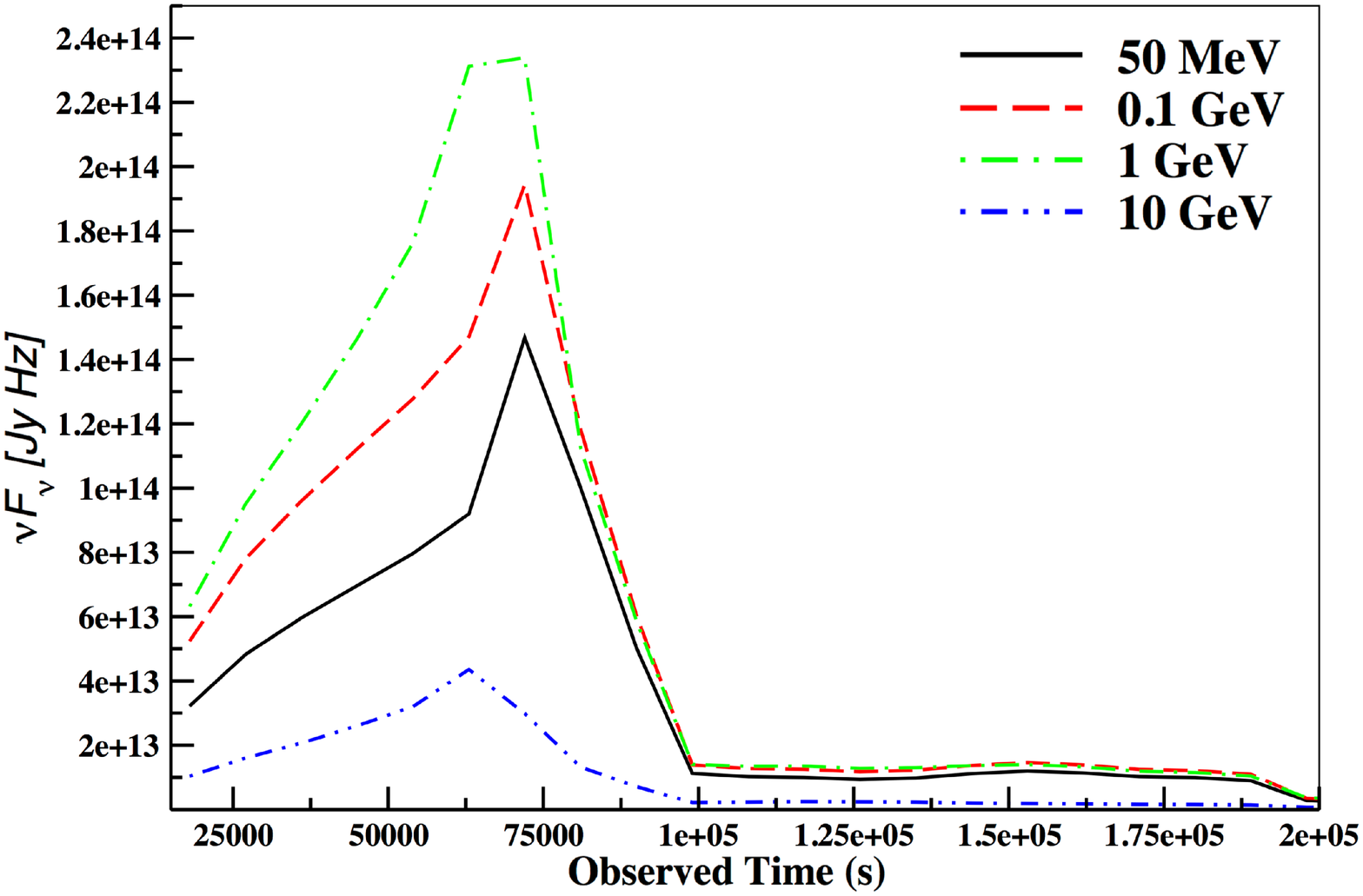}}\hfill
{\includegraphics[width=0.5\textwidth]{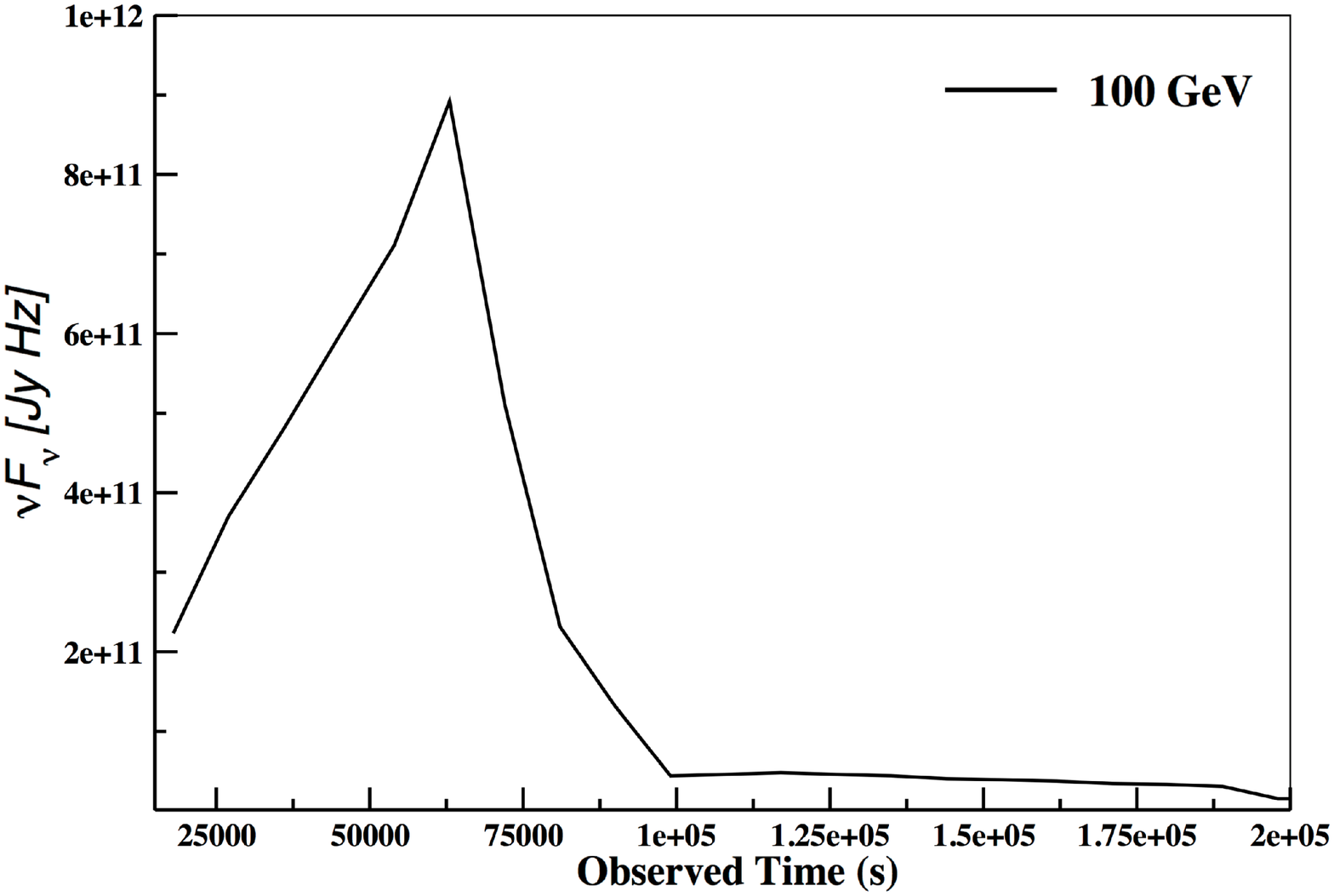}}
\caption{Simulated light curves of the baseline model (Run 1) \textbf{on a linear scale}. Top left: at IR (J), and optical ($R$ and $B$) bands; Top right: at 2.4 and 10 keV X-ray energies; Bottom left: at 50 MeV, 0.1, 1, and 10 GeV $\gamma$-ray energies; Bottom right: at 100 GeV.}
\label{0lcs}
\end{figure}

Figure \ref{0lcs} shows the light curve profiles for the base set calculated for the $J$-band ($\nu = 2.44 \times 10^{14}$~Hz), $R$-band ($\nu = 4.68 \times 10^{14}$~Hz), $B$-band ($\nu = 6.81 \times 10^{14}$~Hz), at 2.4 keV ($\nu = 5.8 \times 10^{17}$~Hz), 10 keV ($\nu = 2.42 \times 10^{18}$~Hz), 50 MeV ($\nu = 1.21 \times 10^{22}$~Hz), 0.1 GeV ($\nu = 2.42 \times 10^{22}$~Hz), 1 GeV ($\nu = 2.42 \times 10^{23}$~Hz), 10 GeV ($\nu = 2.42 \times 10^{24}$~Hz), and 100 GeV ($\nu = 2.42 \times 10^{25}$~Hz). The system is in the acceleration-dominated phase for as long as the shocks are inside the emission region. During that time, respective pulses rise steadily and reach their peaks as particles are constantly being accelerated. For almost all light curves, beyond $\sim 10^{5}$ s, the shocks are out of the system and the pulse profiles are dominated only by the cooling. The outbursts are not strongly asymmetric with $|\xi| < 0.6$.

\onecolumn

\begin{figure}
\centering
\begin{tabular}{cccc}
\includegraphics[width=0.3\textwidth]{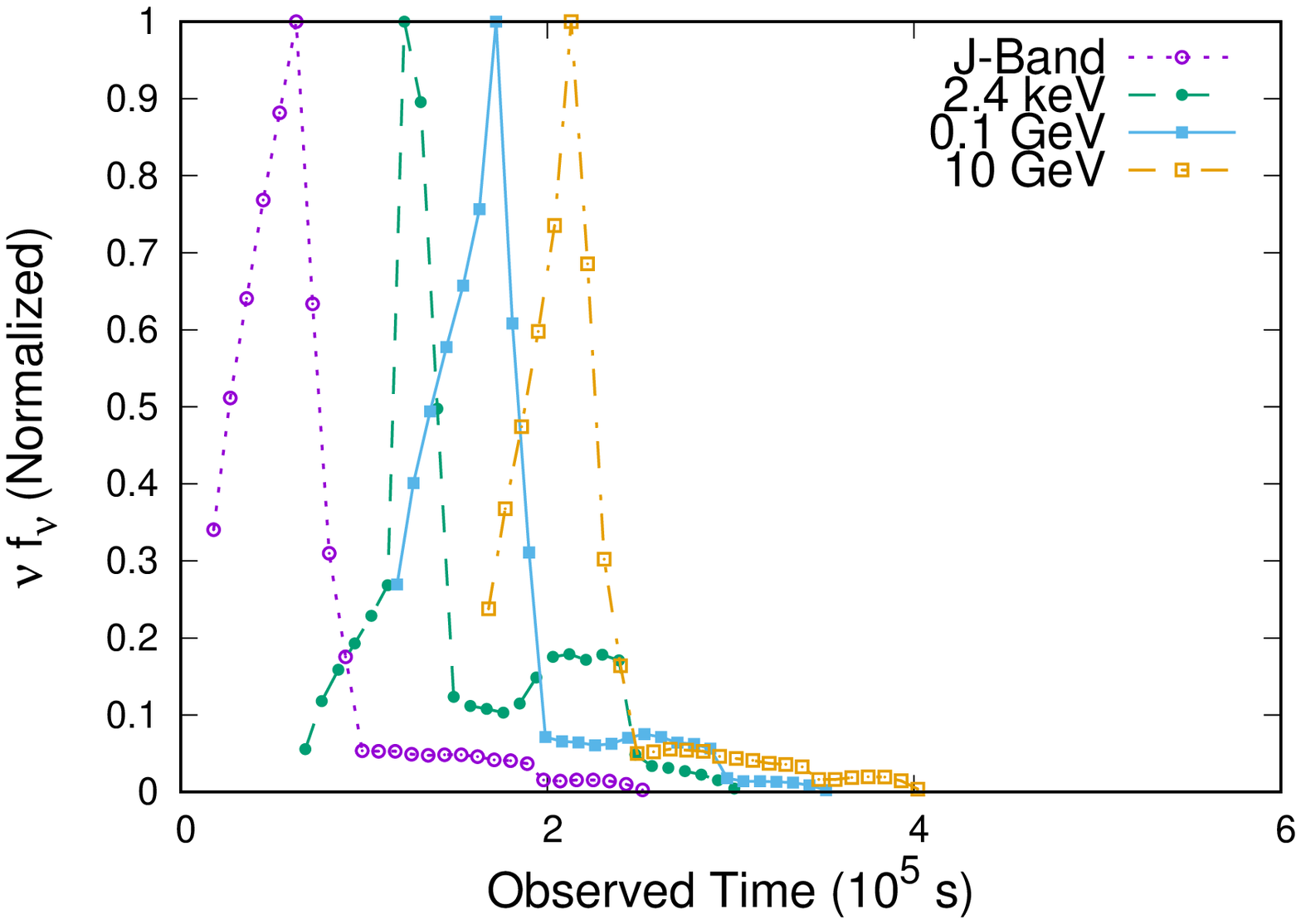} &
\includegraphics[width=0.3\textwidth]{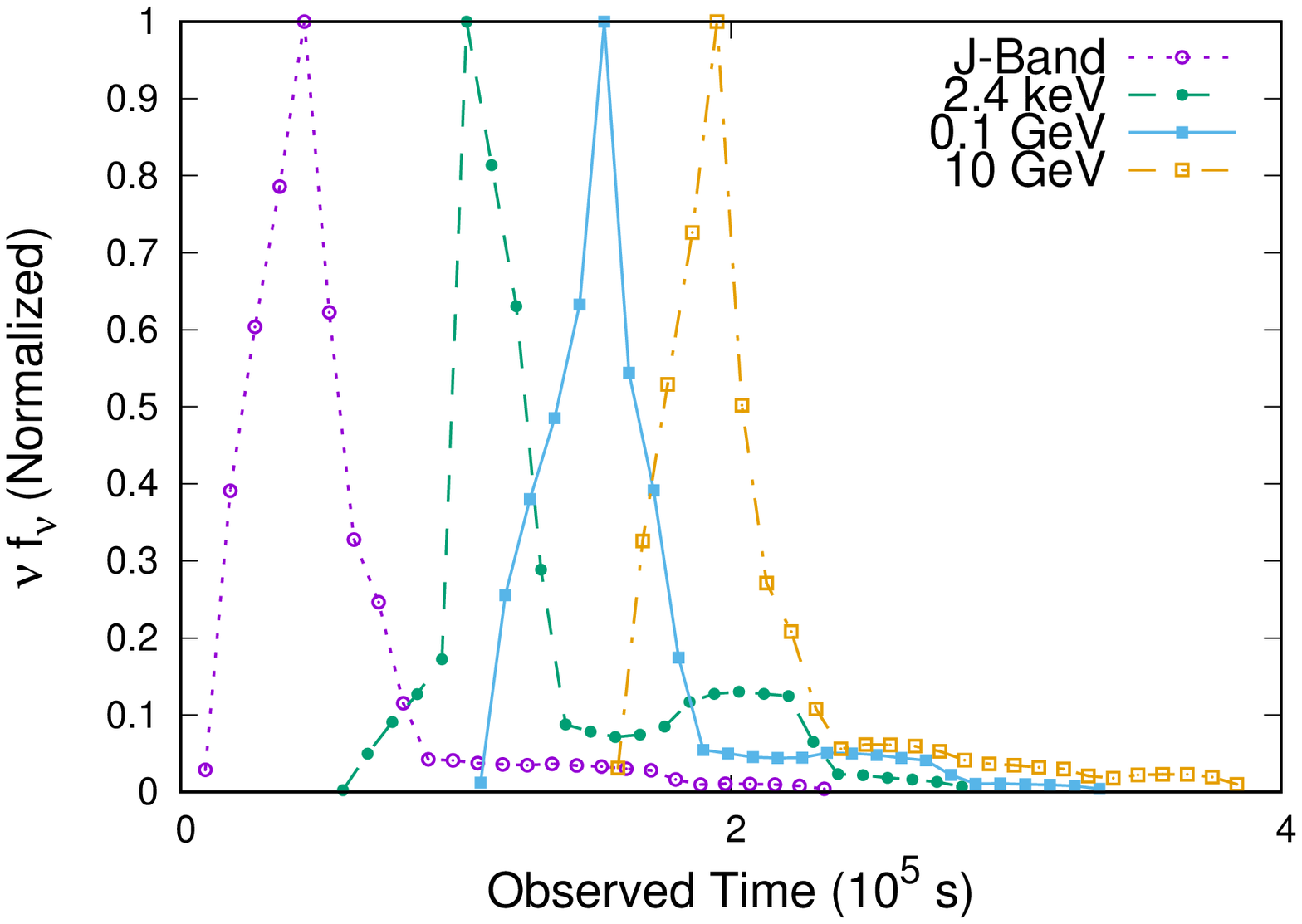} &
\includegraphics[width=0.3\textwidth]{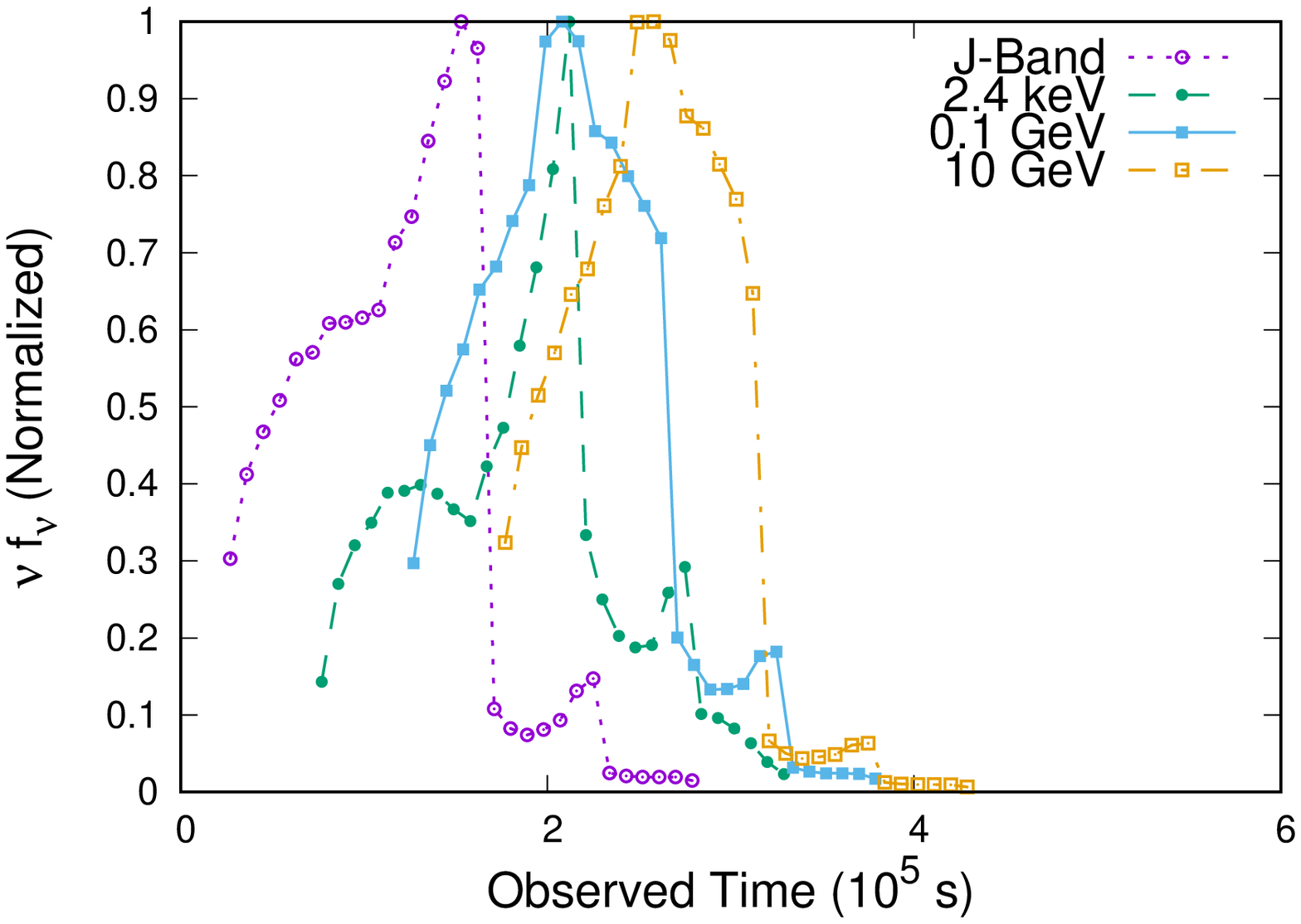} \\
\textbf{Run 1}  & \textbf{Run 2} & \textbf{Run 3}  \\[6pt]
\\
\end{tabular}
\begin{tabular}{cccc}
\includegraphics[width=0.3\textwidth]{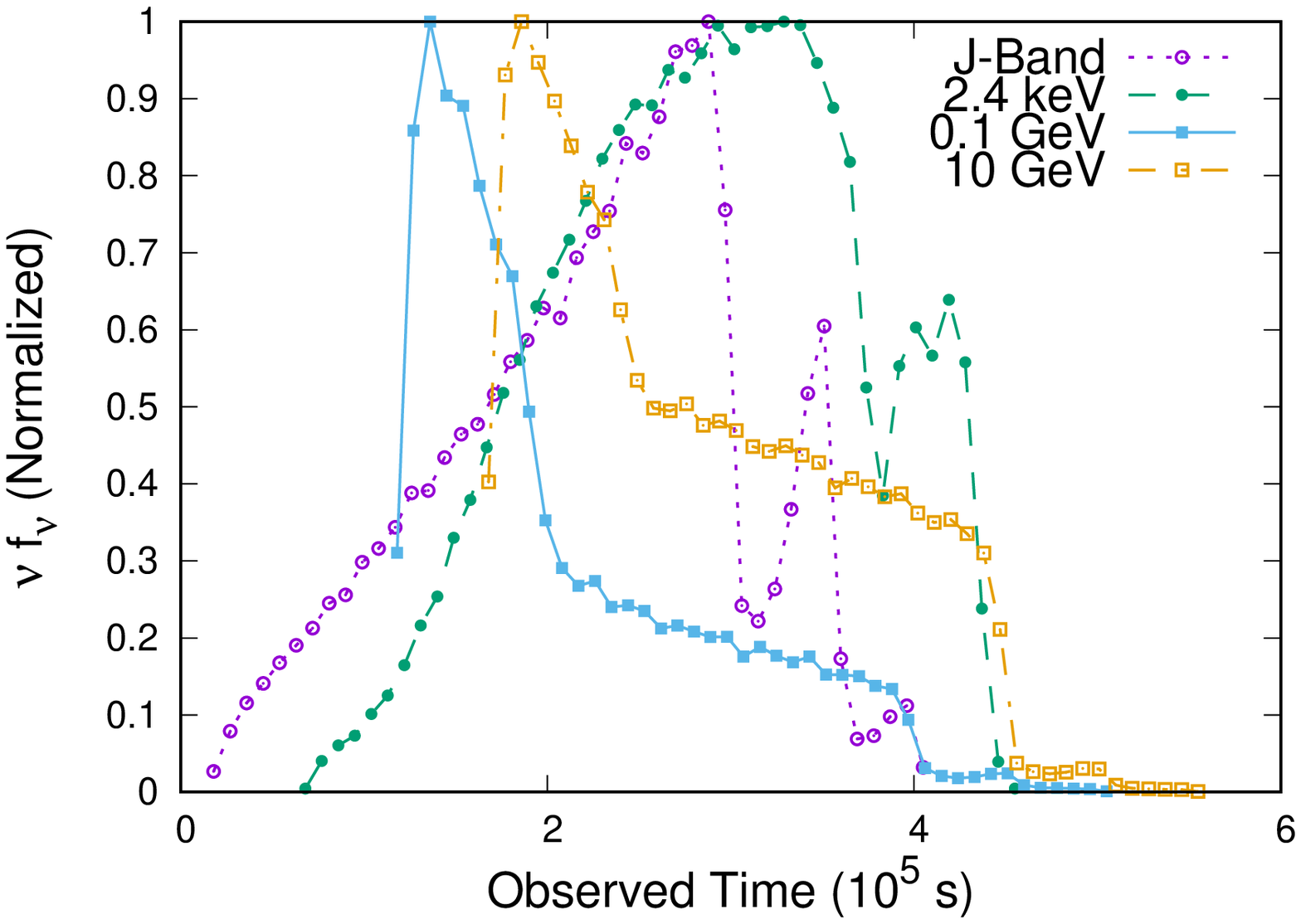} &
\includegraphics[width=0.3\textwidth]{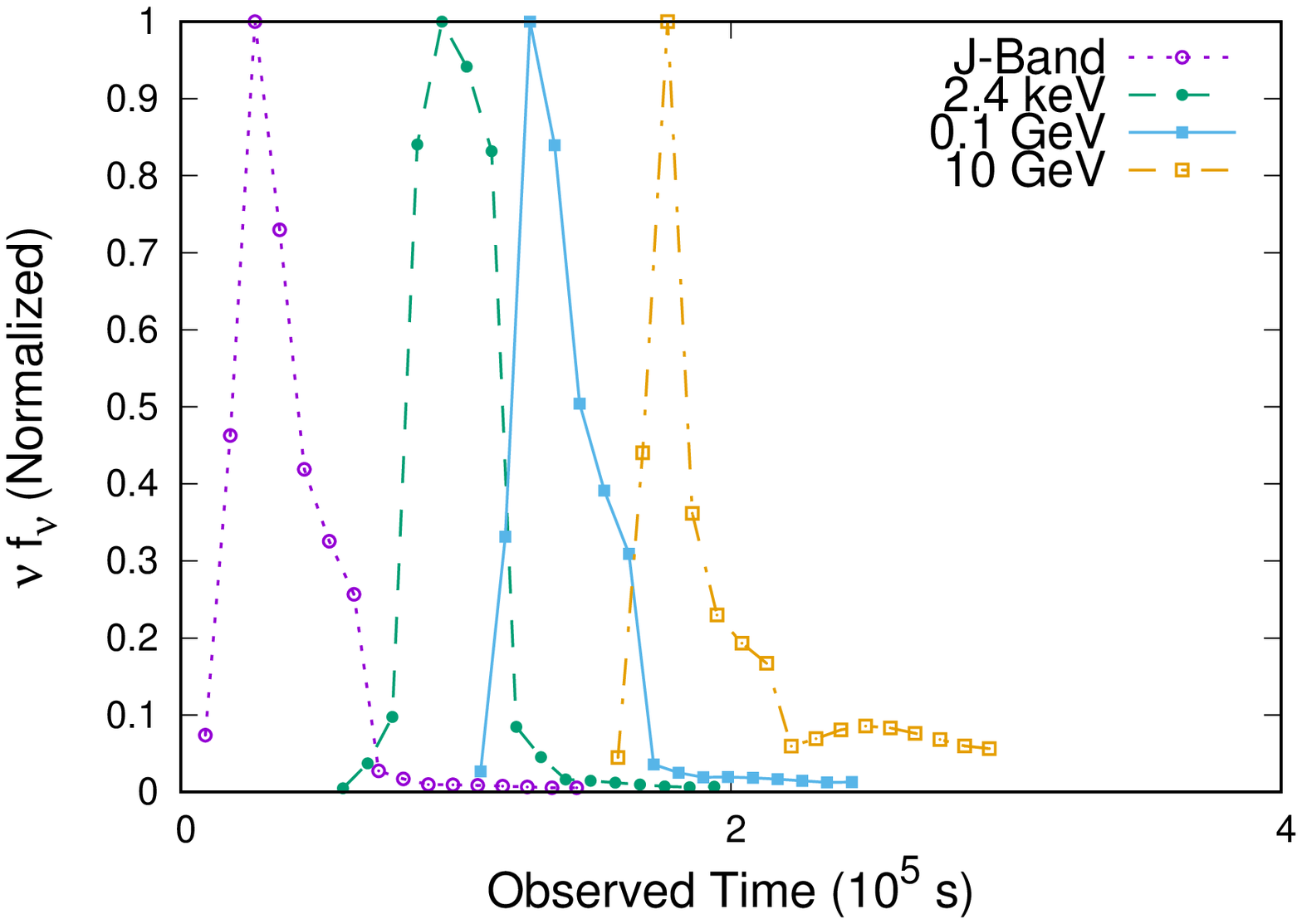} &
\includegraphics[width=0.3\textwidth]{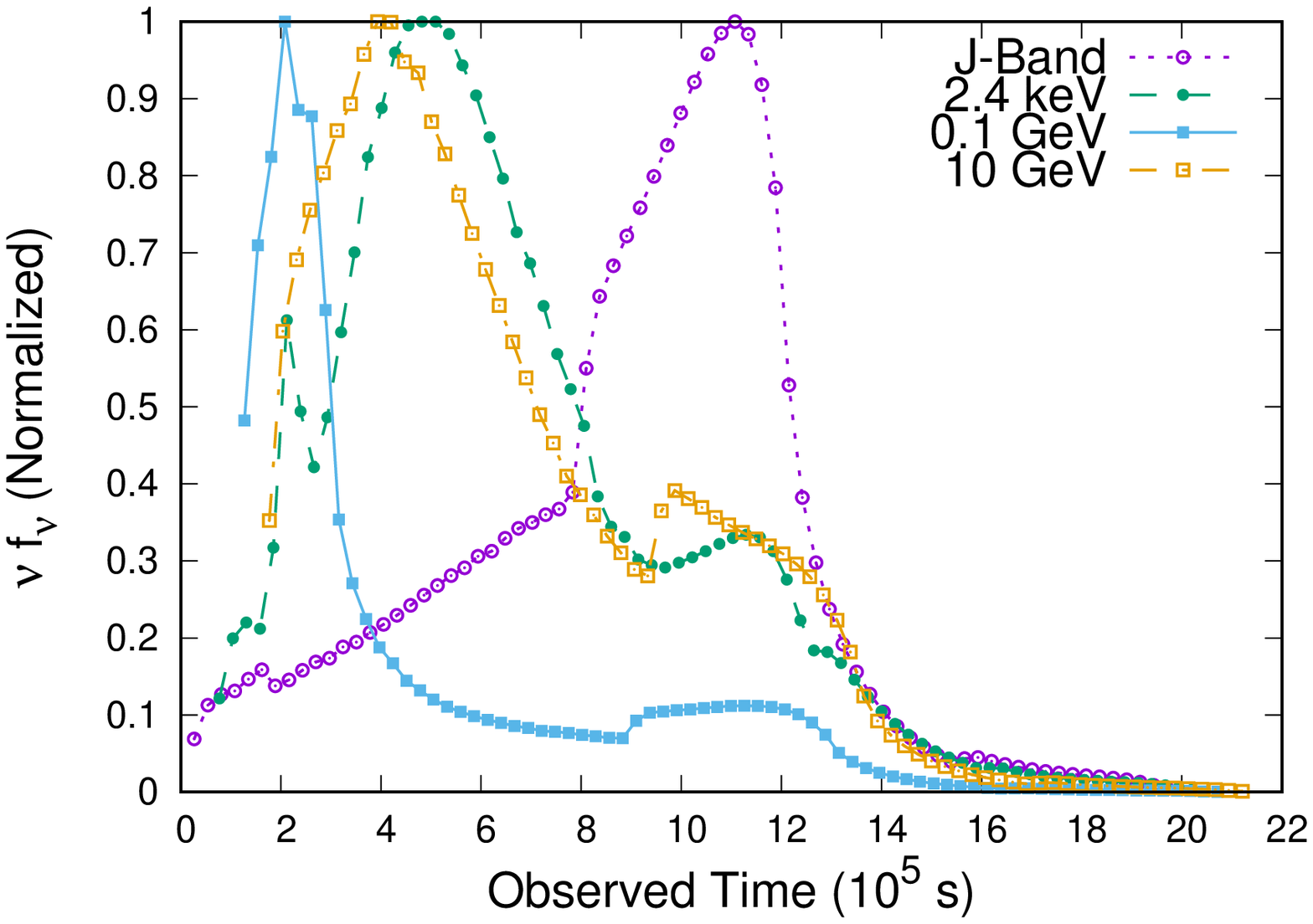} \\
\textbf{Run 4}  & \textbf{Run 5} & \textbf{Run 6}  \\[6pt]
\\
\end{tabular}
\begin{tabular}{cccc}
\includegraphics[width=0.3\textwidth]{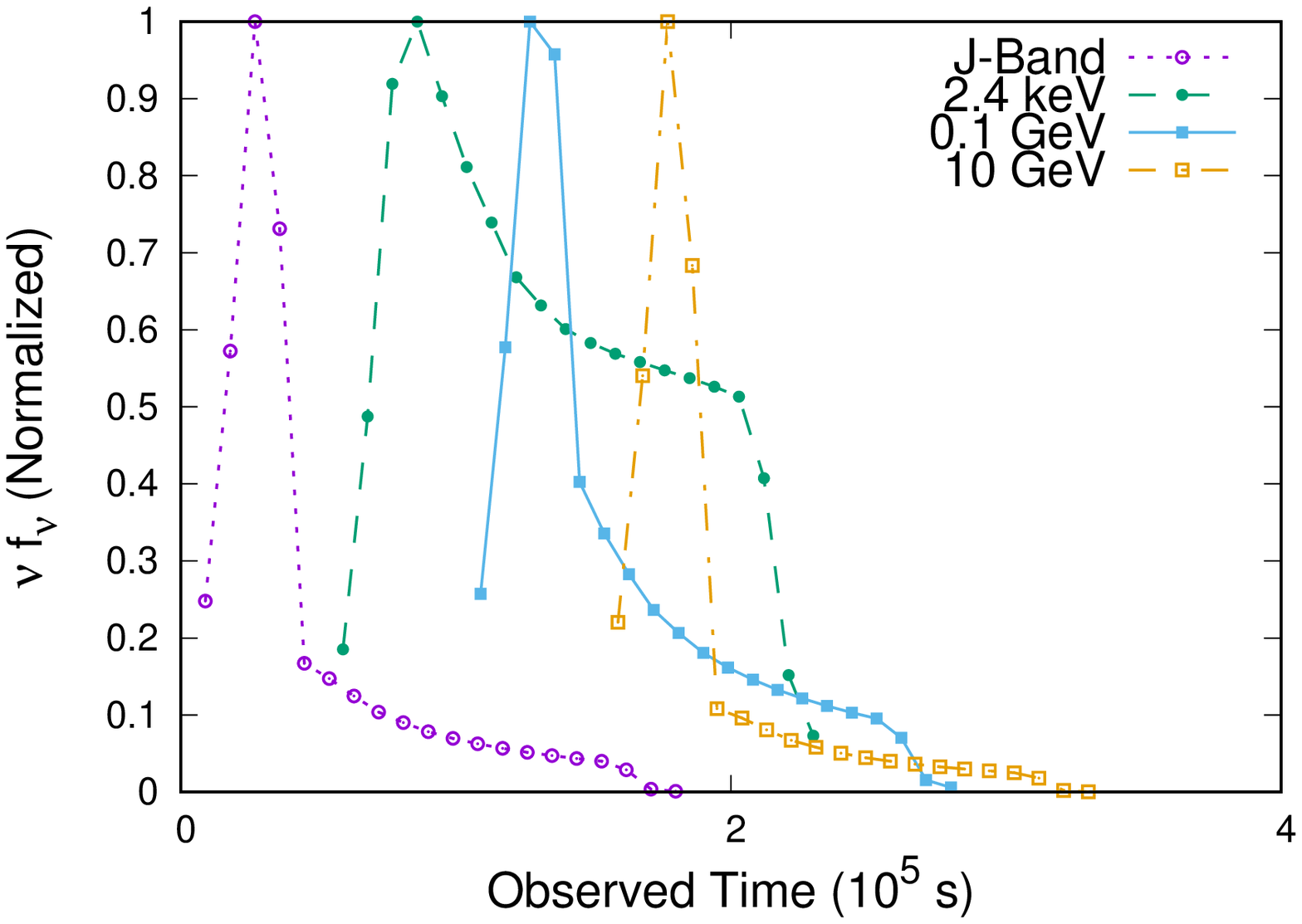} &
\includegraphics[width=0.3\textwidth]{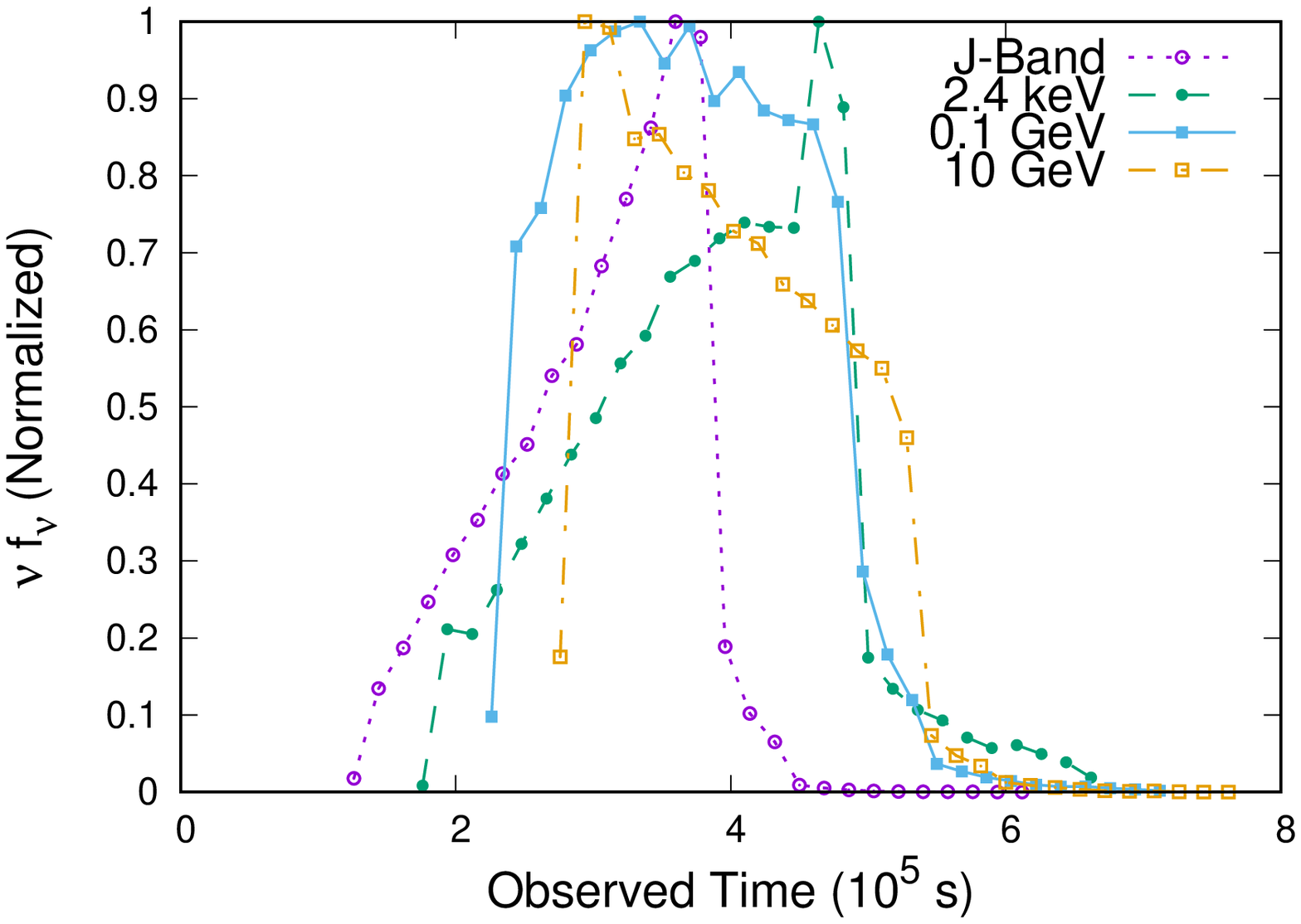} &
\includegraphics[width=0.3\textwidth]{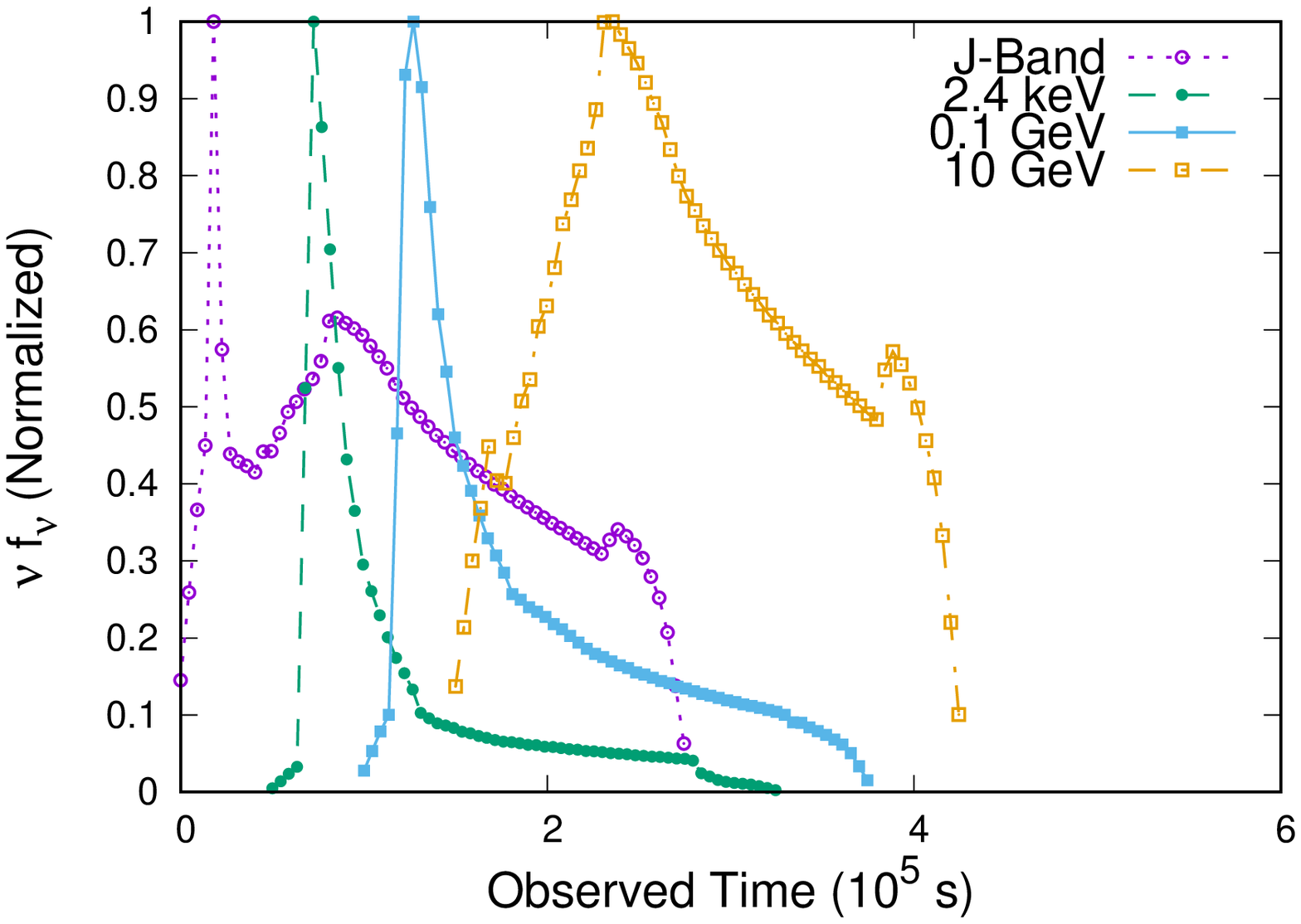} \\
\textbf{Run 7}  & \textbf{Run 8} & \textbf{Run 9}  \\[6pt]
\\
\end{tabular}
\begin{tabular}{cccc}
\includegraphics[width=0.3\textwidth]{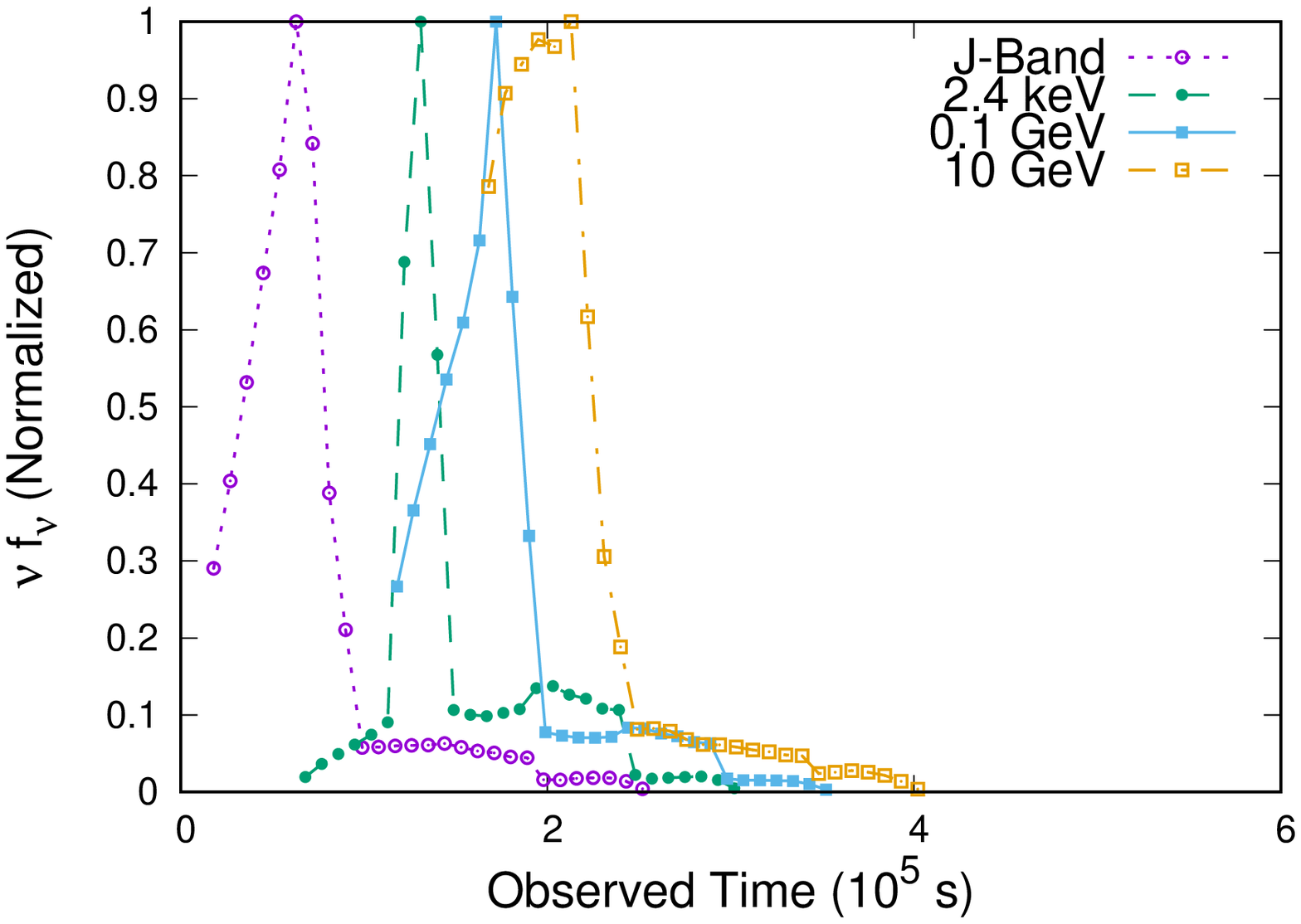} &
\includegraphics[width=0.3\textwidth]{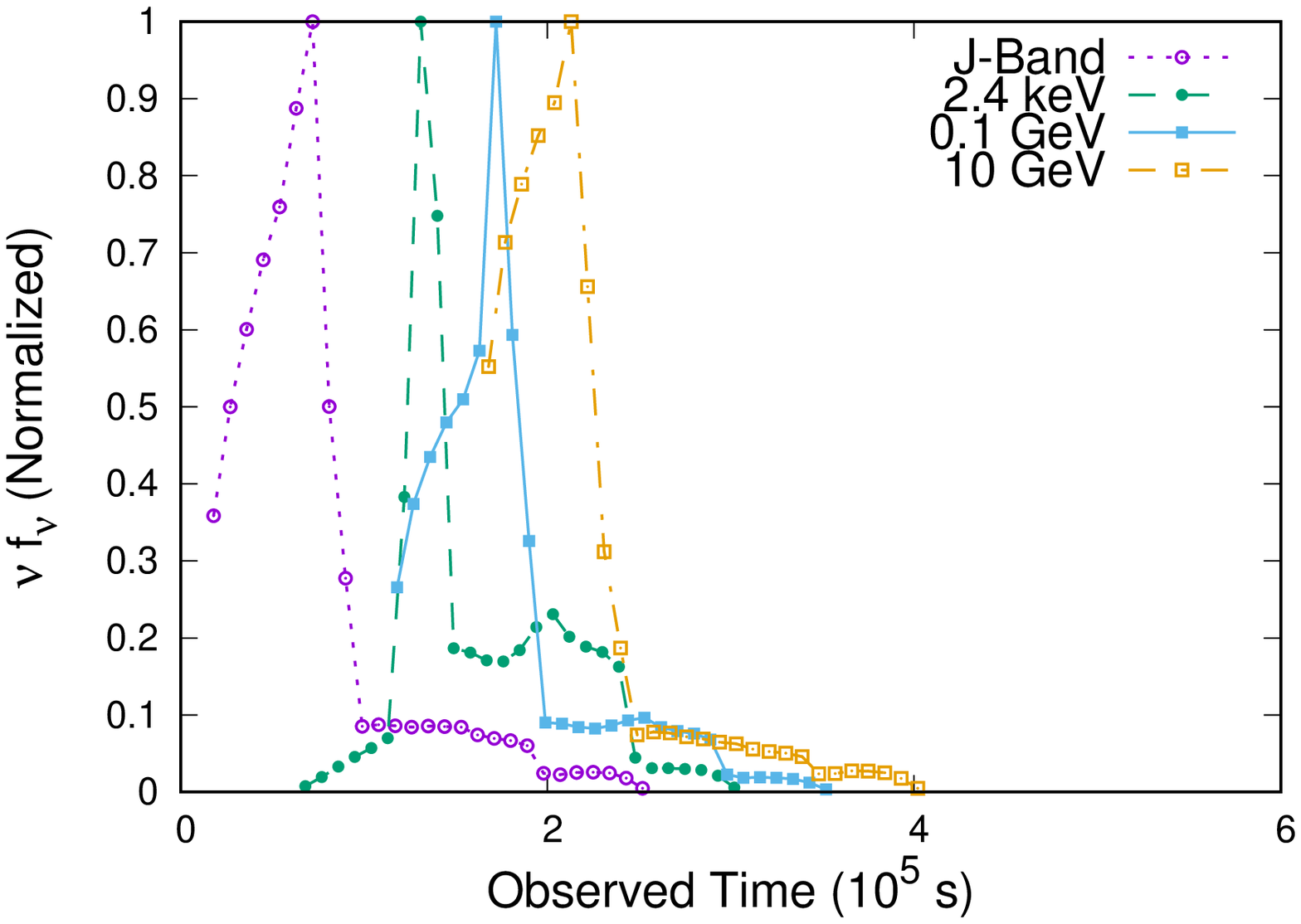} &
\includegraphics[width=0.3\textwidth]{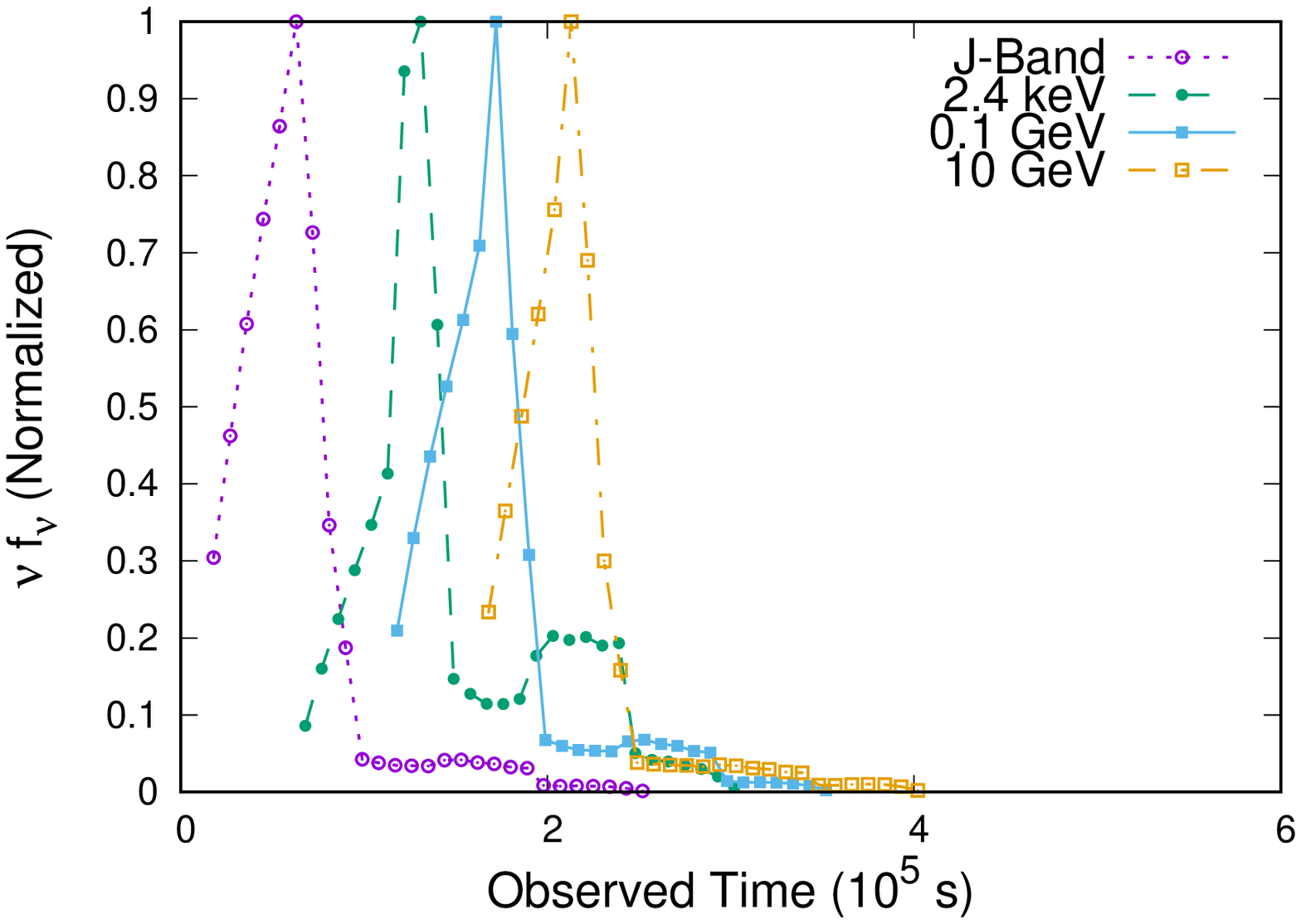} \\
\textbf{Run 10}  & \textbf{Run 11} & \textbf{Run 12}  \\[6pt]
\\
\end{tabular}
\begin{tabular}{cccc}
\includegraphics[width=0.3\textwidth]{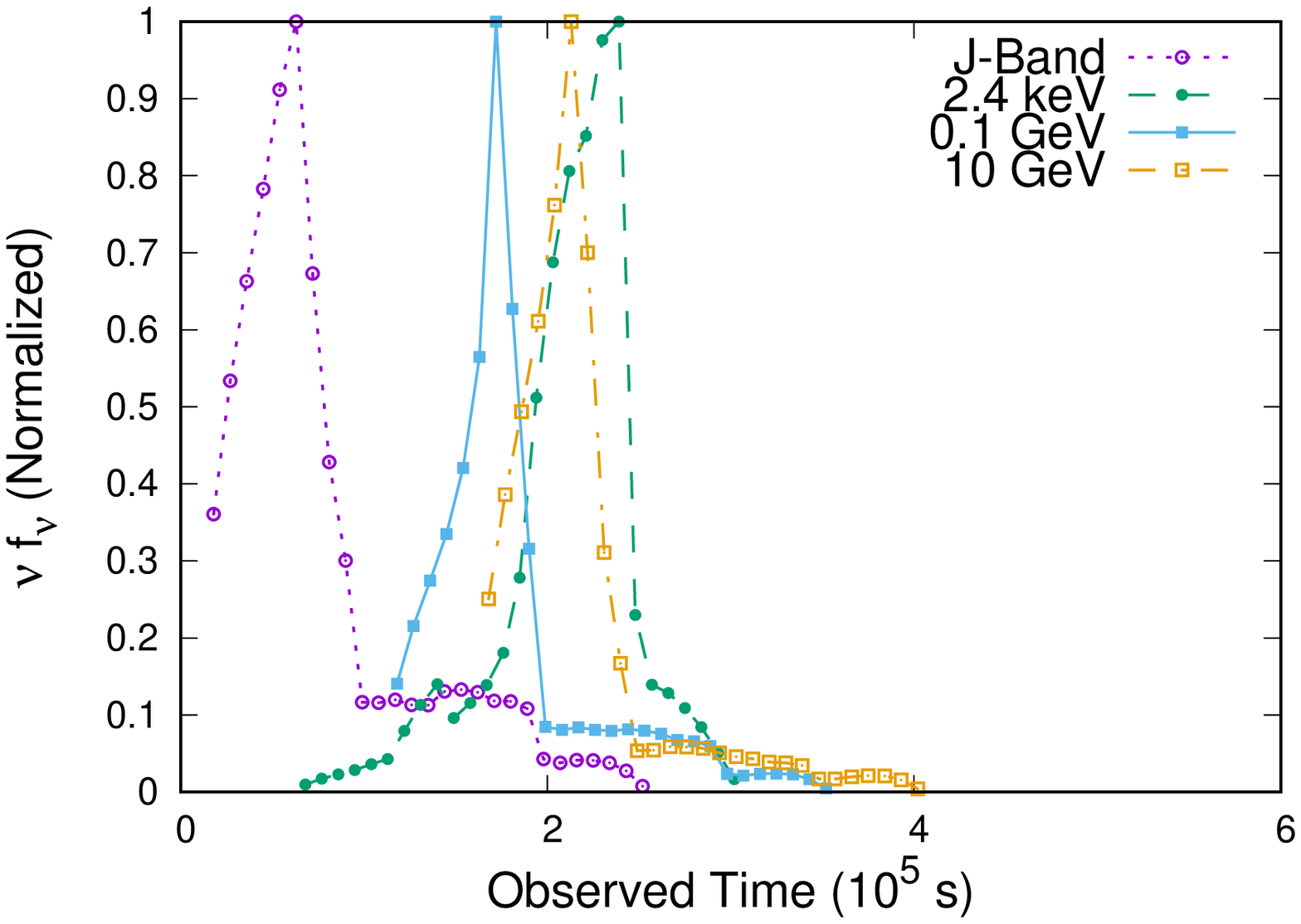} &
\includegraphics[width=0.3\textwidth]{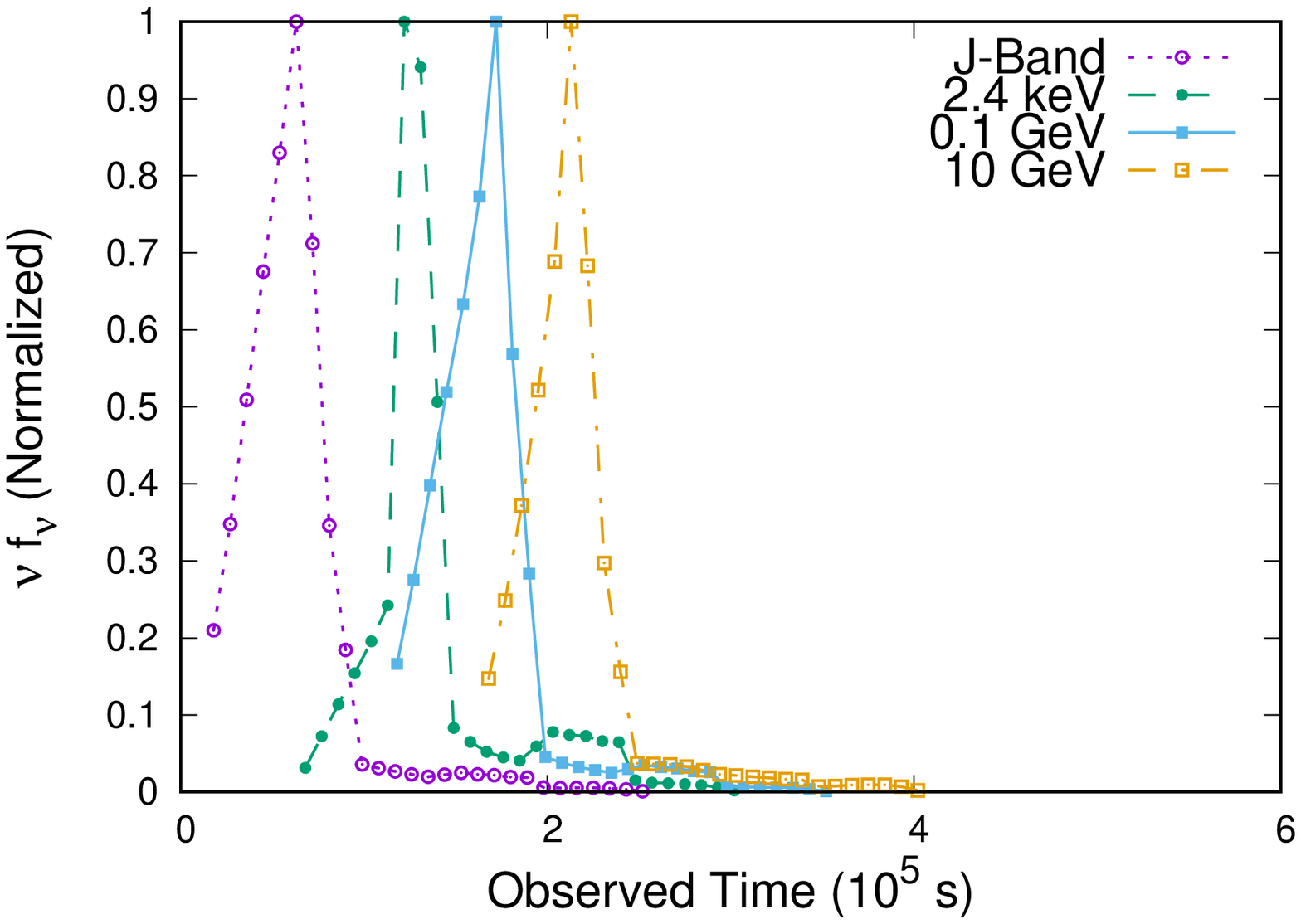} &
\includegraphics[width=0.3\textwidth]{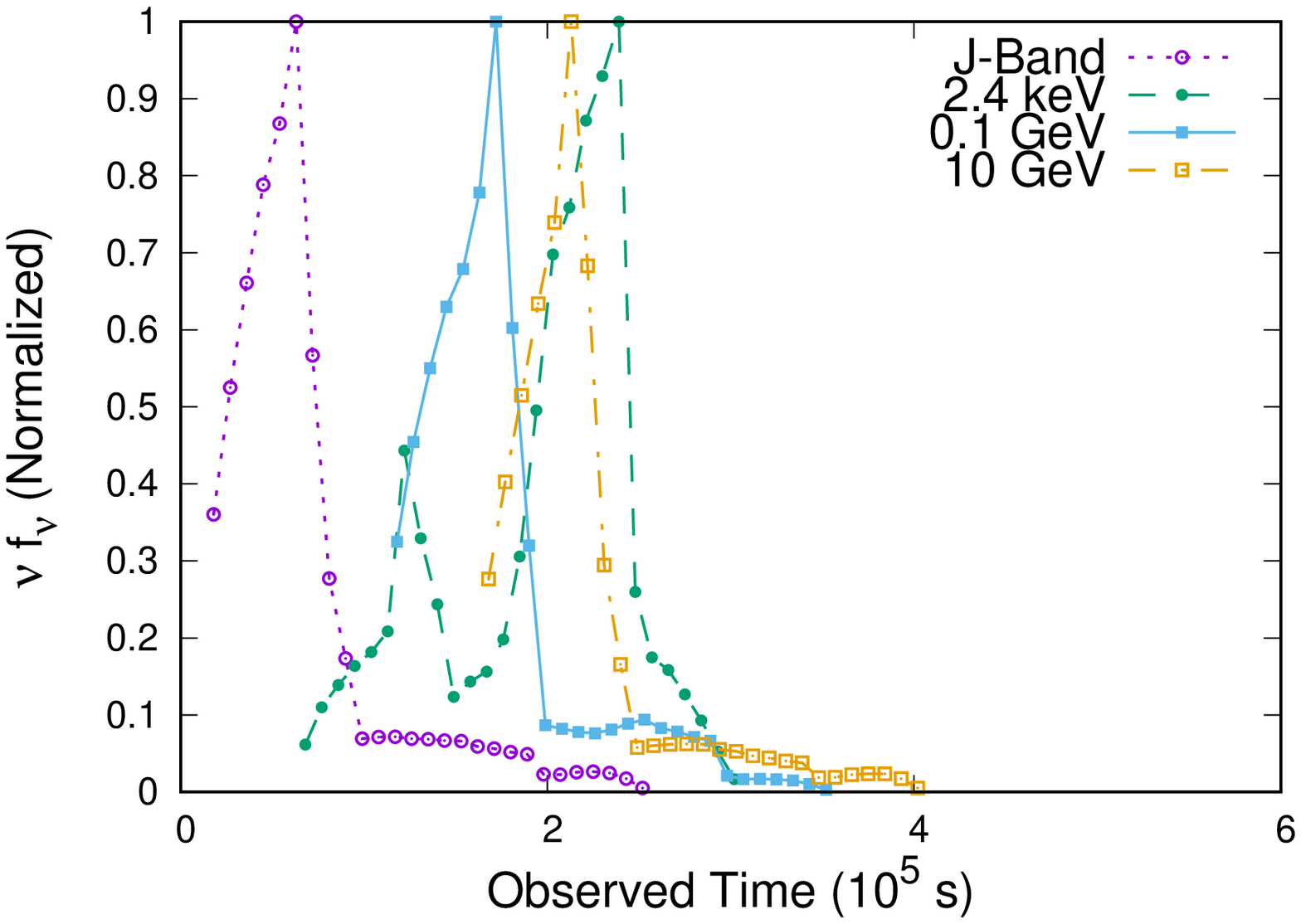} \\
\textbf{Run 13}  & \textbf{Run 14} & \textbf{Run 15}  \\[6pt]
\end{tabular}
\end{figure}

\clearpage
\newpage

\begin{figure}
\centering
\begin{tabular}{cccc}
\includegraphics[width=0.3\textwidth]{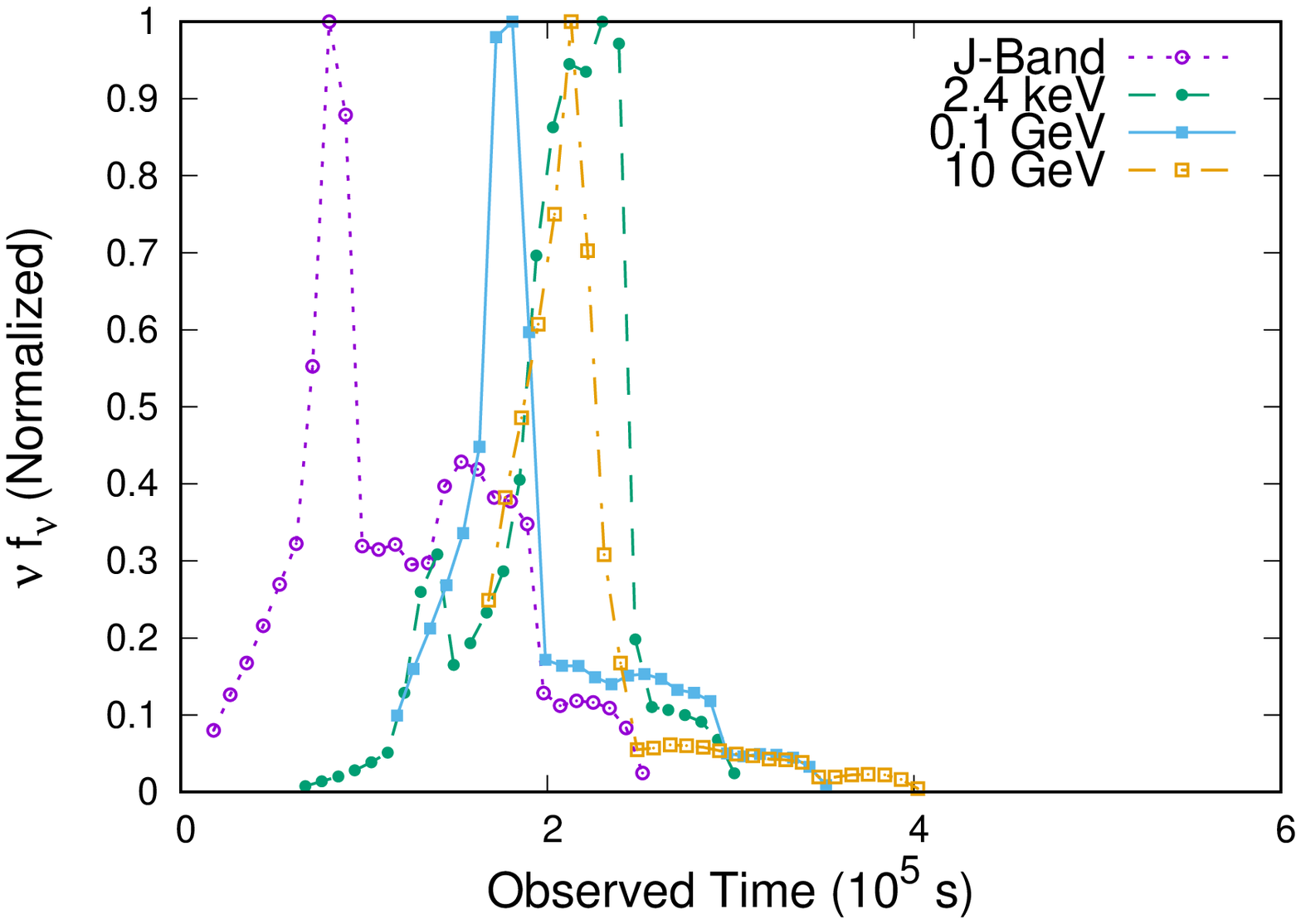} &
\includegraphics[width=0.3\textwidth]{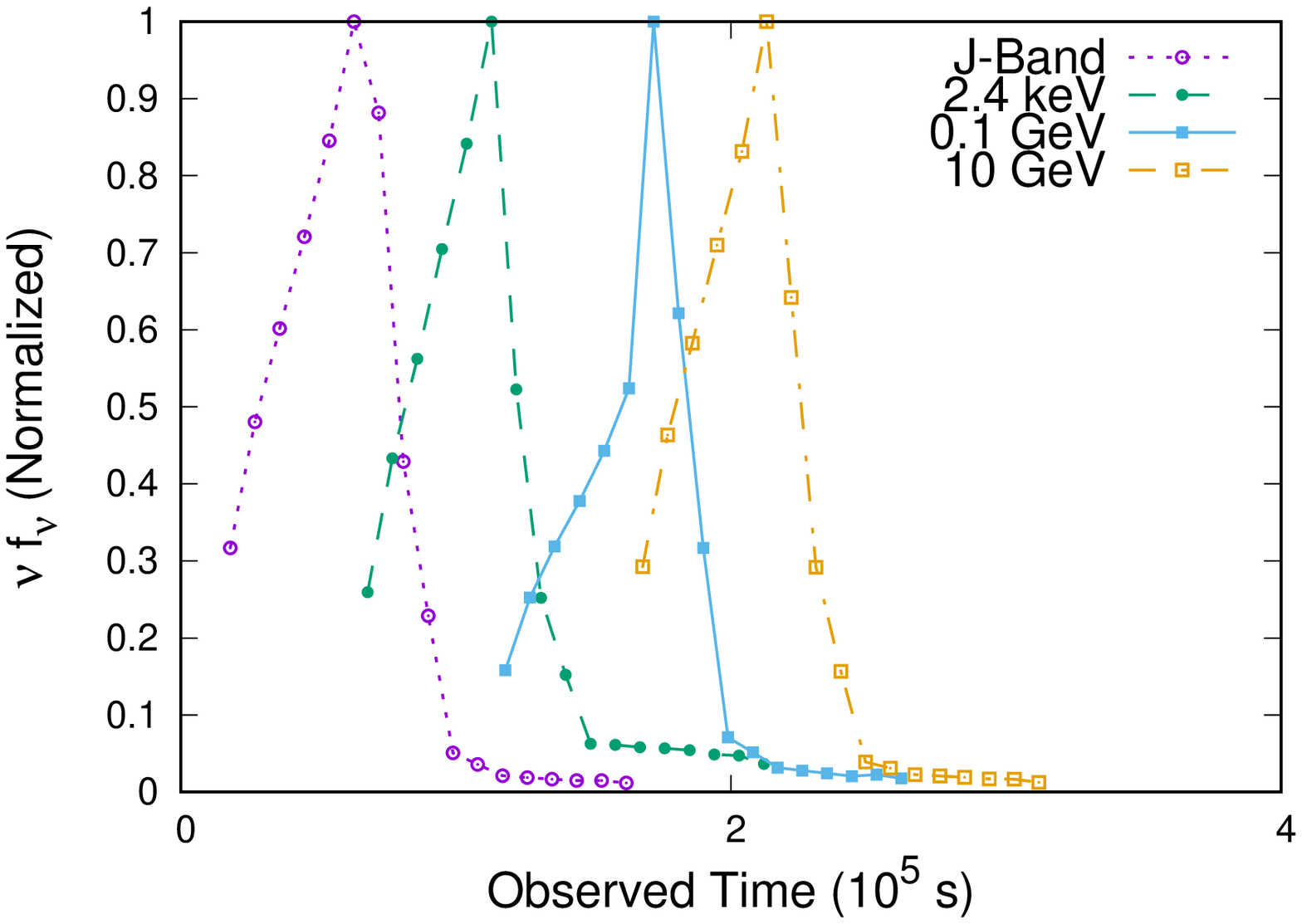} &
\includegraphics[width=0.3\textwidth]{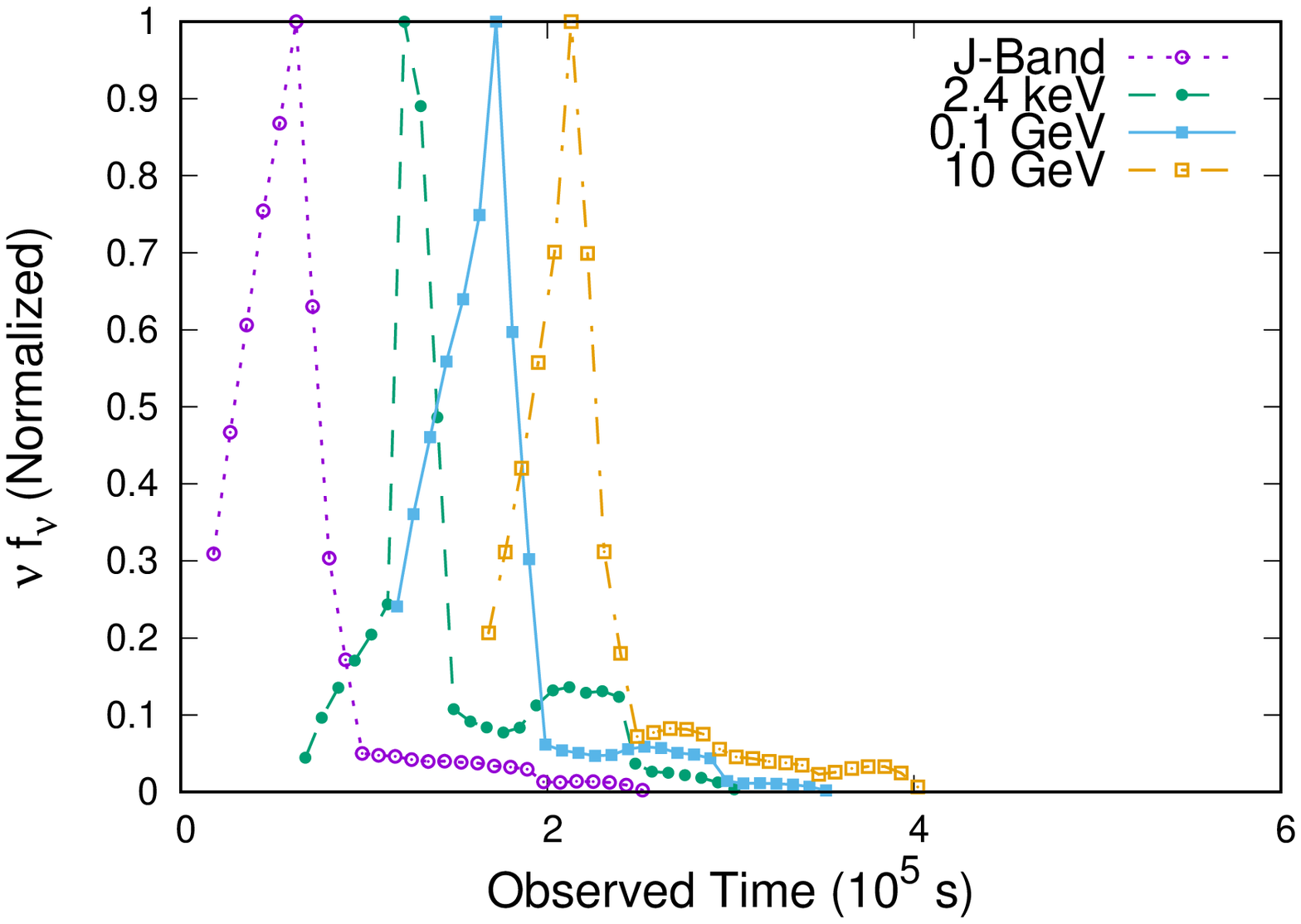} \\
\textbf{Run 16}  & \textbf{Run 17} & \textbf{Run 18}  \\[6pt]
\\
\end{tabular}
\begin{tabular}{cccc}
\includegraphics[width=0.3\textwidth]{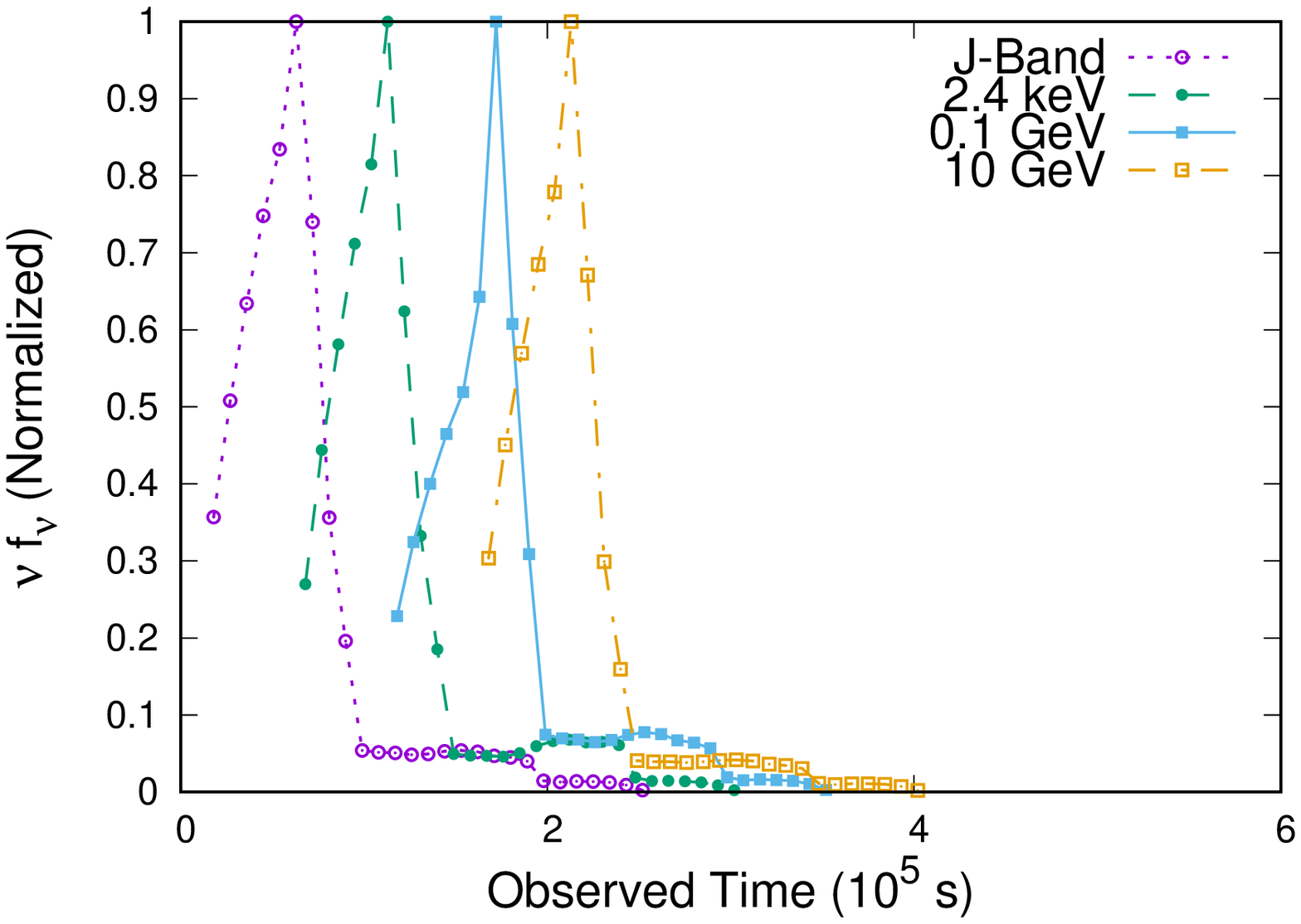} &
\includegraphics[width=0.3\textwidth]{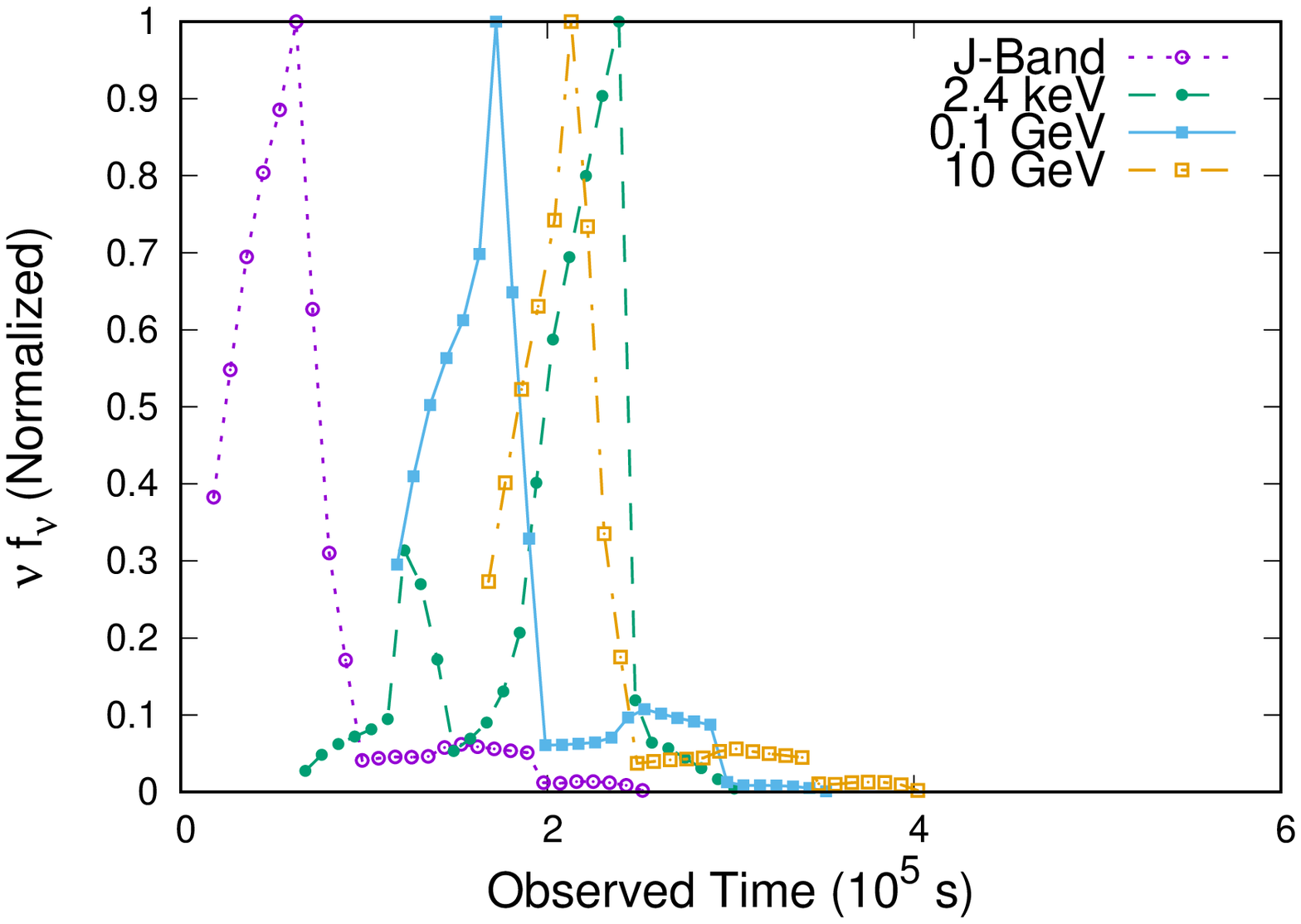} &
\includegraphics[width=0.3\textwidth]{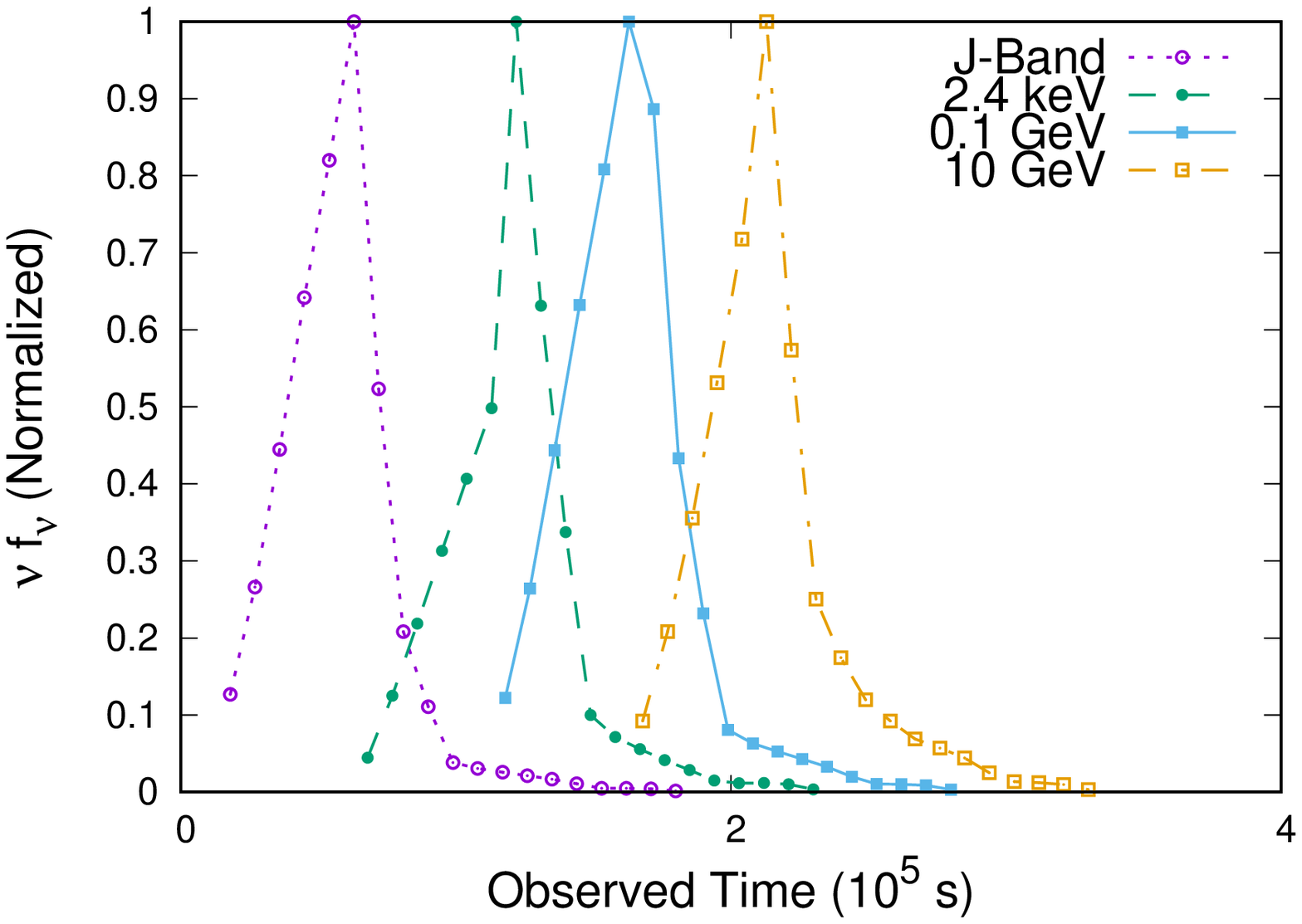} \\
\textbf{Run 19}  & \textbf{Run 20} & \textbf{Run 21}  \\[6pt]
\\
\end{tabular}
\begin{tabular}{cccc}
\includegraphics[width=0.3\textwidth]{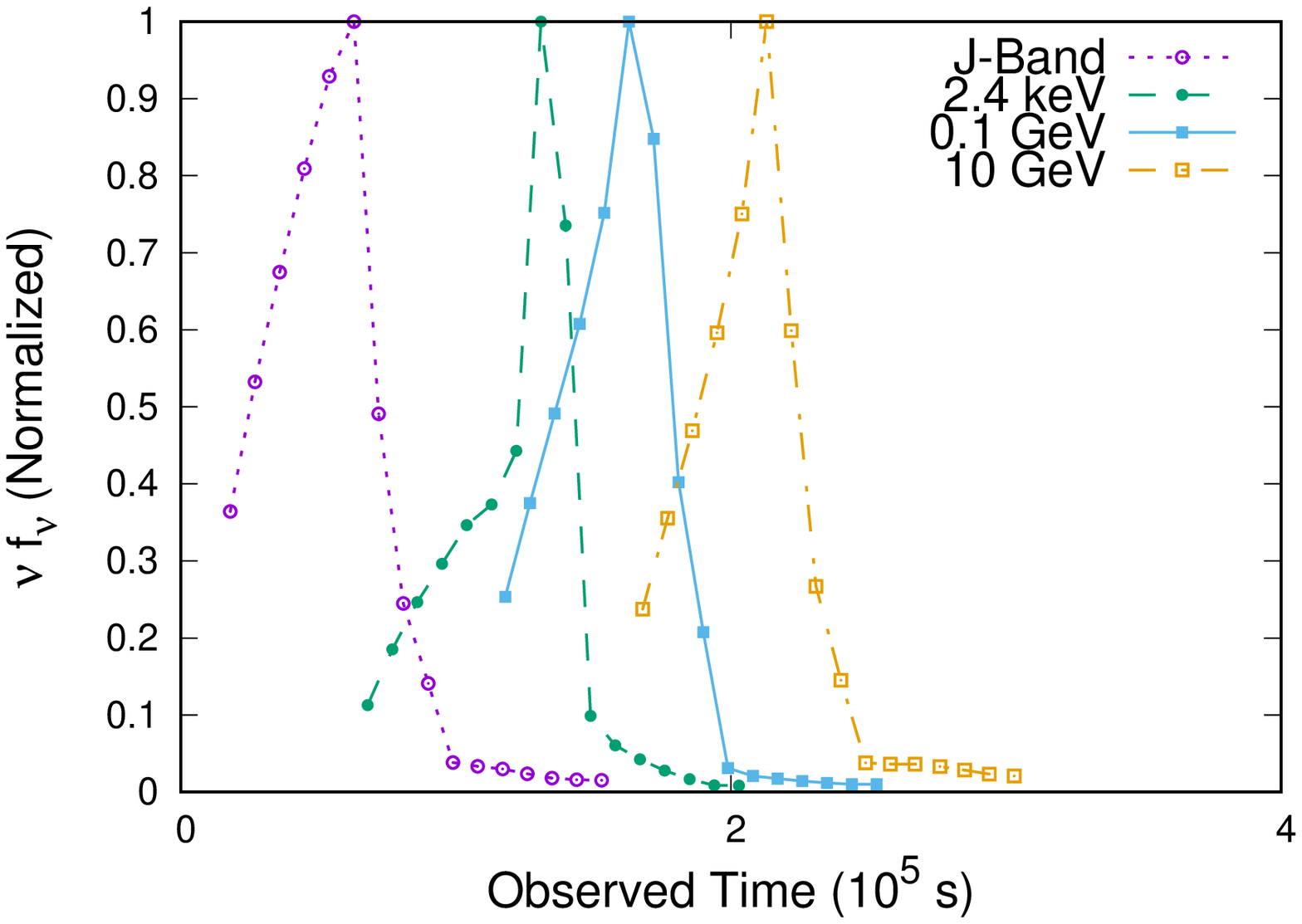} &
\includegraphics[width=0.3\textwidth]{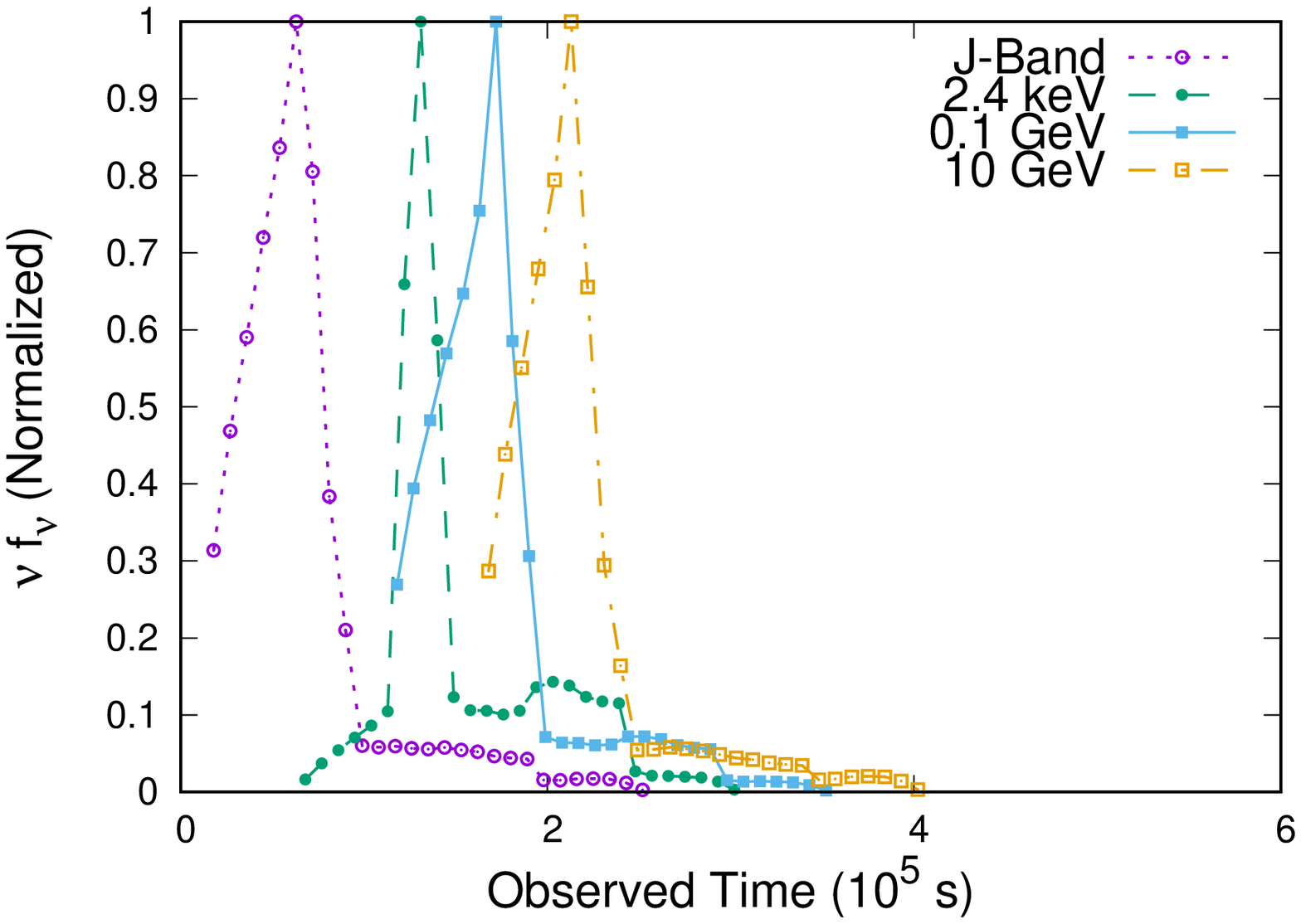} &
\includegraphics[width=0.3\textwidth]{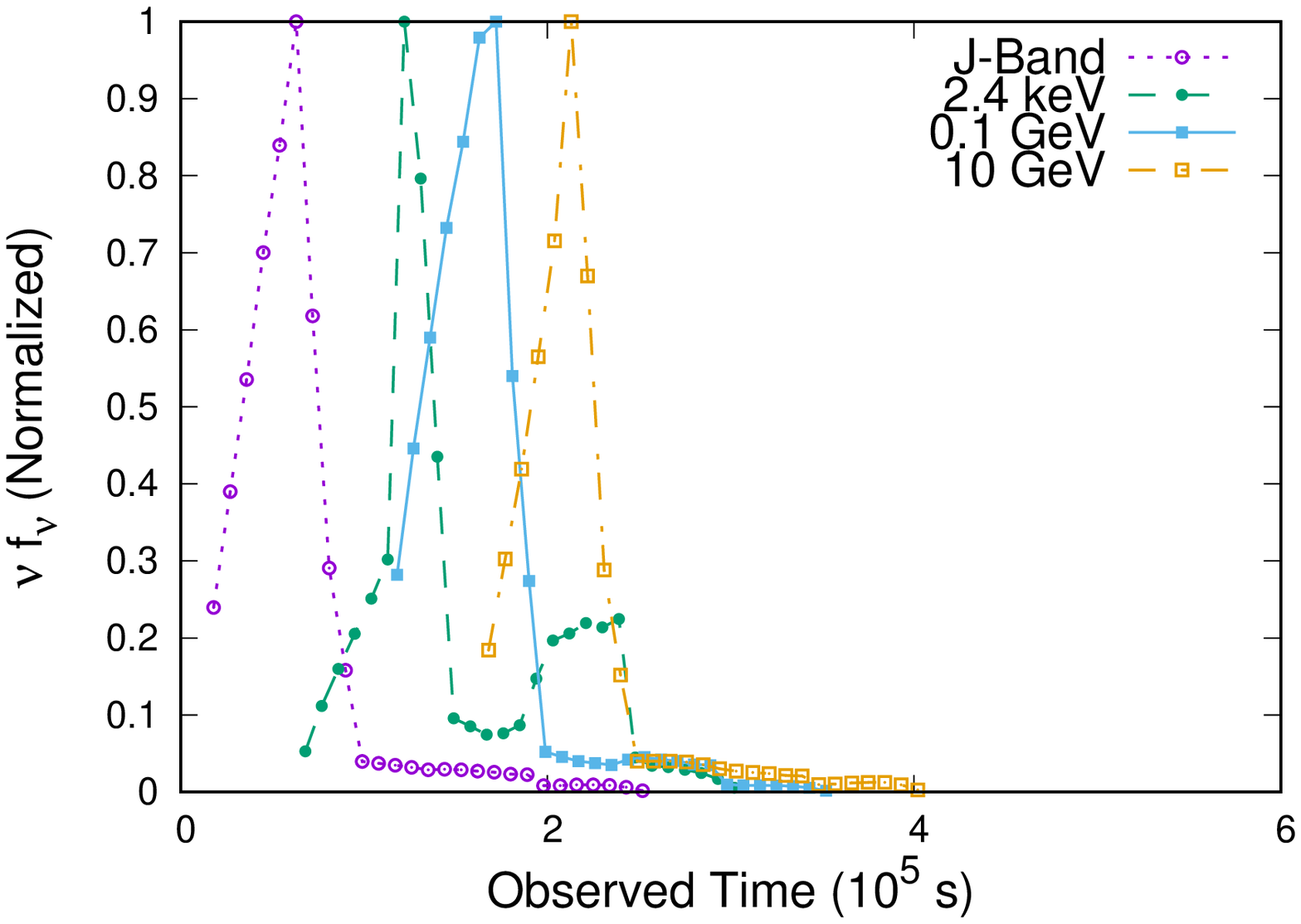} \\
\textbf{Run 22}  & \textbf{Run 23} & \textbf{Run 24}  \\[6pt]
\\
\end{tabular}
\begin{tabular}{cccc}
\includegraphics[width=0.3\textwidth]{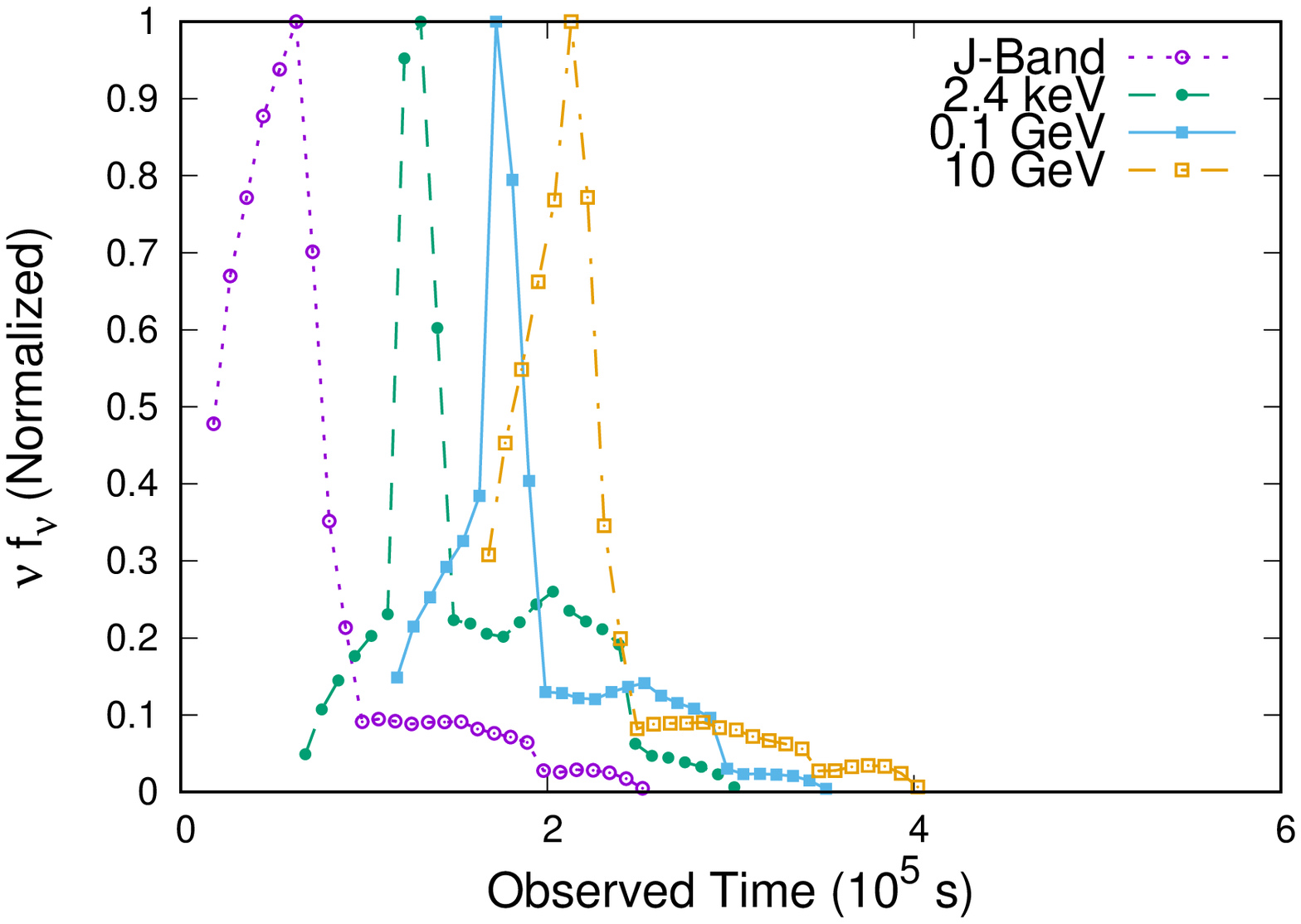} &
\includegraphics[width=0.3\textwidth]{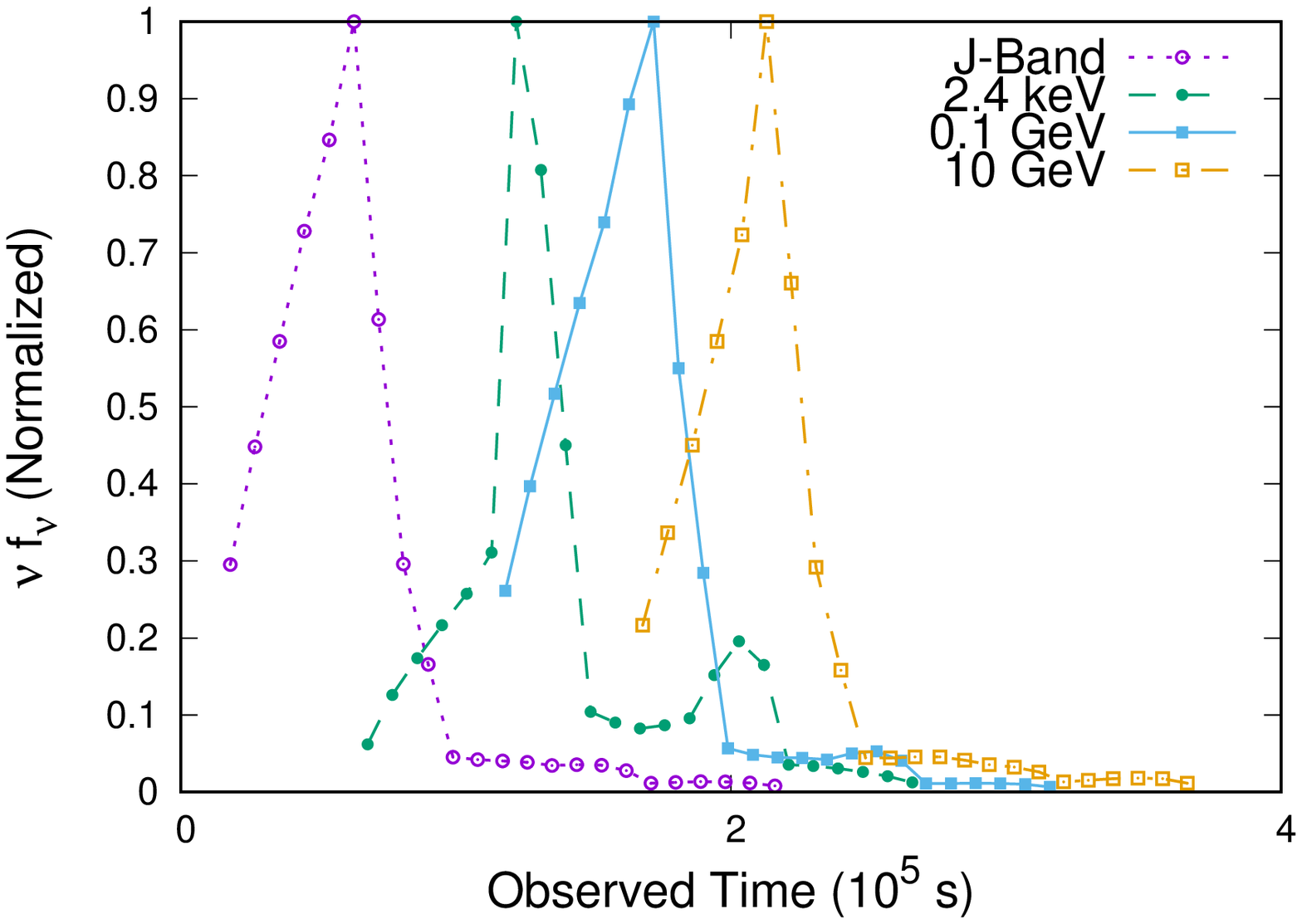} &
\includegraphics[width=0.3\textwidth]{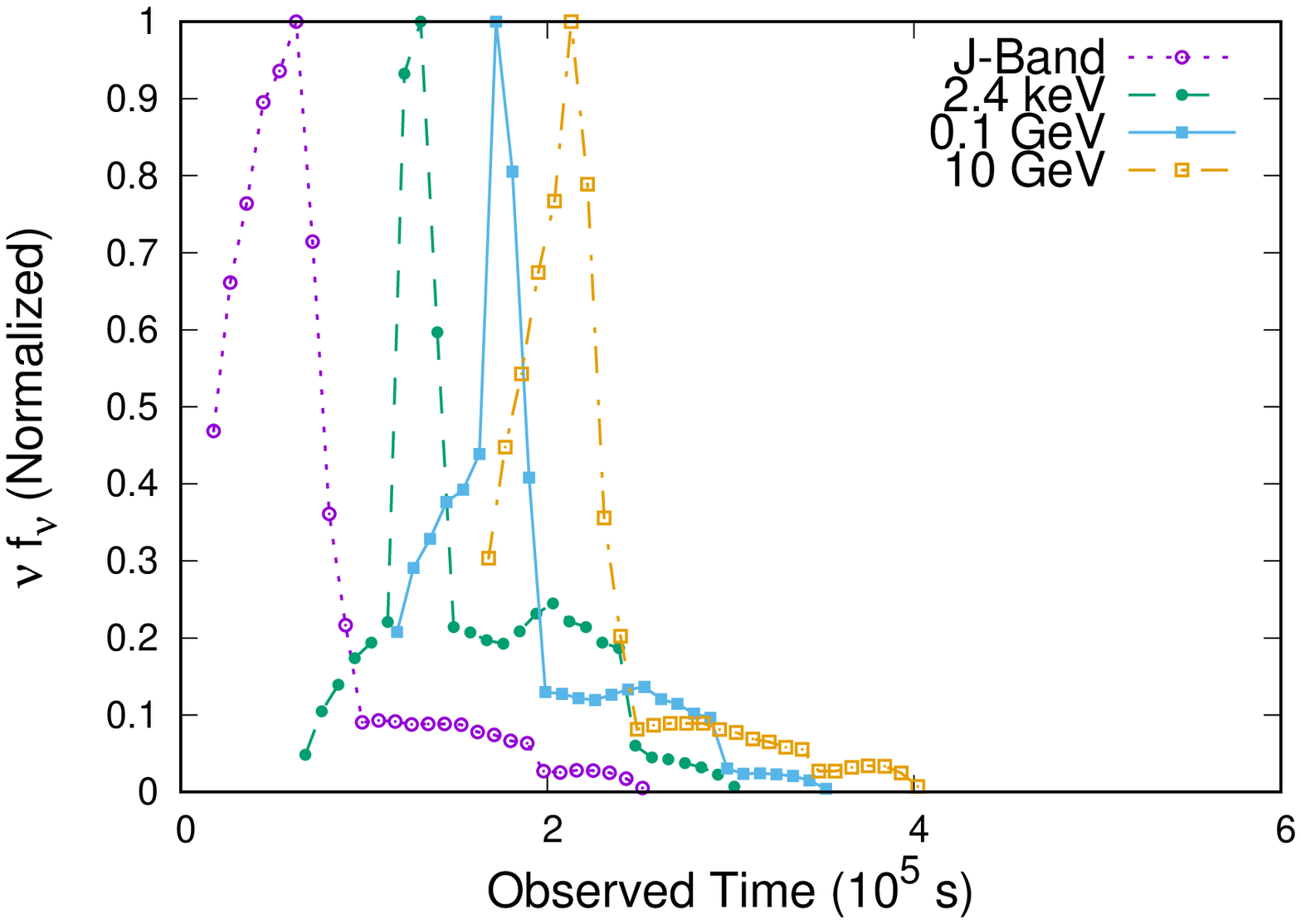} \\
\textbf{Run 25}  & \textbf{Run 26} & \textbf{Run 27}  \\[6pt]
\\
\end{tabular}
\begin{tabular}{cccc}
\includegraphics[width=0.3\textwidth]{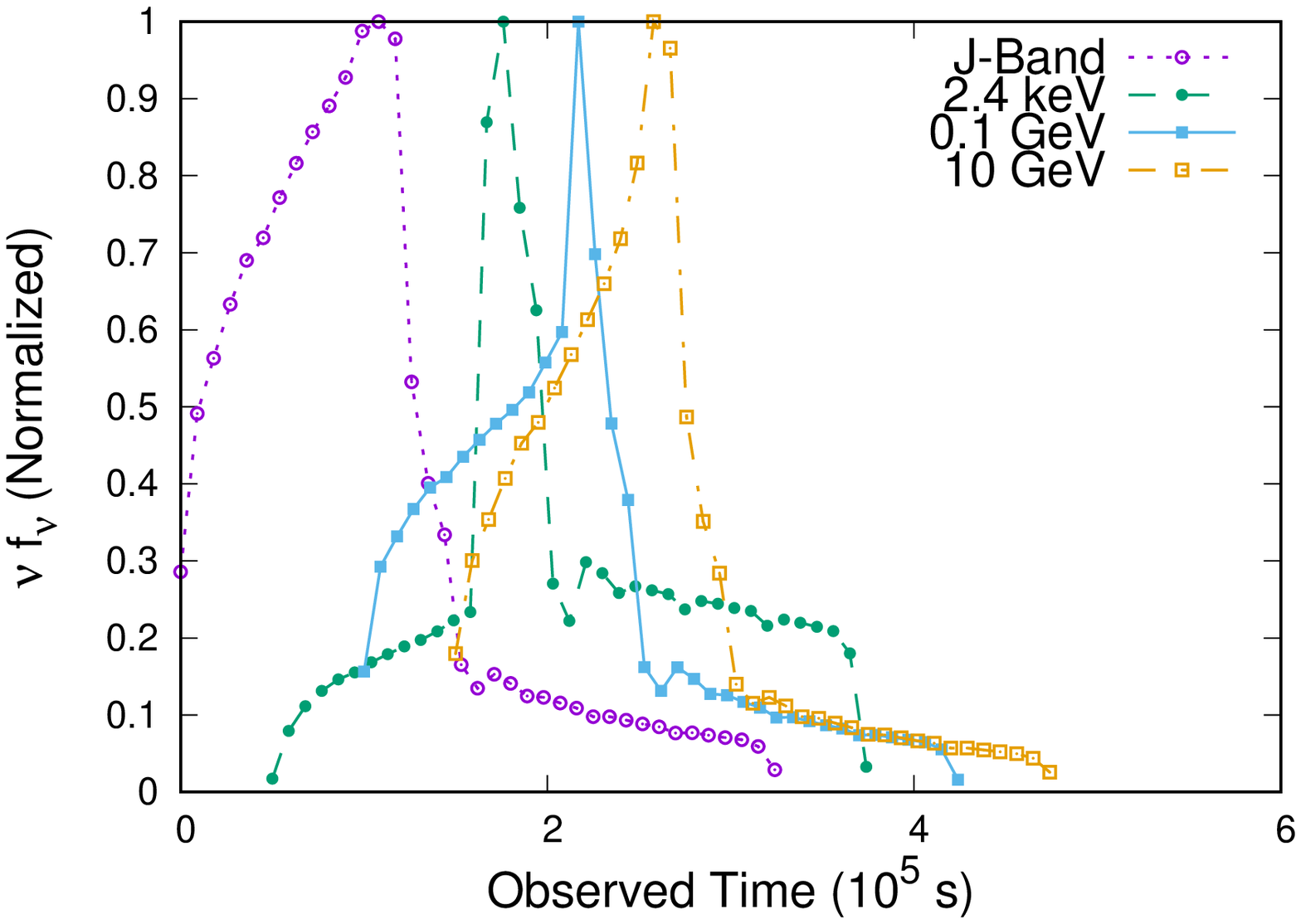} &
\includegraphics[width=0.3\textwidth]{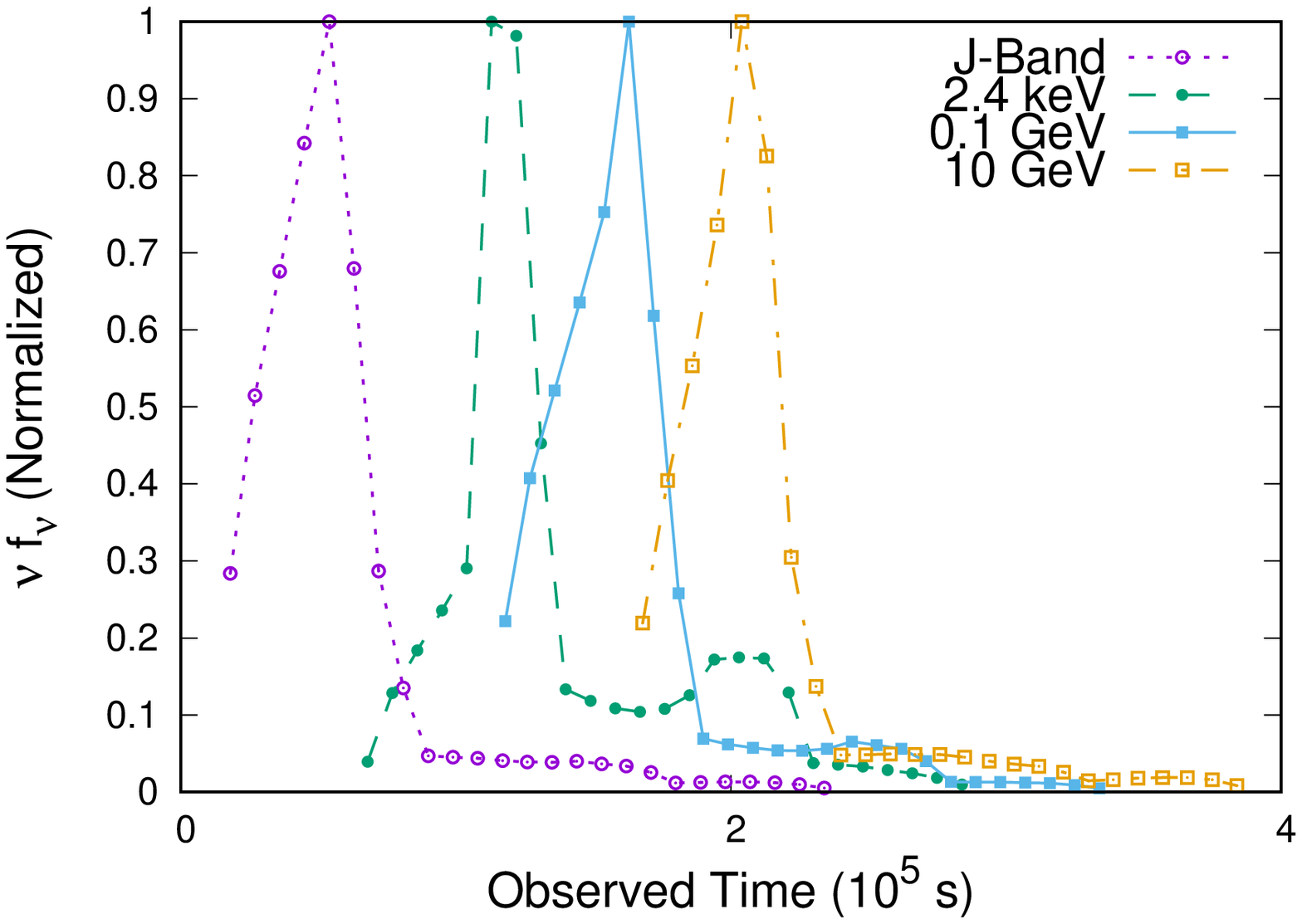} \\
\textbf{Run 28}  & \textbf{Run 29}   \\[6pt]
\end{tabular}
\caption{ Simulated light curves obtained from various runs by varying input parameters as described in Table 4 in the manuscript. The flare profiles are shown in four specific energy bands: $\rm J$-band, 2.4 keV, 0.1 GeV, and 10 GeV for each run.}
\label{fig:lcs}
\end{figure}

\twocolumn


Among the light curve profiles shown in the figure, the synchrotron-dominated optical and inverse Compton (IC)-dominated HE emission is governed completely by the presence of shocks in the system. The pulse peaks around $t^{*}_{\rm peak} = 60$ ks depending on the shock crossing time. On the other hand, X-ray light curves at 2.4 and 10 keV energies are dominated by the rising part of the SSC component. This implies that IR synchrotron photons are involved in the production of these photons through IC scattering off lower energy electrons. Such electrons remain in the system for a longer period of time. As a result, the X-ray light curves peak later than the optical and HE light curves at around $t^{*}_{\rm peak} = 70$ ks. In addition, there is a continuous build-up of late-arriving photons at scattering sites, even after the shocks exit the system, including some contribution from the declining part of the synchrotron component to the 2.4 keV pulse and from the rising part of the ECDT component to the 10 keV pulse. Hence, both pulse profiles exhibit a small hump that lasts for a short period of time after the shocks have completely exited the system. The soft $\gamma$-ray light curves at 50 MeV and 0.1 GeV are dominated by the rising part of the ECDT component with some contribution from the ECBLR emission. Since the seed photon field for this emission constitutes mostly of the IR photons of the dusty torus \citep{jos14}, similar to the X-ray pulse profiles, the soft $\gamma$-ray light curves also peak at around $t^{*}_{\rm peak} = 70$ ks. The flare profiles at these energies peak with a gradual rise and decline faster compared to that of optical and HE light curves. The pulse profile at 1 GeV, on the other hand, is a combination of ECBLR and ECDT components. As a result, it peaks at around the same time as that of optical and HE light curves and attains a plateau for a short period of time before declining rapidly. 

\begin{table}
	\centering
	\scriptsize
	\caption{
	Symmetry Parameter of Outbursts.}
	\label{sym_param}
	\begin{tabular}{ccccc} 
		\hline
		\hline
Run\# & J-Band & 2.4 keV & 0.1 GeV & 10 GeV \\
		\hline
1   &  0.02  &   -0.12   &  -0.44  &   -0.35	\\
2   &   0.34   &    0.40   &    0.19   &   0.20	\\
3   &   -0.58  &    -0.36  &  -0.40  &   -0.40	\\
4   &   -0.70   &    -0.61   &    0.76   &   1.01		\\
5   &    0.29   &    0.23   &    0.46   &     0.47	\\
6   &    -0.48   &    0.63   &   0.31  &   0.65	\\
7   &   0.55  &    0.63  &   -0.02   &    0.17	\\
8   &   -0.46   &   -0.23   &    0.33   &    0.94	\\
9   &    0.97   &    0.68   &    0.59   &    0.58	\\
10   &    0.06   &  -0.20  &   -0.41   &  -0.82	\\
11   &  0.13  &   -0.23   &   -0.38   &  -0.62	\\
12   &  -0.32  &   -0.11   &   -0.38   &   -0.39	\\
13   &   -0.47  &    -0.45   &   -0.21   &   -0.36	\\
14   &   0.09  &   -0.01   &   -0.39   &  -0.27	\\
15   &    -0.42  &     -0.42   &   -0.37   &  -0.37	\\	
16   &   -0.47  &     -0.26   &   -0.07   &   -0.35	\\
17   &   -0.49   &   -0.35   &   -0.38   &   -0.48	\\
18   &   0.14   &   -0.50   &   -0.43   &    -0.24	\\
19   &   -0.48   &  -0.40   &   -0.37   &   -0.46	\\
20   &   -0.53   &    -0.57   &   -0.46   &  -0.43	\\
21   &   -0.26   &  -0.05   &   -0.14   &   -0.00	\\
22   &  -0.45   &   -0.39   &   -0.48   &  -0.40	\\
23   &   0.04   &  -0.41   &  -0.40   &    -0.40	\\
24   &   0.06   &  -0.36   &   -0.31   &   -0.31	\\
25   &  0.02   &  0.15   &   -0.14   &  -0.35	\\
26   &   0.06  &   -0.38   &  -0.34   &  -0.34	\\
27   &   0.02   &   0.14   &   -0.23   &  -0.34	\\
28   &   0.14  &    -0.50   &   -0.48   &   -0.47	\\
29   &  0.10  &    0.19   &   -0.31   &  -0.23	\\

		\hline
	\end{tabular}
\end{table}

\subsection{\label{params}Variation of the Parameters of the Baseline Model}
In this subsection, we particularly look for strongly asymmetric properties in the light curves caused due to the variation of input parameters associated with jet emission. We show the flare profiles at $J$-band, 2.4 keV, 0.1 GeV, and 10 GeV for each run in Figure \ref{fig:lcs}. We select these wave bands because the four primary emission processes, namely, synchrotron, SSC, EC-DT, and EC-BLR, are represented by them. Since we are specifically looking at the symmetry property of the flares, we have normalized the flux at each of these wavelengths by their maximum to make visual comparison easier. We have shifted the 2.4 keV, 0.1 GeV, and 10 GeV light curves by 50 ks, 100 ks, and 150 ks, respectively on the time axis for clarity.
The variation of the parameters are done according to Table \ref{paramlist} and the symmetry parameters for each case at those four wavelengths are given in Table \ref{sym_param}. 

\subsubsection{Bulk Lorentz Factor of Inner and Outer Shells}
Varying $\Gamma_{\rm i}$ or $\Gamma_{\rm o}$ changes the $\Gamma_{\rm sh}$, which in turn affects the Doppler boosting factor (D) and the strength of the magnetic field ($B_{\rm fs/rs}$) of the emission region. Decreasing $\Gamma_{\rm i}$ (run 3) decreases the $\Gamma_{\rm sh}$, D, $B_{\rm fs/rs}$, normalization factor of the electron injection function, and $\Gamma^{\prime}_{\rm fs/rs}$ of the system causing shocks to leave the emission region later. But it simultaneously increases an electron's Larmor radius. 
For a first-order Fermi acceleration scenario, the acceleration timescale is directly proportional to the Larmor radius of the electron \citep{Tam09}. Thus a larger Larmor radius results in slower acceleration of the electrons. Hence, flares across all energy bands have longer rise times than those of run 1. 
The symmetry property is approximately similar to run 1 in all wavelengths except at $J$-band where it is more asymmetric than run 1 although $|\xi| < 0.6$. The IR and optical emission get contribution from the SSC component in addition to the synchrotron radiation, which leads to a second hump in all the profiles.


In the case of increasing $\Gamma_{\rm o}$ (run 4) of the outer shell, the $\Gamma_{\rm sh}$, \textit{D}, width of the emission region, and an electron's Larmor radius increase but the $B_{\rm fs/rs}$, normalization factor of the electron injection function, and $\Gamma^{\prime}_{\rm fs/rs}$ decrease. As a result, shocks propagate through the region slowly, causing electrons to take longer to accelerate to high energies. Hence, optical and X-ray flares last longer compared to their run 1 counterparts, and get a substantial contribution from the SSC component that results in a second peak very similar to run 3. 
However, in the case of $\gamma$-rays, flares have a sharp rise but a very gradual decline compared to run 1. The sharp rise happens due to an increased amount of boosting of external seed photon field. On the other hand, they start to decline shortly after that even though the shocks are still present in the system due to a lower density of highest-energy electrons that are required to produce flares at these energies. At the same time, the decay is very gradual because the relative motion of the shocks through the system is slower, which leads to a scenario where the acceleration phase competes with the cooling phase of the highest-energy electrons leading up to a gradual decline. In this case, $|\xi| > -0.6$ in all bands, making them asymmetric.

\subsubsection{Widths of Inner and Outer Shells}
Increasing the width of the inner shell (run 6) increases the total width of the emission region and an electron's Larmor radius but slightly decreases the $\Gamma_{\rm sh}$, D, and the value of $B_{\rm fs/rs}$. As a result both shocks remain in the system for a slightly longer time than the base set. Consequently, the pulses have a much more gradual decay and a strongly asymmetric profile with $|\xi| > 0.6$ in the 2.4~KeV and 10~GeV energies. 



Flare profiles are mostly symmetric in the case of decreasing the width of the inner shell (run 7). 
At 2.4 keV, the flare is a combination of both synchrotron and SSC processes resulting in a plateau-like profile and a longer decay tail. On the other hand, increasing the width of the outer shell, $\Delta_{\rm o}$, in run 8 increases the overall width and the BLF of the emission region but decreases its magnetic field strength. This combination of parameters are quite similar to that of run 6, and so as expected they produce fairly asymmetric light curve profiles similar to run 6 in $J$-band, 0.1 GeV and 10 GeV. However, in 2.4 KeV, the flare is different since both synchrotron and SSC processes are in effect at this energy.


There is a general asymmetry that is maintained in the pulse profiles at all wavelengths in run 9. 
 in which the width of the outer shell has been decreased. This is because of the amount of time each shock spends in its respective emission region, which is quite disparate for run 9 compared to that of runs 1 \& 7. 


\begin{figure}
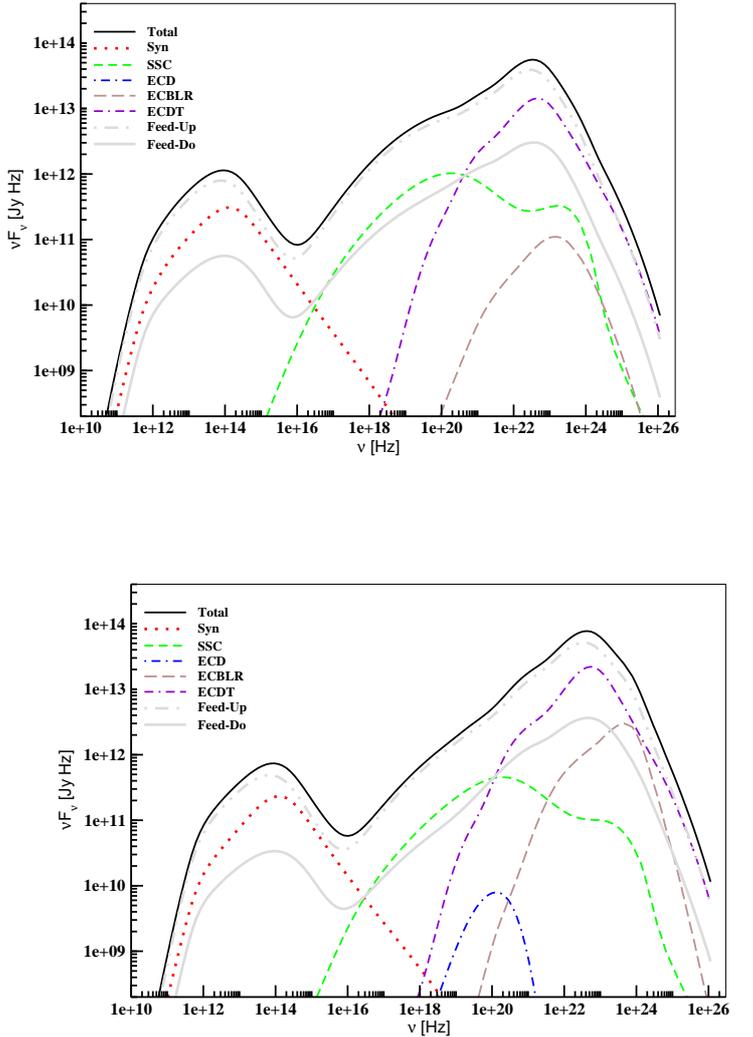

\hspace*{-0.3in}
{\includegraphics[width=0.5\textwidth]{file-figures-Joshi/sed11.eps}}\hfill 

\hspace*{-0.3in}

\vspace{.5in}

{\includegraphics[width=0.5\textwidth]{file-figures-Joshi/sed23.eps}}
\caption{Spectral energy distributions. Top: Run 11; Bottom: Run 23.}
\label{sed1123}
\end{figure}

\subsubsection{Location of the Emission Region and BLR luminosity}
Position of the emission region is within the BLR in run 10 and outside the BLR but within the dusty torus in run 11. On the other hand, in run 22 (Figure \ref{fig:lcs}) the BLR luminosity is increased while it is decreased in run 23. The symmetry property of the optical and X-ray light curves are not affected by these changes as they are not generated by the EC process. However, that of 0.1 GeV pulses do not show significant change either, contrary to expectation. On the other hand, $\xi$ value of the 10 GeV pulse becomes more negative in both run 10 and 11, indicating a faster decay. 
We show the time-averaged SEDs with all the constituent radiative components for the above runs in Figure \ref{sed1123}. It is evident from the figure that when the supply of external seed photons from the BLR decreases, the EC-BLR cooling becomes weaker and the EC-DT and the SSC processes become more dominant. Hence, the decay time does not increase significantly as expected from the analytical calculation in Section \ref{analysis:short} assuming a single emission mechanism at a time. 

As can be seen from Table \ref{sym_param} and Figure \ref{fig:lcs}, 
while changes in the other parameters do cause changes in the value of the symmetry parameter compared to the base values, none of the changes cause strong asymmetry in the resultant light curves in any wave band. In many cases there is a long tail in the light curves at various wave bands, i.e., the emission decays to a few percent of its peak value relatively quickly but it takes much longer for it to become zero or negligible. These tails have not been included in our calculation of the $\xi$ value as such longer decay time will not be observable in most cases due to the very small flux values at the tails.




\section{Conclusion}
In this paper we have analyzed the GeV light curves of 10 blazars and optical R-band light curves of 9 blazars spanning 8 yr --- 2008-2015 --- to identify the long-term outbursts. Furthermore, we have analyzed 26 short-term blazar flares at GeV energies where the flux increased by a factor of a few within $\sim$hr-day. We have particularly studied the symmetry properties of these outbursts at different timescales. In order to interpret the results of our analyses we have employed a theoretical shock-in-jet model of blazar emission. Comparing the properties of the short-term flares at various wave-bands generated by the above model with our results obtained from the observed data, we have inferred the physical parameters of the jet and its dynamics during those outbursts. Below we list our major conclusions: \\
1. Long-term ($\sim$~weeks to months) outbursts are mostly symmetric, i.e., have similar rise and decay timescale, at both GeV energies and optical $R$-band. This result is consistent with what \citet{cha12} found in a smaller sample of blazar outbursts. This indicates that the long-term flares are dominated by the time taken by radiation or a disturbance to pass through the emission region, and not cooling time of radiating particles. \\
2. A larger fraction of the short-term ($\rm \sim day$) flares are asymmetric. However, the fraction with positive values of the $\xi$ parameter (faster rise-slower decay) is similar to that with negative $\xi$ (slower rise-faster decay). \\
3. Outbursts with slower rise than decay time (negative $\xi$) may be due to gradual acceleration of particles, as a result of which radiation cooling timescale is shorter than that for acceleration. In our numerical model, the particle acceleration process is first-order Fermi mechanism, causing the rise of flares to take a few hr to a day. \\
4. Outbursts with faster rise than decay time (positive $\xi$) could be due to larger or smaller than base value of the bulk Lorentz factor (BLF) or width of the inner or outer shell, as demonstrated. The base value or changes in other parameters listed in Table \ref{basesetlist} \& \ref{paramlist} did not produce outbursts with significantly slower decay than rise time. \\
5. In some of the cases, a longer cooling time will occur due to the change of a single parameter if all other parameters stay constant. For example, if the supply of the external photons decreases due to the decrease in the BLR luminosity or increase in the distance between the emission region and BLR. However, our model demonstrates that the effective cooling time remains approximately the same as cooling due to synchrotron self-Compton process or external photons from the dusty torus compensates for the above effect, and consequently the decay time and hence the $\xi$ value does not change significantly. \\ 
6. This work carries out an extensive and systematic study, which can be used to estimate relevant geometric and physical parameters of the jet from the symmetry property of any observed short-term flares in the context of the \textit{MUZORF} model. \\

NR thanks Inter University Center of Astronomy and Astrophysics (IUCAA) for their hospitality and providing computation facilities for four months during which a part of this project was carried out. MJ thanks the Boston University Shared Computing Cluster (SCC) located at The Massachusetts Green High Performance Computing Center (MGHPCC) for providing high-end computational facilities where all the simulations were performed. MJ thanks Dr. Mahito Sasada for helpful discussion. We thank Professor Puragra Guhathakurta for detailed comments and suggestions. RC received support from the UGC start-up grant.












\bsp	
\label{lastpage}
\end{document}